\documentclass[prd,aps,twocolumn,nofootinbib,floatfix,twocolumn]{revtex4-2}

\pdfoutput=1
\usepackage{graphicx}
\usepackage{bm}
\usepackage{amsmath}
\usepackage{amsthm}
\usepackage{amsfonts}
\usepackage{xfrac}
\usepackage{verbatim}
\usepackage{slashed}

\newcommand{\pt}{p_\textrm{T}}
\newcommand{\kt}{k_\textrm{T}}
\newcommand{\Et}{E_\textrm{T}}

\usepackage{array}
\newcommand{\PreserveBackslash}[1]{\let\temp=\\#1\let\\=\temp}
\newcolumntype{C}[1]{>{\PreserveBackslash\centering}p{#1}}
\newcolumntype{R}[1]{>{\PreserveBackslash\raggedleft}p{#1}}
\newcolumntype{L}[1]{>{\PreserveBackslash\raggedright}p{#1}}

\newcommand{\appropto}{\mathrel{\vcenter{
  \offinterlineskip\halign{\hfil$##$\cr
    \propto\cr\noalign{\kern2pt}\sim\cr\noalign{\kern-2pt}}}}}

\usepackage{tikz}
\usetikzlibrary{shapes.geometric, arrows.meta, positioning}
\tikzstyle{startstop} = [rectangle, rounded corners, minimum width=15mm, minimum height=7.5mm,text centered, draw=black, fill=none]
\tikzstyle{io} = [trapezium, trapezium left angle=70, trapezium right angle=110, minimum width=15mm, minimum height=7.5mm, text centered, draw=black, fill=none]
\tikzstyle{process} = [rectangle, minimum width=15mm, minimum height=7.5mm, text centered, draw=black, fill=none]
\tikzstyle{decision} = [diamond, minimum width=15mm, minimum height=7.5mm, text centered, draw=black, fill=none]
\tikzstyle{arrow} = [thick,->,>=stealth]

\makeatletter
\let\ftype@table\ftype@figure
\makeatother

\begin{document}

\title{Jet SIFT-ing:\\a new scale-invariant jet clustering algorithm for the substructure era}

\author{Andrew J.~Larkoski$^{1}$}
\author{Denis Rathjens$^{2}$}
\author{Jason Veatch$^{3}$}
\author{Joel W. Walker$^{4}$}
\affiliation{$^1$~Department of Physics and Astronomy, University of California, Los Angeles, CA 90095, USA}
\affiliation{$^2$~Mitchell Institute for Fundamental Physics and Astronomy, \\
Department of Physics and Astronomy, Texas A\&M University, College Station, TX 77843, USA}
\affiliation{$^3$~Department of Physics, California State University East Bay, Hayward, CA 94542, USA}
\affiliation{$^4$~Department of Physics and Astronomy, Sam Houston State University, Huntsville, TX 77341, USA}

\begin{abstract}
We introduce a new jet clustering algorithm named SIFT (Scale-Invariant Filtered Tree)
that maintains the resolution of substructure for collimated decay products at large boosts.
The scale-invariant measure combines properties of $\kt$ and anti-$\kt$ by preferring early association of
soft radiation with a resilient hard axis, while avoiding the specification of a fixed cone size.
Integrated filtering and variable-radius isolation criteria block assimilation of soft wide-angle radiation and provide a halting condition.
Mutually hard structures are preserved to the end of clustering, automatically generating a tree of subjet axis candidates.
Excellent object identification and kinematic reconstruction for multi-pronged resonances are realized across more than an order of magnitude in transverse energy.
The clustering measure history facilitates high-performance substructure tagging, which we quantify with the aid of supervised machine learning.
These properties suggest that SIFT may prove to be a useful tool for the continuing study of jet substructure.
\end{abstract}

\maketitle

\section{Introduction\label{sct:introduction}}

The collider production of an isolated partonic object bearing uncanceled strong nuclear charge
is immediately followed by a frenzied showering of soft and collinear radiation with 
a complex process of recombination into meta-stable color-singlet hadronic states.  In order to compare
theoretical predictions for hard scattering events against experimental observations, it is necessary to systematically
reassemble these ``jets'' of fragmentary debris into a faithful representation of their particle source.
The leading clustering algorithm serving this purpose at the Large Hadron Collider (LHC) is anti-$\kt$~\cite{Cacciari:2008gp},
which is valued for yielding regular jet shapes that are simple to calibrate.

It can often be the case that clustering is complicated by early decays of a heavy unstable state into
multiple hard prongs, e.g., as $(W^+ \to u\, \bar{d})$, or $(t \to W^+ b \to u\, \bar{d}\, b)$.
This additional structure can be of great benefit for tagging presence of the heavy initial state.
However, identification will be confounded if essential constituents of distinct prongs
either remain uncollected, or are merged together into a joint assemblage.
Standard practice is to err in the latter direction, via construction of a large-radius
jet that is intended to encapsulate all relevant showering
products, and to subsequently attempt recovery of the lost ``substructure''
using a separate algorithm such as $N$-subjettiness~\cite{Thaler:2010tr}.

The casting of this wide net invariably also sweeps up extraneous low-energy radiation at large
angular separations and successful substructure tagging typically hinges on secondarily
``grooming'' large-radius jets with a technique such as Soft Drop~\cite{Larkoski:2014wba}.
Furthermore, the appropriate angular coverage is intrinsically dependent
upon the process being investigated, since more boosted parents will
tend to produce more collimated beams of children.
Accordingly, such methods are conventionally tuned to target
specific energy scales for maximal efficacy.

In this paper, we introduce a new jet clustering algorithm called
SIFT (Scale-Invariant Filtered Tree) that is engineered
to avoid losing resolution of structure in the first place.
Similar considerations previously motivated development
of the exclusive XCone~\cite{Stewart:2015waa,Thaler:2015xaa} algorithm
based on minimization of $N$-jettiness~\cite{Stewart:2010tn}.
Most algorithms in common use reference a fixed angular cone
size $R_0$, inside of which all presented objects
will cluster, and beyond which all exterior objects will be excluded.
We identify this parameter, and the conjugate momentum scale which it imprints,
as primary culprits responsible for the ensuing loss of resolution.

In keeping, our proposal rejects any such references
to an external scale, while asymptotically recovering successful
angular and kinematic behaviors of algorithms in the $\kt$-family.
Specifically, the SIFT prioritizes pairing objects that have hierarchically dissimilar transverse momentum scales
(i.e., one member is soft \emph{relative} to its partner),
and narrow angular separation (i.e., the members are collinear).
The form we are lead to by these considerations is quite similar to a 
prior clustering measure named Geneva~\cite{BETHKE1992310}, although
our principal motivation (retention of substructure) differs
substantively from those that applied historically.

We additionally resolve a critical fault that otherwise precludes
the practical application of radius-free measures such as Geneva,
namely the tendency to gather uncorrelated soft radiation at wide
angular separations.  This is accomplished with a novel filtering
and halting prescription that is motivated by Soft Drop,
but defined in terms of the scale-invariant measure itself and
applied to each candidate merger during the initial clustering.

The final-state objects retained after algorithm termination are isolated
variable-large-radius jets that dynamically bundle decay products of
massive resonances in response to the \emph{process} scale.
Since mutually hard prongs tend to merge last (this behavior
is very different from that of anti-$\kt$), the end stages
of clustering densely encode information indicating
the presence of substructure.  Accordingly, we introduce the concept
of an $N$-subjet tree, which represents the ensemble of a
merged object's projections onto $N$ axes,
as directly associated with the clustering progression from $N$ prongs to $1$.

We demonstrate that the subjet axes obtained
in this manner are effective for applications such as
the computation of $N$-subjettiness.  Additionally,
we confirm that they facilitate accurate and sharply-peaked
reconstruction of associated mass resonances.
Finally, we show that the sequential evolution history
of the SIFT measure across mergers is itself an excellent
substructure discriminant.  As quantified with the
aid of a Boosted Decision Tree (BDT), we find that
it significantly outperforms the benchmark approach to
distinguishing one-, two-, and three-prong events using
$N$-subjettiness ratios, especially in
the presence of a large transverse boost.

The outline of this work is as follows.
Section~\ref{sct:algorithm} presents a master sequential jet clustering algorithm into which SIFT and the $\kt$ prescriptions may be embedded.
Sections~\ref{sct:siftmeasure}--\ref{sct:njettree} define the SIFT algorithm in terms of its scale-invariant measure
(with transformation to a coordinate representation), filtering and isolation criteria, and final-state $N$-subjet tree objects.
Section~\ref{sct:comparison} visually contrasts clustering priorities, soft radiation catchment, and halting status against common algorithms.
Sections~\ref{sct:reconstruction}--\ref{sct:tagging} comparatively assesses SIFT's performance in various applications,
relating firstly to jet resolution and mass reconstruction, and secondly to the tagging of structure.
Section~\ref{sct:theory} addresses computability, infrared and collinear safety, and $\mathcal{O}(1)$ deviations from recursive safety.
Section~\ref{sct:conclusion} concludes and summarizes.
Appendix~\ref{sct:collidervars} provides a pedagogical review of hadron collider coordinates.
Appendix~\ref{sct:software} describes available software implementations of the SIFT algorithm
along with methods for reproducing current results and plans for integration
with the {\sc FastJet}~\cite{Cacciari:2011ma,Cacciari:2005hq} contributions library.

\section{The Master Algorithm\label{sct:algorithm}}

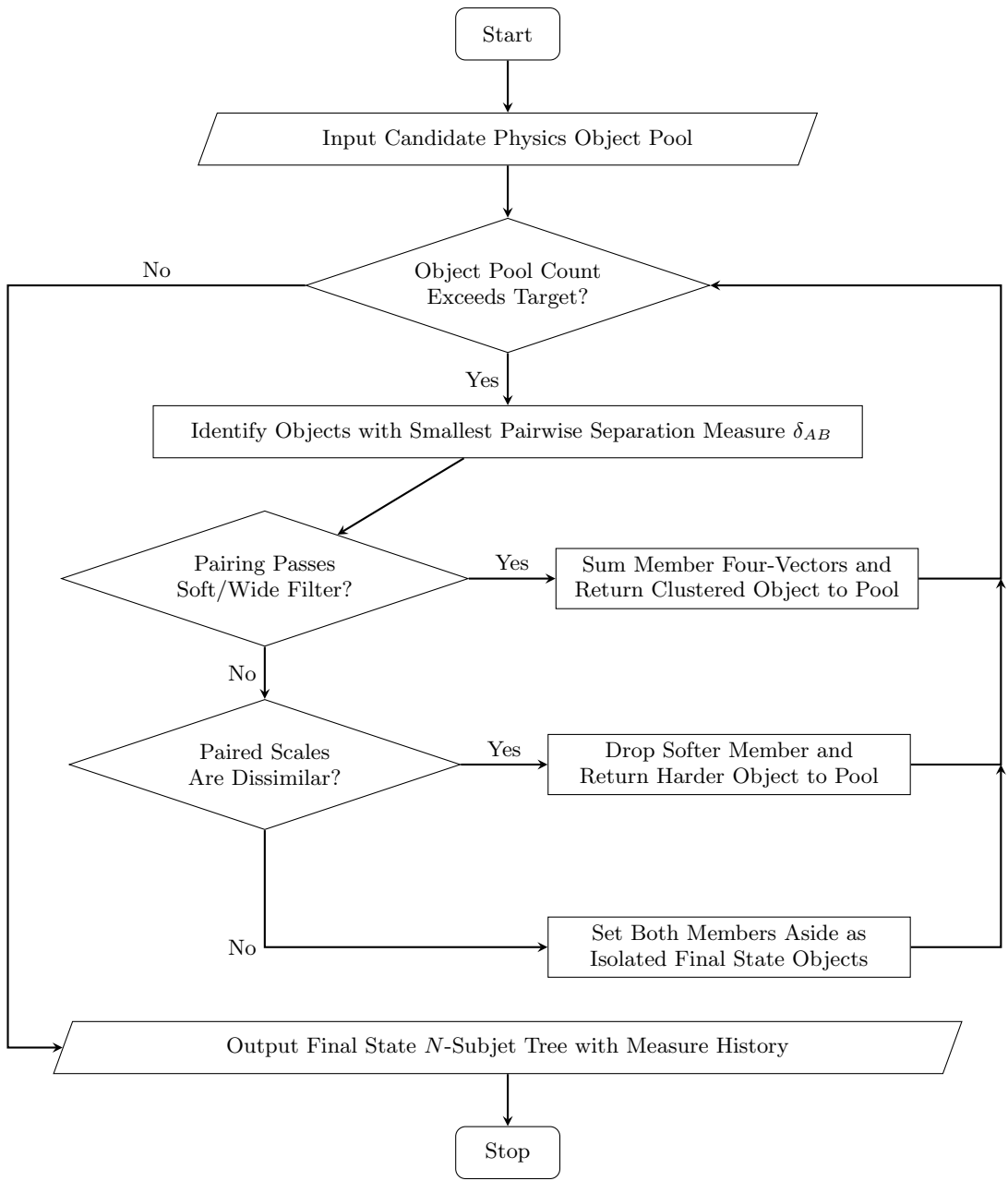
\begin{figure*}[ht!]
\centering
\begin{tikzpicture}[node distance=7.5mm]
\node (start) [startstop] {Start};
\node (in1) [io, below=of start] {Input Candidate Physics Object Pool};
\node (dec1) [decision, aspect=3, text width=30mm, below=of in1, yshift=0mm, xshift=0mm] {Object Pool Count\\Exceeds Target?};
\node (pro1) [process, below=of dec1] {~~~~Identify Objects with Smallest Pairwise Separation Measure $\delta_{AB}$~~~~};
\node (dec2) [decision, aspect=3, text width=30mm, below=of pro1, yshift=0mm, xshift=-35mm] {Pairing Passes\\Soft/Wide Filter?};
\node (pro2) [process, text width=50mm, right=of dec2, yshift=0mm, xshift=5mm] {Sum Member Four-Vectors and\\Return Clustered Object to Pool};
\node (dec3) [decision, aspect=3, text width=30mm, below=of dec2, yshift=0mm, xshift=0mm] {Paired Scales\\Are Dissimilar?};
\node (pro3) [process, text width=50mm, right=of dec3, yshift=0mm, xshift=5mm] {Drop Softer Member and\\Return Harder Object to Pool};
\node (pro4) [process, text width=50mm, below=of pro3, yshift=-10mm, xshift=0mm] {Set Both Members Aside as\\Isolated Final State Objects};
\node (out1) [io, below=of dec3, yshift=-20mm, xshift=+35mm] {Output Final State $N$-Subjet Tree with Measure History};
\node (stop) [startstop, below=of out1] {Stop};
\draw [arrow] (start) -- (in1);
\draw [arrow] (in1) -- (dec1);
\draw [arrow] (dec1) -- node[anchor=east] {Yes} (pro1);
\draw [arrow] (dec1) -- node[anchor=south] {No} ++(-72mm,0) |- (out1);
\draw [arrow] (pro1) -- (dec2);
\draw [arrow] (dec2) -- node[anchor=south] {Yes} (pro2);
\draw [arrow] (dec2) -- node[anchor=east] {No} (dec3);
\draw [arrow] (dec3) -- node[anchor=south] {Yes} (pro3);
\draw [arrow] (dec3) |- node[anchor=east] {No} (pro4);
\draw [arrow] (pro2) -- ++(+38mm,0) coordinate (m1) |- (dec1);
\draw [arrow] (pro3) -| coordinate(m2) (m1);
\draw [arrow] (pro4) -| (m2);
\draw [arrow] (out1) -- (stop);
\end{tikzpicture}
\caption{\footnotesize
Generalized logical flow chart representing a family of sequential jet clustering algorithms
defined by a distance measure $\delta_{AB}$, along with a specification for
filtering and isolation criteria and/or a target final-state jet count for exclusive clustering.}
\label{fig:flow}
\end{figure*}

This section summarizes the global structure of a general sequential
jet clustering algorithm, as illustrated in FIG.~\ref{fig:flow}.
Examples are provided of how standard algorithms fit into this framework.
These examples will be referenced subsequently to motivate the SIFT
algorithm and to comparatively assess its performance.

To begin, a pool of $N$ low-level physics objects
(e.g., four-vector components of track-assisted calorimeter hits)
is populated, typically including reconstructed hadrons as well as
photons and light leptons that fail applicable hardness, identification, or isolation criteria.

The main loop then begins by finding the two most-proximal objects ($A,B$),
as defined via minimization of a specified distance measure $\delta_{AB}$ over each candidate pairing.
For example, the anti-$\kt$ algorithm is a member (with index $n = -1$) of a broader class
of algorithms that includes the earlier $\kt$~\cite{Catani:1993hr,Ellis:1993tq} (with $n = +1$)
and Cambridge-Aachen~\cite{Dokshitzer:1997in,Wobisch:1998wt} (with $n = 0$) formulations,
corresponding to the measure $\delta_{AB}^{k_{\rm T},n}$ defined as:
\begin{equation}
\delta_{AB}^{k_{\rm T},n} \equiv
{\rm min} \bigg[ \, {\left(\pt^A\right)}^{2n} \!\!,\, {\left(\pt^B\right)}^{2n} \, \bigg] \times
{\left( \frac{\Delta R_{AB}}{R_0} \right)}^{\!2}
\label{eq:ktmeasure}
\end{equation}

The quantity ${( \Delta R_{AB} )}^2 \equiv {( \Delta \eta)}^2 + {( \Delta \phi )}^2$
expresses ``geometric'' adjacency as a Cartesian norm-square of differences
in pseudo-rapidity $\eta$ and azimuthal angle $\phi $ (cf. Appendix~\ref{sct:collidervars})
relative to the maximal cone radius $R_0$.
The $(n = 0)$ scenario prioritizes small values of $\Delta R_{AB}$ without reference to the transverse momentum $\pt$.
A positive momentum exponent $(n = +1)$ first associates pairs wherein at least one member is very soft,
naturally rewinding the chronology of the showering process.
A negative momentum exponent $(n = -1)$ favors pairs wherein at least one
member represents hard radiation directed away from the beamline.

Once an object pair has been selected, there are various ways
to handle its members.  Broadly, three useful alternatives are
available, identified here as clustering, dropping, and isolating.
The criteria for distinguishing between these actions are 
an important part of any algorithm's halting condition.
Clustering means that the objects ($A,B$) are merged, usually
via a four-vector sum ($p^\mu_{AB} = p^\mu_{A} + p^\mu_{B}$),
and replaced by this merged object in the pool.
Typically, clustering proceeds unless the applicable
angular separation is too great and/or the momentum scales
of the members are too dissimilar.

For pairings failing this filter,
two options remain.  Dropping, wherein the softer member is
set aside (it may literally be discarded or rather
reclassified as a final-state object) while the harder
member is returned to the object pool, sensibly applies
when momentum scales are hierarchically imbalanced.
Conversely, a symmetric treatment of both members
is motivated when momentum scales are similar,
and isolating involves mutual reclassification 
as final-state objects.  Exclusive algorithms
(which guarantee a fixed ending count $N_{\rm exc}$ of jets)
typically always cluster, although any construction
that reduces the net object count by
one unit per iteration is consistent\footnote{
A valid example is clustering plus dropping without isolation.}.

To complete the prior example, conventional implementations of $\kt$-family
clustering do not discard objects or collectively isolate final-state pairs.
However, they do singly reclassify all objects with no partners
nearer than $R_0$ in $\Delta R$ as final-state jets.
This behavior may be conveniently embedded into the master algorithm flow
by also allowing each object in the active pool to pair one at a time with the ``beam''
and associating it with the alternative measure $\delta_{\rm beam}^{k_{\rm T},n}=\pt^{2n}$.
The dropping criterion is then adapted to set aside any object for which
this beam distance is identified as the global minimum.
Note that such objects are indeed guaranteed to have
no neighbors inside a centered cone of size $R_0$.

The described loop repeats until the number of objects remaining in the
active pool reaches a specified threshold.  In the context of an exclusive
clustering mode this would correspond to a target count $N_{\rm exc}$,
whereas continuation otherwise simply requires the presence of multiple ($N>1$) objects.
Many clustering algorithms, including SIFT, may be operated in either of these modes.
Upon satisfaction of the halting criteria, the list of final-state objects
is returned along with a record of the clustering sequence and associated measure history.

The SIFT algorithm will now be systematically developed
by providing specific prescriptions for each element of the master algorithm.
The scale-invariant measure $\delta_{AB}$ is defined in Section~\ref{sct:siftmeasure}
and provided with an intuitive geometric form in Section~\ref{sct:geometricform}.
The filtering and isolation criteria are itemized in Section~\ref{sct:filtering}.
The $N$-subjet tree is introduced in Section~\ref{sct:njettree}.

\section{The Scale-Invariant Measure~\label{sct:siftmeasure}}

This section establishes the SIFT clustering measure $\delta_{AB}$ and
places it in the context of similar constructs from the literature.
Our principal objective is to develop an approach that is intrinsically
resilient against loss of substructure in boosted event topologies,
i.e., which maintains resolution of collimated radiation
associated with distinct partonic precursors in the large-$\pt$ limit.

We identify specification of an angular size parameter $R_0$ as the primary culprit imposing
a conjugate momentum scale dependence on the performance of traditional approaches to jet clustering.
In pursuit of a scale-independent algorithm,
we require that the clustering measure be free of any such factor.
Nevertheless, it is desirable to asymptotically
recover angular and kinematic characteristics
of existing successful approaches such as anti-$\kt$ (cf. Eq.~\ref{eq:ktmeasure} with $n = -1$),
and we thus seek out proxies for its dominant behaviors.
Specifically, these include a preference for pairs with close angular proximity,
and a preference for pairs where one member carries a large transverse boost.
For the former purpose, we invoke the mass-square difference, defined as follows:
\begin{align}
\Delta m_{AB}^{2} &\equiv (p^{\mu}_A + p^{\mu}_B)^2 - m_{A}^{2} - m_{B}^{2}
\,=\, 2 p^{\mu}_A p_{\mu}^{B}
\label{eq:dmsq}\\
& \simeq 2 E^{A}E^{B}\times(1-\cos\Delta \theta_{AB})
\simeq E^{A} E^{B}\Delta \,\theta_{AB}^{2} \nonumber
\end{align}

The property that small mass-square changes correlate with collinearity of
decay products was similarly leveraged by the early JADE~\cite{Bethke:1988zc} algorithm.
For the latter purpose, we turn to suppression in a denominator by the summed transverse energy-square:
\begin{align}
{\textstyle \sum}\, \Et^2 &\equiv (\Et^{A})^{2} + (\Et^{B})^{2}
\label{eq:etsq} \\
\Et^2 &\equiv \pt^2 + m^2 = E^2 - p_z^2
\nonumber
\end{align}

The summation plays a role similar to that of the ``min'' criterion in Eq.~(\ref{eq:ktmeasure}),
in the case that the pair of objects under consideration is very asymmetrically boosted.
The choice of $\Et^2$ over simply $\pt^2$ prevents a certain
type of divergence, as it is possible for the vector quantity $\vec{p}_{\rm T}$
to cancel during a certain phase of the clustering, but not without generation of mass.
All together, the simple prescription for the proposed algorithm is to sequentially
cluster indexed objects $A$ and $B$ corresponding to the smallest pairwise value 
of the following expression:
\begin{equation}
\delta_{AB} \equiv
\frac{\Delta m_{AB}^2}{(\Et^{A})^{2}+(\Et^{B})^{2}}
\label{eq:siftmeasure}
\end{equation}

Since this measure represents our default context,
we write it without an explicit superscript label for simplicity.
We note that it consists of the dimensionless ratio of a Lorentz invariant
and an invariant under longitudinal boosts, and that it manifestly possesses the desired
freedom from arbitrary externally-specified scales.
For comparison, the JADE clustering measure is:
\begin{equation}
\delta_{AB}^{\rm JADE} \equiv \frac{E_{A}E_{B}}{s}\times\left(1-\cos\Delta \theta_{AB}\right)
\label{eq:JADE}
\end{equation}

Having been designed for application at a lepton collider,
JADE~\cite{JADE:1986kta} references total energies ($E_A,E_B$)
and spherically symmetric angular separations $\Delta \theta_{AB}$
rather than cylindrical quantities.
It also treats all merger candidates as massless,
i.e., it uniformly factors ($\vert\vec{p}\vert\Rightarrow E$)
out of the angular dependence.
Although Eq.~(\ref{eq:JADE}) is dimensionless and avoids referencing
a fixed cone size, it is not scale-invariant, due to normalization
against the global Mandelstam center-of-momentum energy $\sqrt{s}$.
This distinction is amplified in a hadron collider context,
where the partonic $\sqrt{\hat{s}}$ and laboratory center-of-momentum
frames are generally not equivalent.

However, the historical measure most alike to SIFT is Geneva,
which was developed as a modification to JADE:
\begin{equation}
\delta_{AB}^{\rm Geneva} \equiv \frac{8}{9} \frac{E_{A}E_{B}}{{(E_A+E_B)}^2}\times\left(1-\cos\Delta \theta_{AB}\right)
\label{eq:geneva}
\end{equation}

Like Eq.~(\ref{eq:siftmeasure}), Eq.~(\ref{eq:geneva}) achieves scale-invariance
by constructing its denominator from dimensionful quantities local to the clustering process.
There are also several differences between the two forms, which are summarized here and explored further
in Section~\ref{sct:geometricform}.  Variations in the dimensionless normalization are irrelevant\footnote{
Geneva was introduced with a coefficient of $8/9$ to match the normalization of JADE in the limit of three hard prongs.}.
The substitution of cylindrical for spherical kinematic quantities is relevant, albeit trivial.
Exchanging the squared sum for a sum of squares causes the SIFT measure fall off more sharply than the Geneva measure as the momentum scales of the clustering candidates diverge.  But, the most
critical distinguishing feature is that SIFT is sensitive to accumulated mass.

Despite core similarities at measure level, the novel filtering and isolation criteria 
described in Section~\ref{sct:filtering} are responsible for essential divergences
in the properties of final-state objects produced by SIFT relative to those from Geneva.
In particular, we will show that the $N$-subjet tree introduced in Section~\ref{sct:njettree}
is effective for reconstructing hard objects and tagging the presence of substructure
at large transverse boosts without the need for additional de/reclustering or post-processing.

\begin{figure*}[ht!]
\centering
\hspace{7pt}\includegraphics[width=.95\textwidth]{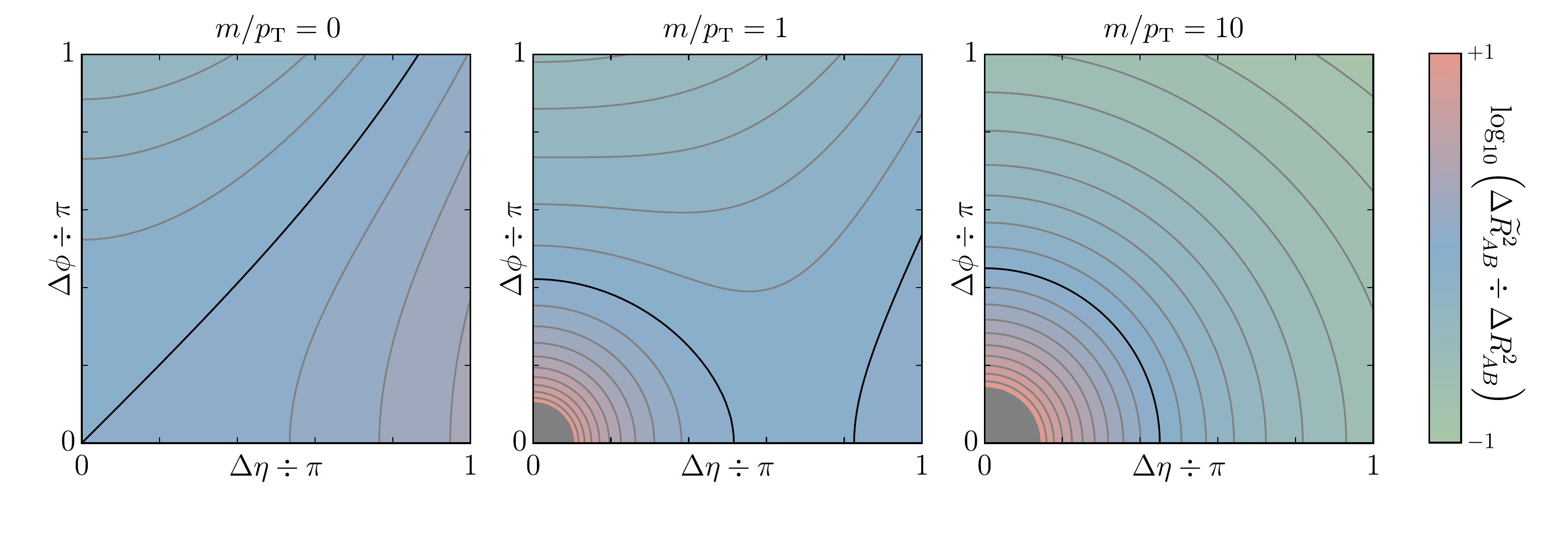}
\vspace{-12pt}
\caption{\footnotesize
The factor $\Delta \widetilde{R}^2_{AB}$ is compared to
the traditional angular measure $\Delta R^2_{AB}$,
as a function of $\Delta\eta$ and $\Delta \phi$,
and for various values of ($m/\pt$).
Regions where the SIFT measure enhances (suppresses)
clustering are shown in green (red).
Blue indicates similar behavior, with
black contour placed at unity.  Grey contours mark increments
of $0.1$ in $\log_{10}(\Delta \widetilde{R}^2_{AB}\div \Delta R^2_{AB})$.}
\label{fig:drtilde}
\end{figure*}

\section{The Geometric Form\label{sct:geometricform}}

This section presents a transformation of the SIFT clustering 
measure into the geometric language of coordinate differences.
While the underlying physics is invariant under this change
of variables, the resulting expression is better suited for developing
intuition regarding clustering priorities and how they compare to those of traditional algorithms.
It will also be used to motivate and express the filtering and isolation criteria in Section~\ref{sct:filtering},
and to streamline calculations touching on computational safety in Section~\ref{sct:theory}.
Additionally, it is vital for realizing fast numerical implementations of the SIFT
algorithm that employ optimized data structures based on geometric coordinate adjacency.

The canonical form of distance measure on a manifold is
a sum of bilinear coordinate differentials $dx^{i} dx^{j}$
with general coordinate-dependent coefficients $g_{ij}$.
Measures applicable to jet clustering must be integrated,
referencing finite coordinate separations $\Delta x^i$.
The measure expressed in Eq.~(\ref{eq:siftmeasure}) does not apparently
refer to coordinate differences at all, although Eq.~(\ref{eq:dmsq})
provides a hint of how an implicit dependence of this type might manifest.
We start from the mass-square difference:
\begin{equation}
\Delta m_{AB}^2
= 2 \times \left( E^{A}E^{B} - p_z^{A} p_z^{B} - \pt^{A} \pt^{B} \cos\Delta \phi_{AB} \right)
\label{eq:dmsqdiffs}
\end{equation}

To proceed, recall that that Lorentz transformations are generated
by hyperbolic ``rotation'' in rapidity $y$.  In particular, we may
boost via matrix multiplication from the transverse frame with $y=0$,
$p_z = 0$, and $E = E_T$ to any longitudinally related frame.
\begin{equation}
\begin{pmatrix}
E \\ p_z
\end{pmatrix}
=
\begin{pmatrix}
\cosh y & \sinh y \\
\sinh y & \cosh y
\end{pmatrix}
\begin{pmatrix}
\Et \\ 0
\end{pmatrix}
=
\begin{pmatrix}
\Et \cosh y \\
\Et \sinh y \\
\end{pmatrix}
\label{eq:hyperbolic_rot}
\end{equation}

This may be used to reduce Eq.~(\ref{eq:dmsqdiffs}),
using the standard hyperbolic difference identity.
\begin{align}
& \Et^{A}\Et^{B} - p_z^{A} p_z^{B}
\label{eq:hyperbolic_diffs} \\
\quad=\,\,& \Et^{A} \Et^{B} \times \left( \cosh y^{A} \cosh y^{B} - \sinh y^{A} \sinh y^{B} \right)
\nonumber \\
\quad=\,\,& \Et^{A} \Et^{B} \times \cosh \Delta y_{AB}
\nonumber
\end{align}

The transverse energy will factor perfectly out of the
mass-square difference if the constituent four-vectors
are individually massless, i.e.,~if $(m_{A}=m_{B}=0)$.
Otherwise, there are residual coefficients $(\xi_A,\xi_B)$ defined as follows:
\begin{align}
\Delta m_{AB}^{2} &=2\, \Et^{A} \Et^{B} \, \times \left( \cosh \Delta y_{AB} -\xi^{A} \xi^{B} \cos\Delta \phi_{AB} \right)
\nonumber \\
\xi &\equiv\, \frac{\pt}{\Et}
\,=\, {\bigg(1-\frac{m^2}{\Et^2}\bigg)}^{\hspace{-2pt}+\sfrac{1}{2}}
\!\!\!=\, {\bigg(1+\frac{m^2}{\pt^2}\bigg)}^{\hspace{-2pt}-\sfrac{1}{2}}
\label{eq:dmsqdiffs2}
\end{align}

The role of the $\xi$ in Eq.~(\ref{eq:dmsqdiffs2})
is to function as a ``lever arm'' deemphasizing
azimuthal differences in the non-relativistic limit and at low $\pt$.
The precise relationship between $\Delta m_{AB}^2$ and the angular
separation $\Delta R_{AB}^2$ can also now be readily established,
referencing Taylor expansions for cosine and hyperbolic cosine in the limit.
\begin{align}
\Delta \widetilde{R}_{AB}^2
&\equiv \frac{\Delta m_{AB}^2}{\Et^{A} \Et^{B}}
\label{eq:drsqtilde} \\
&=2\, \times \left( \cosh \Delta y_{AB} -\xi^{A} \xi^{B} \cos\Delta \phi_{AB} \right)
\nonumber \\
&\simeq\, \Delta \eta_{AB}^2 + \Delta \phi_{AB}^2
\,\equiv\, \Delta R_{AB}^2
\nonumber
\end{align}

\begin{figure*}[ht!]
\centering
\hspace{7pt}\includegraphics[width=.95\textwidth]{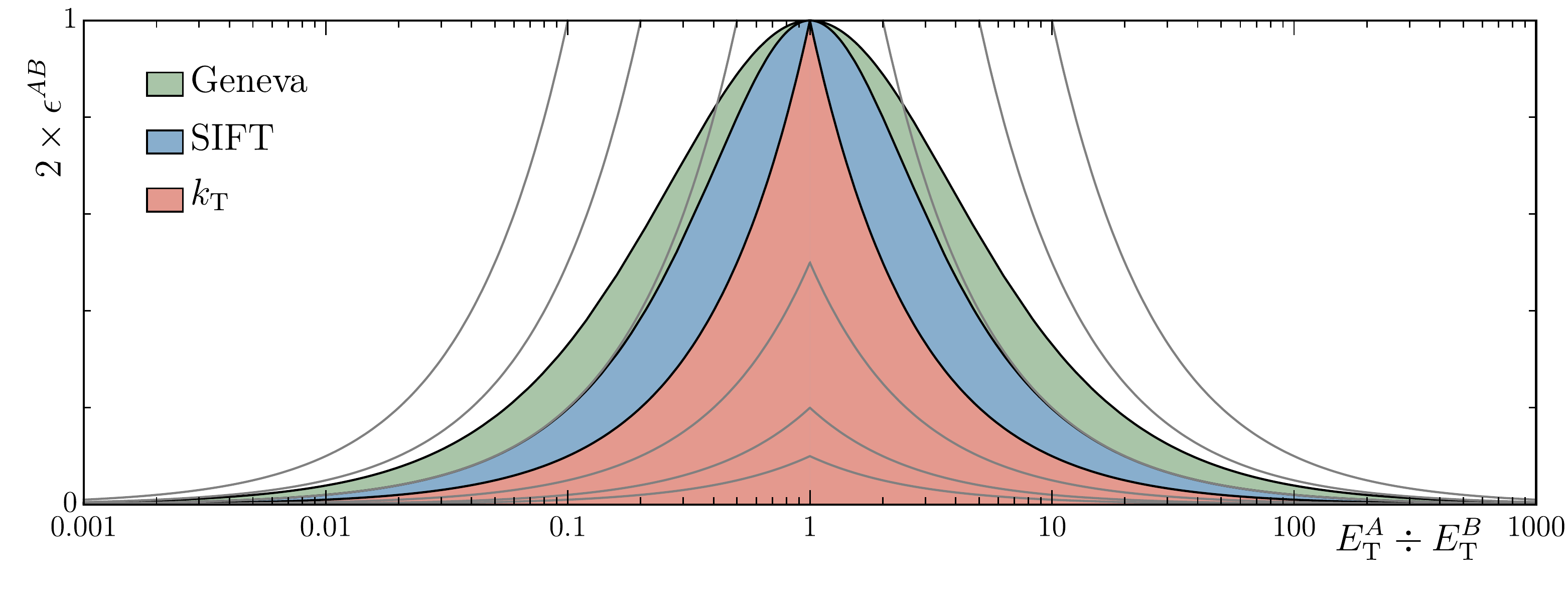}
\vspace{-12pt}
\caption{\footnotesize
The factor $2\times\epsilon^{AB}$ is log-symmetric
in the ratio $\Et^A/\Et^B$ of transverse energies,
becoming small whenever candidate scales are hierarchically dissimilar.
SIFT (blue) is compared against analogous behaviors for the Geneva (green) and
$\kt$-family (red) algorithms, the latter at ($\,\beta \equiv \Et^{A}\Et^{B}/E_0^2 \Rightarrow 1\,$).
Grey contours illustrate scaling of the $\kt$ measures with a power $n=\pm 1$ of
($\,\beta \Rightarrow \sfrac{1}{10},\, \sfrac{1}{5},\, \sfrac{1}{2},\, 2,\, 5,\,10\,$).}
\label{fig:epsilon}
\end{figure*}

The indicated correspondence becomes increasingly exact as one approaches
the collinear $(\Delta R^2 \ll 1)$ and massless
($\Delta y \Rightarrow \Delta \eta$, $\xi \Rightarrow 1$) limits.
The fact that $\cosh \Delta y$ is unbounded
and has purely positive series coefficients,
whereas $\cosh \Delta \phi$ is bounded
and its terms are of alternating sign, 
implies that $\Delta\widetilde{R}^2$ is more sensitive to separations
in rapidity than separations in azimuth.

The ratio ($\Delta\widetilde{R}^2 \div \Delta R^2$) is explored
graphically in FIG.~\ref{fig:drtilde} as a function of $\Delta\eta$
and $\Delta\phi$, for various values of ($m/\pt$).  For purposes of
illustration, deviations from ($\eta = 0$) are applied symmetrically
to the candidate object pair,
and the quoted mass ratio applies equivalently to both objects.
Black contours indicate unity, bisecting regions of
neutral bias, which are colored in blue.
Grey contours are spaced at increments of $0.1$
in the base-$10$ logarithm, and regions where SIFT
exhibits enhancement (suppression) of clustering
are colored green (red).

In the massless limit (lefthand panel) a near-symmetry
persists under ($\Delta \eta \Leftrightarrow \Delta\phi$),
with corrections from subleading terms as described previously.
When $m$ approaches $\pt$ (center panel), the pseudo-rapidity versus azimuth
symmetry is meaningfully broken at large $\Delta R$ and a
strong aversion emerges to the clustering of massive states
at small $\Delta R$.  The latter effect dominates for
highly non-relativistic objects (righthand panel),
and extends to larger separations, such that the
distinction between $\Delta \eta$ and $\Delta\phi$
is again washed out.  SIFT binds objects
approaching ($\Delta R \sim \pi$) significantly more tightly
than the $\kt$ algorithms in this limit.  Intuition
for that crossover in sign can be garnered from leading
terms in the multi-variate expansion shown
following, as developed from
Eqs.~(\ref{eq:dmsqdiffs2},~\ref{eq:drsqtilde},~\ref{eq:rap}):
\begin{align}
\label{eq:rsqeffdiff}
\Delta \widetilde{R}_{AB}^2 &\Rightarrow \Delta R_{AB}^2 \\
&+ \Bigg\{1-\frac{\Delta R_{AB}^2}{2}\Bigg\}\times
\Bigg\{\left(\frac{m_A}{\pt^A}\right)^2\!\!+\left(\frac{m_B}{\pt^B}\right)^2\Bigg\}
+ \cdots
\nonumber
\end{align}

We turn attention next to the denominator from Eq.~(\ref{eq:siftmeasure}),
defining a new quantity $\epsilon^{AB}$ in conjunction
with the transverse energy product factored out of $\Delta m_{AB}^2$.
\begin{equation}
\epsilon^{AB}
\equiv
\frac{\Et^{A} \Et^{B}}{(\Et^{A})^{2} + (\Et^{B})^{2}}
= \left\{\left(\frac{\Et^{A}}{\Et^{B}}\right) + \left(\frac{\Et^{B}}{\Et^{A}}\right)\right\}^{-1}
\label{eq:eps}
\end{equation}

This expression has a symmetry under the transformation
$(\alpha \equiv \Et^A/\Et^B \Rightarrow \alpha^{-1})$.  It is maximized
at $\alpha = 1$, where $(\epsilon^{AB} \Rightarrow \sfrac{1}{2})$,
and minimized at $\alpha = (0,+\infty)$, where $(\epsilon^{AB} \Rightarrow 0)$.
The response is balanced by a change of variables
$\Delta u = \ln \alpha$, i.e.,~$\alpha = e^{\Delta u}$,
where the logarithm converts ratios into differences.
\begin{align}
u &\equiv \ln {\Big( \Et / {\rm [GeV]} \Big)}
\label{eq:cosh_et} \\
\epsilon^{AB} &= {\Big( e^{+\Delta u_{AB}} + e^{-\Delta u_{AB}}\Big)}^{-1}
= {\Big( 2\cosh \Delta u_{AB}\Big)}^{-1}
\nonumber
\end{align}

Putting everything together, we arrive at a
formulation of the measure from Eq.~(\ref{eq:siftmeasure})
that is expressed almost entirely in terms of coordinate differences
of the rapidities, azimuths, and log-transverse energies,
excepting the coefficients $\xi$ from Eq.~(\ref{eq:dmsqdiffs2}), which depend on the ratios $m/\pt$.
Of course, it is possible to adopt a reduction of the measure where $(\xi \Rightarrow 1)$ by fiat,
which is equivalent to taking a massless limit in the fashion of JADE and Geneva,
as described in Section~\ref{sct:siftmeasure}.
\begin{align}
\delta_{AB} &= \epsilon^{AB} \times \Delta\widetilde{R}_{AB}^2
\label{eq:measure_diffs} \\
&= \frac{ \cosh \Delta y_{AB} -\xi^{A} \xi^{B} \cos\Delta \phi_{AB} }{\cosh \Delta u_{AB}}
\nonumber
\end{align}

The scale-invariance of Eq.~(\ref{eq:measure_diffs}) is explicit, in two regards.
By construction, there is no reference to an external angular cutoff $R_0$.
In addition, the fact that transverse energies are referenced
only via ratios, and never in absolute terms, is emergent.
The measure is additionally observed to smoothly blend attributes of
$\kt$ and anti-$\kt$ jet finding, insomuch as
the former prioritizes clustering when one member of a pair is soft,
the latter when one member of a pair is hard,
and SIFT when the transverse energies are logarithmically disparate.

\begin{figure*}[ht!]
\centering
\includegraphics[width=.95\textwidth]{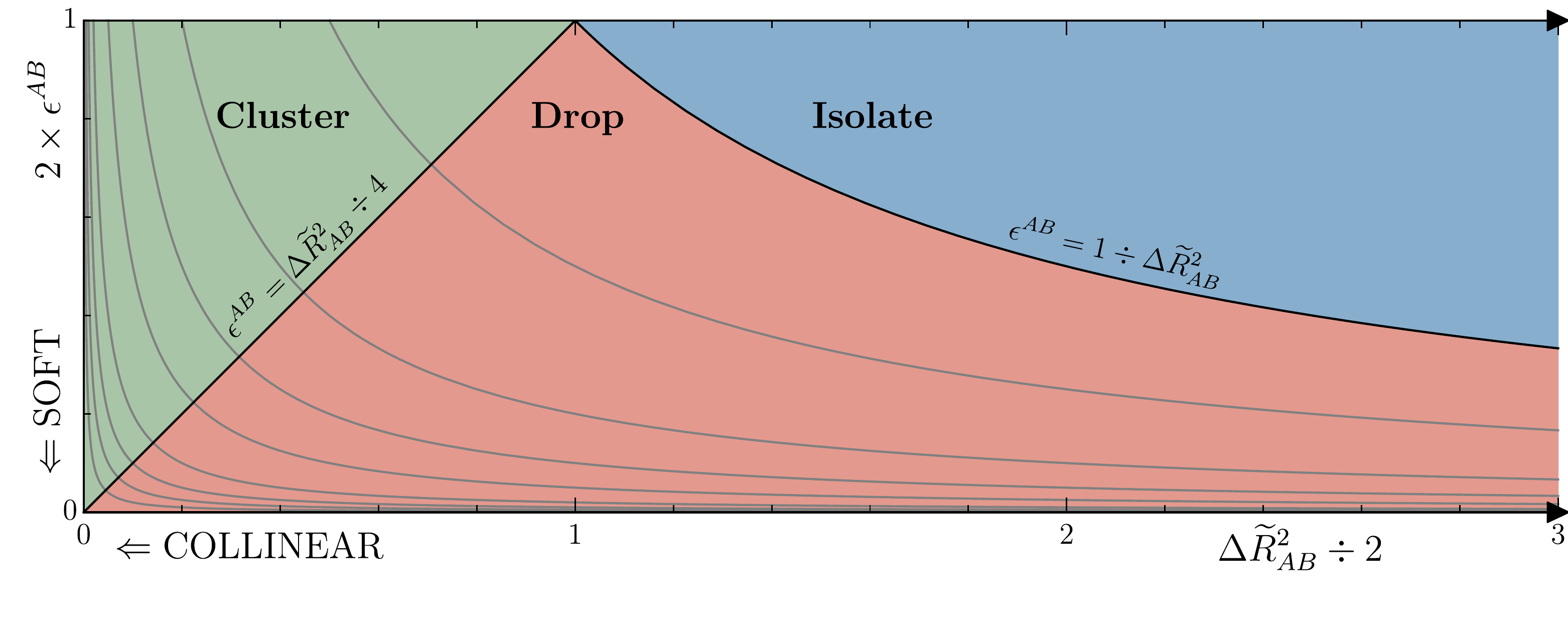}
\vspace{-12pt}
\caption{\footnotesize
Phase diagram for the separation of object merging, filtering, and isolation responses.}
\label{fig:phase}
\end{figure*}

This behavior is illustrated in FIG.~\ref{fig:epsilon}, with
$2\times\epsilon^{AB}$ plotted in blue
as a function of ($\alpha \equiv \Et^A/\Et^B$).
For comparison, the analogous momentum-dependent factor 
for Geneva from Eq.~(\ref{eq:geneva}) is shown in
green on the same axes, taking ($E \Rightarrow \Et$)
and normalizing to unity at ($\alpha = 1$).
Like SIFT, Geneva is symmetric with respect to variation of
the absolute event scale $\beta$.
However, the extra cross term appearing in its denominator
produces heavier tails when the scale ratio $\alpha$ is unbalanced.
The cusped red region similarly represents the $\kt$ and anti-$\kt$ measures.
Specifically, the following expression is proportional
to Eq.~(\ref{eq:ktmeasure}) for ($n=\pm1$) in the massless limit:
\begin{equation}
\delta_{AB}^{k_{\rm T},n} \,\appropto\,
{\left(\frac{\Et^{A}\Et^{B}}{E_0^2}\right)}^{\!\!n}\!\!
\times
{\rm min} \left[ \frac{\Et^{A}}{\Et^{B}}, \frac{\Et^{B}}{\Et^{A}} \right]
\label{eq:measureprop}
\end{equation}

The selected normalization agrees with $2\times\epsilon^{AB}$
in the further limits ($\beta \equiv \Et^{A}\Et^{B}/E_0^2 \Rightarrow 1$)
and ($\alpha \Rightarrow 1$), where $E_0$ is an arbitrary constant
reference energy.  The distinction between $\kt$
and anti-$\kt$ clustering amounts to an enhancement
versus suppression by the product (squared geometric mean)
of transverse momenta.  This is illustrated with the grey
contours in FIG.~\ref{fig:epsilon}, which rescale
by $\beta = (\sfrac{1}{10},\,\sfrac{1}{5},\,\sfrac{1}{2},\,2,\,5,\,10)$
from inner-lower to outer-upper for 
($n=+1$), or in the reverse order for ($n=-1$).
The $\beta$-invariant case ($n=0$) is also potentially of interest,
but it is a new construction that is not to be confused with
the Cambridge-Aachen algorithm, which has no energy dependence at all.

We conclude this section with a transformation that identifies 
the coordinate $u$ introduced in Eq.~(\ref{eq:cosh_et})
as a sort of ``dual'' to the rapidity $y$ from Eq.~(\ref{eq:rap}).
The log-transverse momentum $\ln (\,\pt\,/{\rm [GeV]})$ is similarly linked
to pseudo-rapidity $\eta$ in the massless limit.
\begin{alignat}{2}
\Omega_\pm &\equiv \phantom{\frac{1}{2}} \ln {\bigg( \frac{E \pm p_z}{{\rm [GeV]}} \bigg)} &&
\label{eq:rap_loget} \\
\frac{\Omega_+ +\hspace{2pt} \Omega_-}{2} &= \frac{1}{2} \ln {\bigg( \frac{E^2 - p_z^2}{{\rm [GeV]}^2} \bigg)} &&\,=\, u
\nonumber \\
\frac{\Omega_+ -\hspace{2pt} \Omega_-}{2} &= \frac{1}{2} \ln {\bigg( \frac{E + p_z}{E - p_z} \bigg)} &&\,=\, y
\nonumber
\end{alignat}

\section{Filtering, Isolation, and Halting\label{sct:filtering}}

This section establishes the SIFT filtering and isolation criteria,
which are used in conjunction to formulate a suitable halting
condition for the non-exclusive clustering mode.
Direct integration of a grooming stage effectively rejects stray radiation and pileup.
SIFT's behavior will be visualized with and without filtering
in Section~\ref{sct:comparison}, and compared against each
$\kt$-family algorithm in the presence of a soft ``ghost'' radiation background.

In conjunction, the two factorized terms in Eq.~(\ref{eq:measure_diffs})
ensure that clustering prioritizes the merger of object pairs that
have a hierarchically soft member and/or that are geometrically collinear,
mimicking fundamental poles in the matrix element for QCD showering.
This behavior is illustrated by the ``phase diagram'' in FIG.~\ref{fig:phase},
where the product of horizontal ``$x$'' and vertical ``$y(x)$'' coordinates on that plane
is equal to $\delta_{AB}$.  Grey ``$y(x) = 1/x$'' contours trace constant values
of the measure, equal to ($.002,\,.005,\, .01,\, .02,\, .05,\, .1,\, .2,\, .5$),
with minimal values gathered toward the lower-left.
As a consequence, SIFT successfully preserves mutually hard structures with tight angular adjacency,
maintaining their resolution as distinct objects up to the final stages of clustering\footnote{
Mutually soft pairings tend not to occur, since such objects are typically
gathered up by a harder partner at an early stage.}.

However, iterative application of the SIFT (or Geneva) measure
does not offer an immediately apparent halting mechanism,
and it will ultimately consume any presented
objects into a single all-encompassing jet if left to run.
These measures are additionally prone to sweeping up uncorrelated soft
radiation onto highly-boosted partners at wide angular separation.
The solution to both problems turns out to be related.
For inspiration, we turn to the Soft Drop procedure,
wherein a candidate jet is recursively declustered and
the softer of separated constituents is discarded
until the following criterion is satisfied:
\begin{equation}
\frac{\min\, (\pt^A,\pt^B)}{\pt^A+\pt^B} > z_{\rm cut} \left(\frac{\Delta R_{AB}}{R_0}\right)^\beta
\label{eq:soft_drop}
\end{equation}
The dimensionless $z_{\rm cut}$ coefficient is typically $\mathcal{O}(0.1)$.
The exponent $\beta$ can vary for different applications, although we focus here on $\beta = 2$.
We first attempt to recast the elements of Eq.~(\ref{eq:soft_drop}) into
expressions with asymptotically similar behavior that adopt the vocabulary of Eq.~(\ref{eq:siftmeasure}).
The factor $\Delta \widetilde{R}_{AB}^2$ can be carried over directly.  Likewise,
$\epsilon^{AB}$ behaves similarly to a minimized ratio of transverse energies,
with the advantage of analyticity.

\begin{align}
\epsilon^{AB}
&= \left\{\min \left(\frac{\Et^{A}}{\Et^{B}}\right) + \max \left(\frac{\Et^{A}}{\Et^{B}}\right)\right\}^{-1}
\nonumber \\
&\simeq \left\{\max \left(\frac{\Et^{A}}{\Et^{B}}\right)\right\}^{-1}
\nonumber \\
&= \min \left(\frac{\Et^{A}}{\Et^{B}}\right)
\simeq \frac{ \min ( \Et^{A}, \Et^{B} )}{ \Et^{A} + \Et^{B}}
\label{eq:min_proxy}
\end{align}

Paired factors of $2$ previously emerged ``naturally''
in Eqs.~(\ref{eq:drsqtilde},~\ref{eq:cosh_et}),
and we group them here in a way that could be interpreted
as setting ``$z_{\rm cut} = \sfrac{1}{4}$''
in the context of a large-radius $(R_0 = 1)$ jet.
Putting all of this together, the suggested analog to Eq.~(\ref{eq:soft_drop}) is shown following.
\begin{equation}
\textrm{Cluster:} \hspace{12pt}
\frac{\Delta \widetilde{R}_{AB}^{2}}{2} <
\big\{ \left( 2\, \epsilon^{AB}\right) \le 1 \big\}
\label{eq:sd_proxy}
\end{equation}

An important distinction from conventional usage is
that this protocol is to be applied during the initial clustering cycle
itself, at the point of each candidate merger.
This is similar in spirit to Recursive Soft Drop~\cite{Dreyer:2018tjj}.

Intuition may be garnered by turning again to FIG.~\ref{fig:phase},
where clustering consistent with the Eq.~(\ref{eq:sd_proxy}) prescription
is observed to occur only in the upper-left (green) region of the phase diagram,
above the ``$y(x)=x$'' diagonal cross-cutting contours of the measure.
The upper bound on $(2\,\epsilon^{AB})$ precludes
clustering if $(\Delta\widetilde{R}_{AB} \ge \sqrt{2})$.
However, this angular threshold is \emph{dynamic},
and the capturable cone diminishes in area
with increasing imbalance of the transverse scales.
This condition may be recast into a limit
on the clustering measure from Eq.~(\ref{eq:measure_diffs}).
\begin{equation}
\textrm{Cluster:} \hspace{12pt}
\delta_{AB} < \big\{ \left( 2\, \epsilon^{AB} \right)^2 \le 1 \big\}
\label{eq:sd_measure}
\end{equation}

The question now is what becomes of those objects rejected
by the Eq.~(\ref{eq:sd_measure}) filter.
Specifically, are they discarded, or are they classified
for retention as isolated final-state jets?
One possible solution is to simply determine this based on the magnitude
of the transverse energy, but that runs somewhat counter to
current objectives.  Seeking a simple scale-invariant
criterion for selecting between isolation and rejection,
we observe that there are distinct two ways in which Eq.~(\ref{eq:sd_proxy}) may fail.
Namely, the angular opening may be too wide, and/or the transverse scales
may be too hierarchically separated. This is illustrated again by
FIG.~\ref{fig:phase}, wherein one may exit the clustering region (green)
by moving rightward (wider angular separation) or downward (more scale disparity).

If the former cause is dominant,
e.g., if $(\Delta \widetilde{R}_{AB}^{2} \gg 1)$ and $(\epsilon^{AB} \sim 1)$
such that both members are on equal footing and $(\delta_{AB}\gg 1)$,
then collective isolation as a pair of distinct final states is appropriate.
Conversely, if the latter cause primarily applies,
e.g., if $(\epsilon^{AB} \ll 1)$ and $(\Delta \widetilde{R}_{AB}^2 \sim 1)$ such that $(\delta_{AB}\ll 1)$,
then the pertinent action is to asymmetrically set aside just the softer candidate\footnote{
Although we treat ``drop'' here as a synonym for ``discard'', a useful
alternative (cf. Section~\ref{sct:algorithm}) is single reclassification
as a final state, since softness is not guaranteed in an absolute sense.}.
These scenarios may be quantitatively distinguished in a manner that
generates a balance with and continuation of Eq.~(\ref{eq:sd_proxy}), as follows.
\begin{alignat}{2}
\textrm{Drop:} &&\hspace{3pt} \big\{ \left( 2\, \epsilon^{AB}\right) \le 1 \big\} &\le \frac{\Delta \widetilde{R}_{AB}^{2}}{2} < \big\{ 1 \le \left( 2\, \epsilon^{AB}\right)^{-1} \big\}
\nonumber \\
\textrm{Isolate:} &&\hspace{3pt} \big\{ 1 \le \left( 2\, \epsilon^{AB}\right)^{-1} \big\} &\le \frac{\Delta \widetilde{R}_{AB}^{2}}{2}
\label{eq:iso_drop}
\end{alignat}

Each criterion may also be recast in terms of the clustering measure,
with isolation always and only indicated against an absolute reference
value ($1$) of the SIFT measure $\delta_{AB}$ that heralds a substantial mass-gap.
Observe that the onset of object isolation necessarily culminates in
global algorithmic halting since all residual pairings must correspond
to larger values of $\delta_{AB}$.
\begin{alignat}{2}
\textrm{Drop:}&&\hspace{8pt} \big\{ \left( 2\, \epsilon^{AB}\right)^2 \le 1 \big\} &\le \delta_{AB} < \big\{ 1 \big\}
\nonumber \\
\textrm{Isolate:}&&\hspace{8pt} \big\{ 1 \big\} &\le \delta_{AB}
\label{eq:iso_drop_measure}
\end{alignat}

FIG.~\ref{fig:phase} again provides visual intuition,
where isolation occurs in the upper-right (blue) regions for
values of the measure above unity, and soft wide radiation
is dropped in the lower-central (red) regions.
The clustering and isolation regions are fully separated
from each other, making contact only at the zero-area ``triple point''
with ($2\,\epsilon^{AB}=1$, $\Delta \widetilde{R}_{AB}^{2}/2=1$).
We will provide additional support from simulation for triggering isolation
at a fixed $\mathcal{O}(1)$ value of the measure in Section~\ref{sct:tagging}.

Note that non-relativistic
or beam-like objects at very large rapidity
that approach the ($\pt \Rightarrow 0$, $m \ne 0$) limit
will never cluster, since vanishing of the lever arm
($\xi = 0$) in Eq.~(\ref{eq:dmsqdiffs2}) implies
via Eq.~(\ref{eq:drsqtilde}) that ($\Delta \widetilde{R}_{AB}^{2}/2\ge1$).
Also, objects of equivalent transverse energy
with angular separation $(\Delta\widetilde{R}_{AB} \ge \sqrt{2})$
will always be marked for isolation,
although this radius is again \emph{dynamic}.
The phase gap between clustering and isolation opens
up further when scales are mismatched, and
the effective angular square-separation
exceeds its traditional counterpart
as mass is accumulated during clustering
(cf. Eq.~\ref{eq:rsqeffdiff}).

\begin{figure*}[ht!]
\centering
\includegraphics[width=0.8\textwidth]{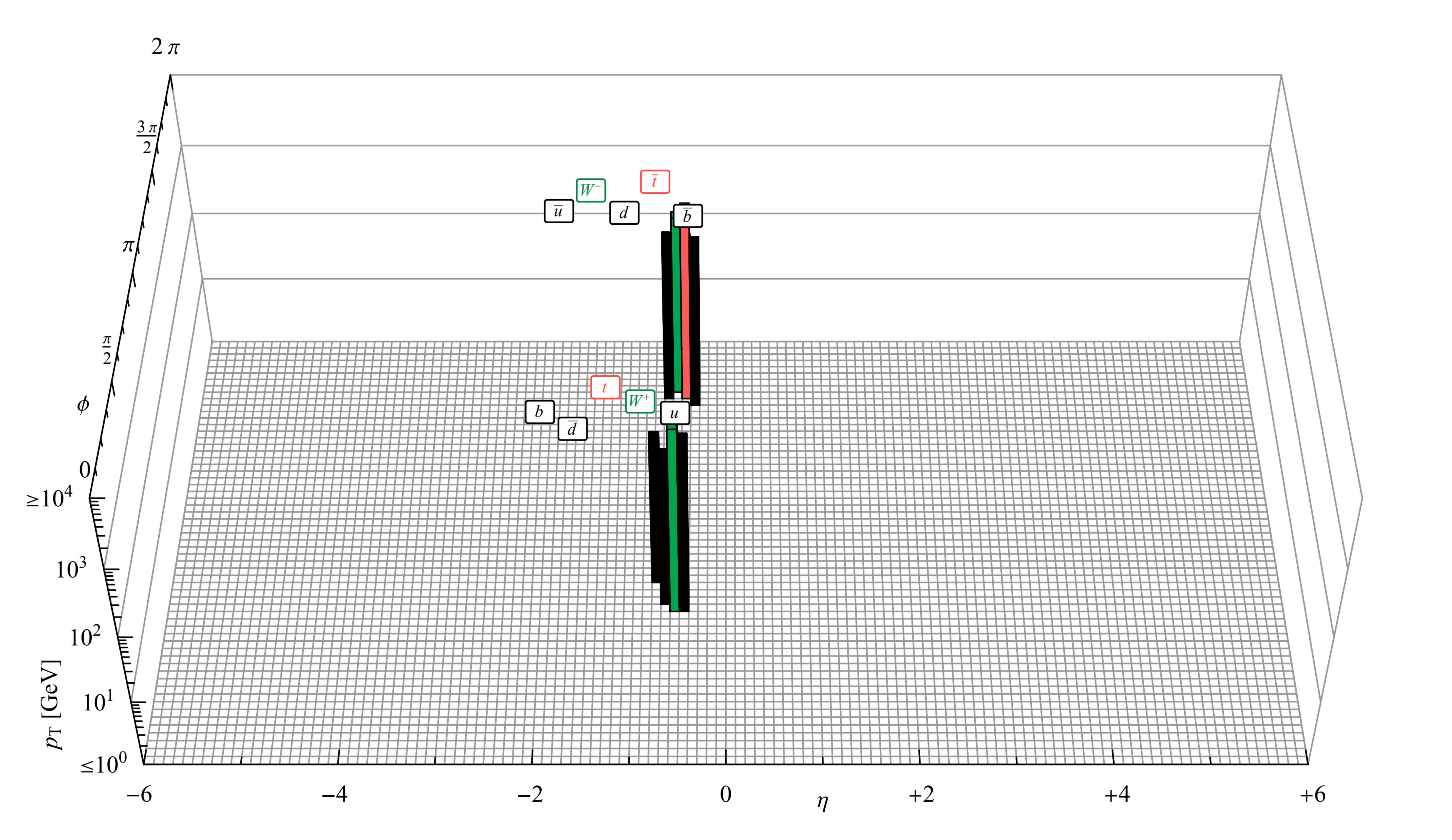} \\ 
\includegraphics[width=0.8\textwidth]{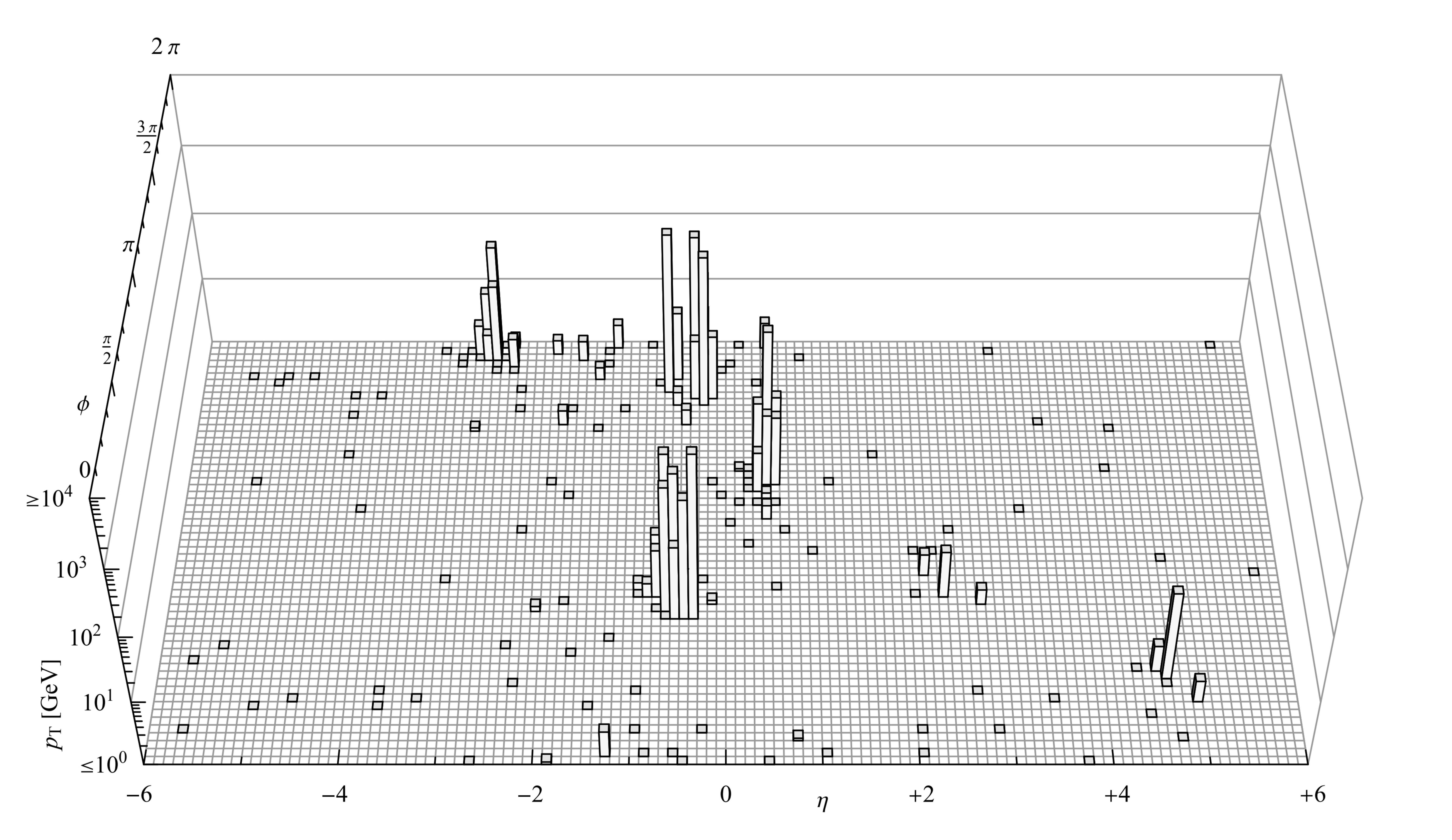}
\caption{\footnotesize
Upper: A simulated LHC scattering event with $\sqrt{s} = 14$~TeV is visualized at the partonic level.
Top quark pair production $p p \to t \bar{t}$ (red, red) is followed by the
decays $t \to W^+ b$ (green, black) and $W^+ \to u\, \bar{d}$ (black, black), plus conjugates.
A transverse boost of $\pt \simeq 800$~GeV for each top quark produces narrow collimation of decay products.
Lower: Generator-level radiation deposits resulting from showering, hadronization,
and decay of the partonic event appearing in the upper frame.}
\label{fig:filmframesEVT}
\end{figure*}

\begin{figure*}[ht!]
\centering
\includegraphics[width=0.8\textwidth]{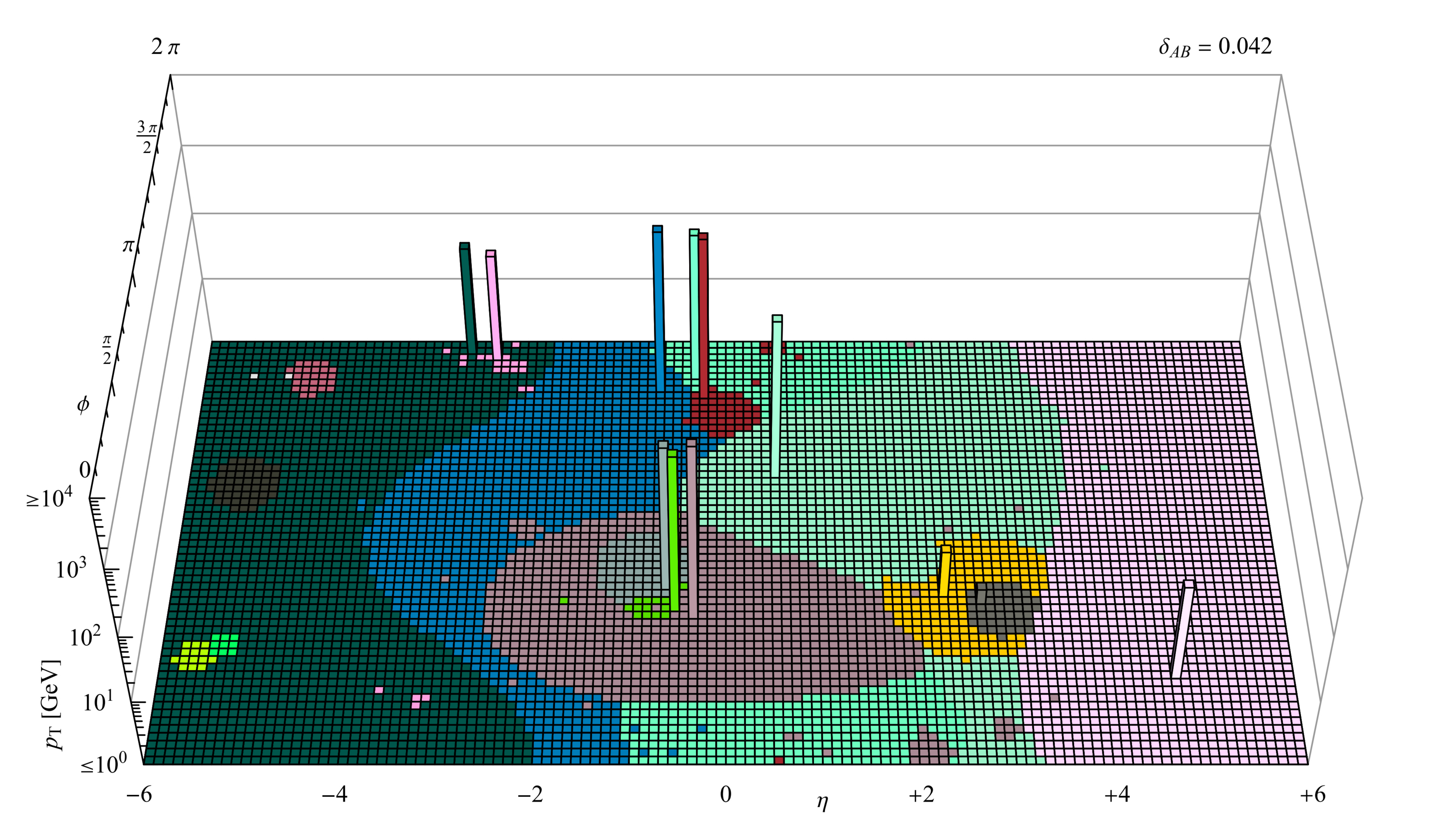} \\ 
\includegraphics[width=0.8\textwidth]{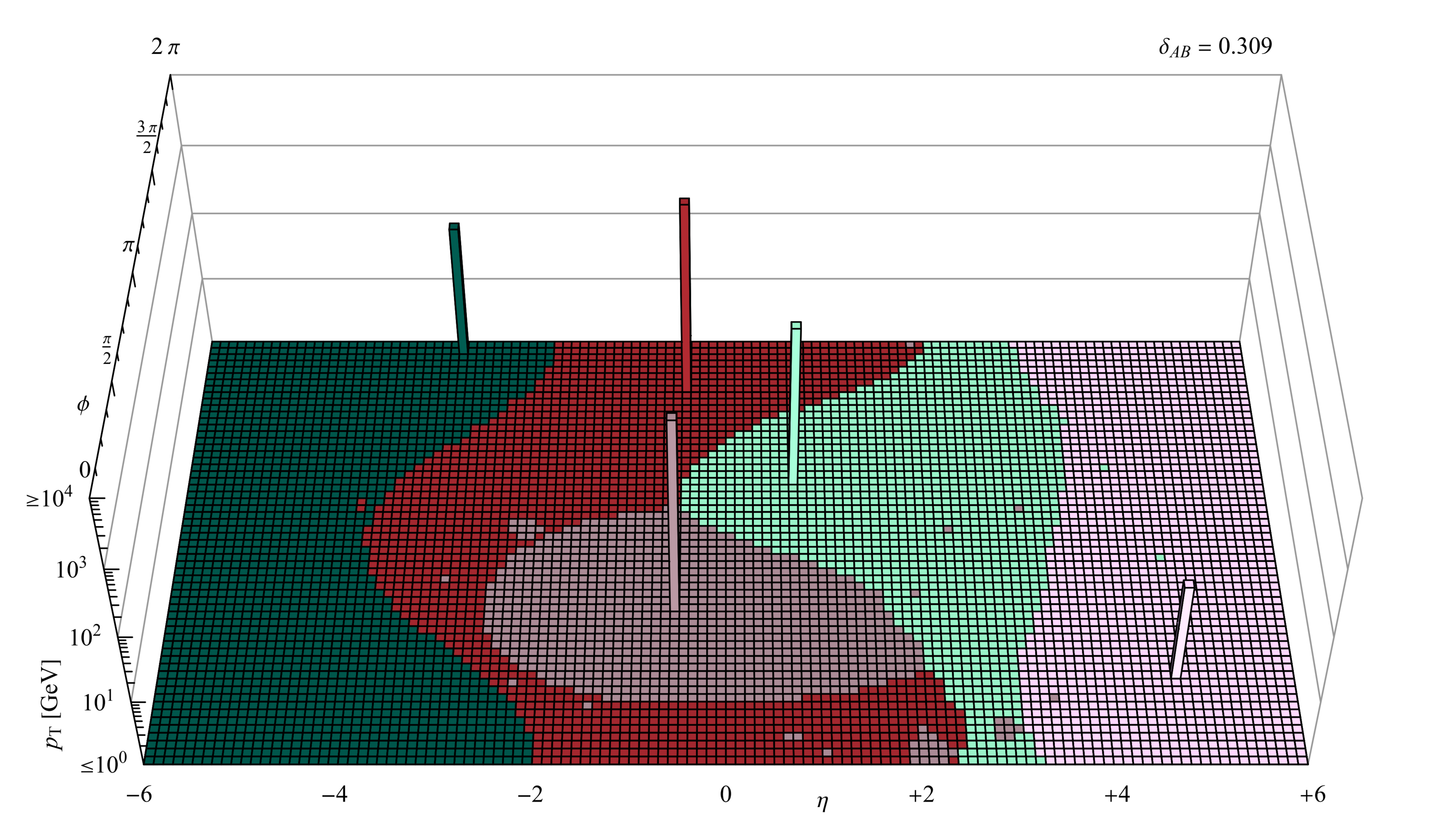}
\caption{\footnotesize
Frames representing sequential clustering of the FIG.~\ref{fig:filmframesEVT} event
using the exclusive $N_{\rm exc}=1$ SIFT algorithm
without the associated filtering or isolation criteria
(cf. FIG.~\ref{fig:filmframesE} for non-exclusive clustering with both criteria enabled).
Upper: Mutually hard prongs with narrow angular separation remain unmerged up
to the final stages of clustering.  However, hard objects are likely to sweep
up soft radiation at wide angles.  Lower: An image of the initial pair
production is reconstructed just prior to termination.  In the absence of a supplementary
halting criterion these structures will subsequently merge to completion, accompanied
by a large discontinuity in the measure $\delta_{AB}$.}
\label{fig:filmframesA}
\end{figure*}

\begin{figure*}[ht!]
\centering
\includegraphics[width=0.8\textwidth]{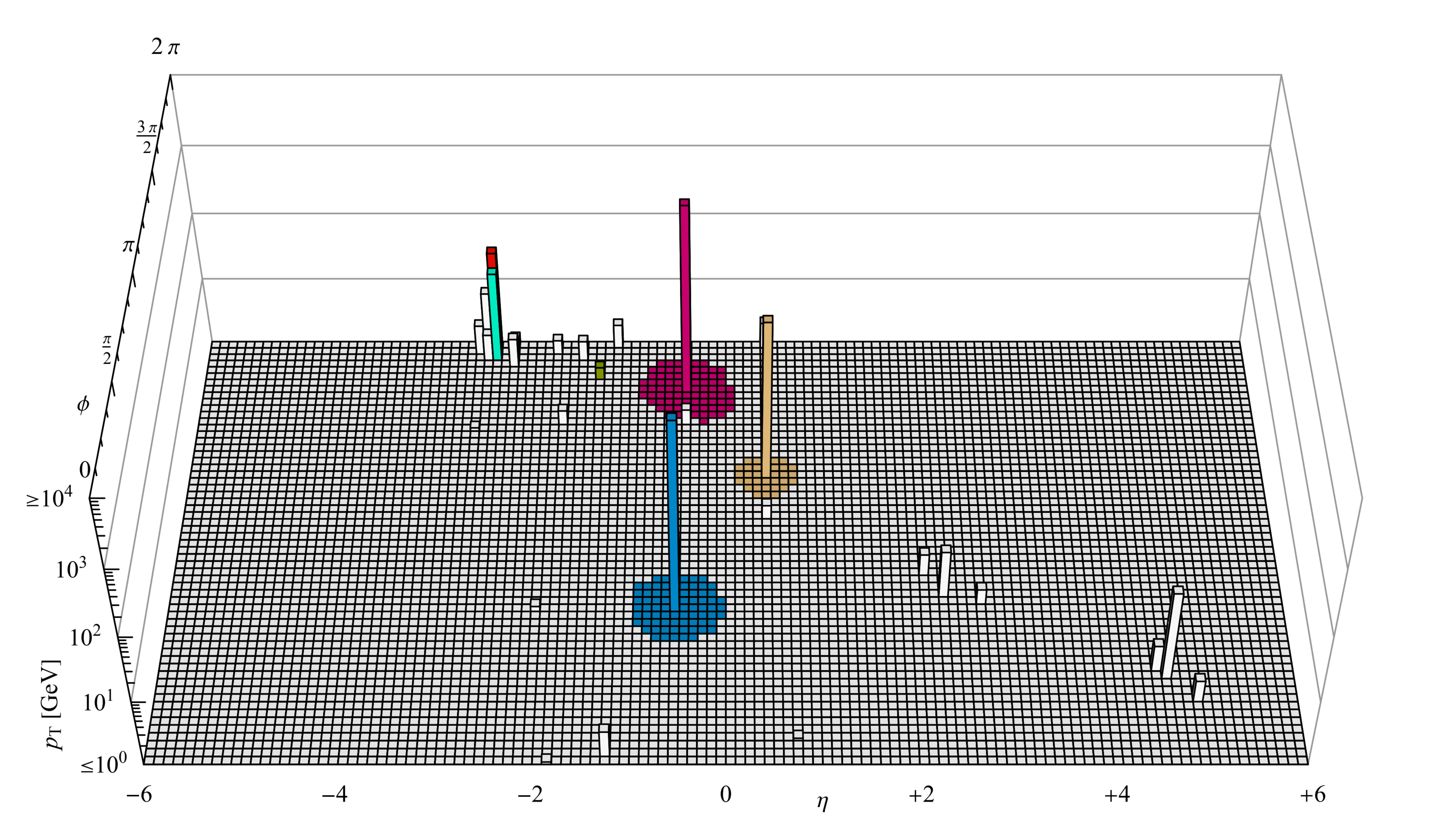} \\ 
\includegraphics[width=0.8\textwidth]{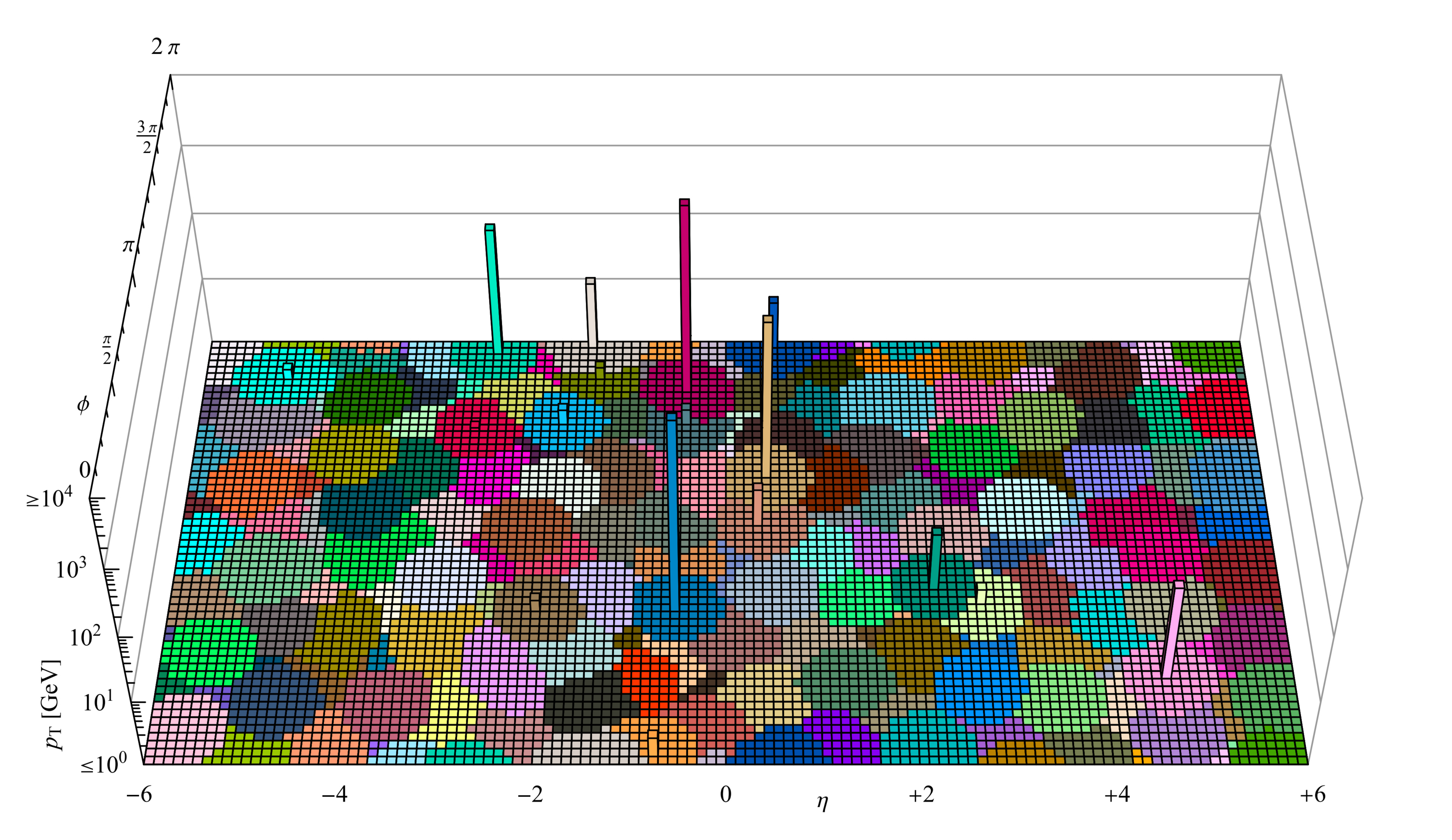}
\caption{\footnotesize
Frames representing sequential clustering of the FIG.~\ref{fig:filmframesEVT} event
using the anti-$\kt$ algorithm.
Upper: Priority is given to the hardest radiation, which immediately captures surrounding territory.
Hard substructure at angular scales smaller than the clustering radius
will be washed out rapidly.  Lower: The final state is characterized by
regular jet shapes with uniform expected areas.}
\label{fig:filmframesB}
\end{figure*}

\begin{figure*}[ht!]
\centering
\includegraphics[width=0.8\textwidth]{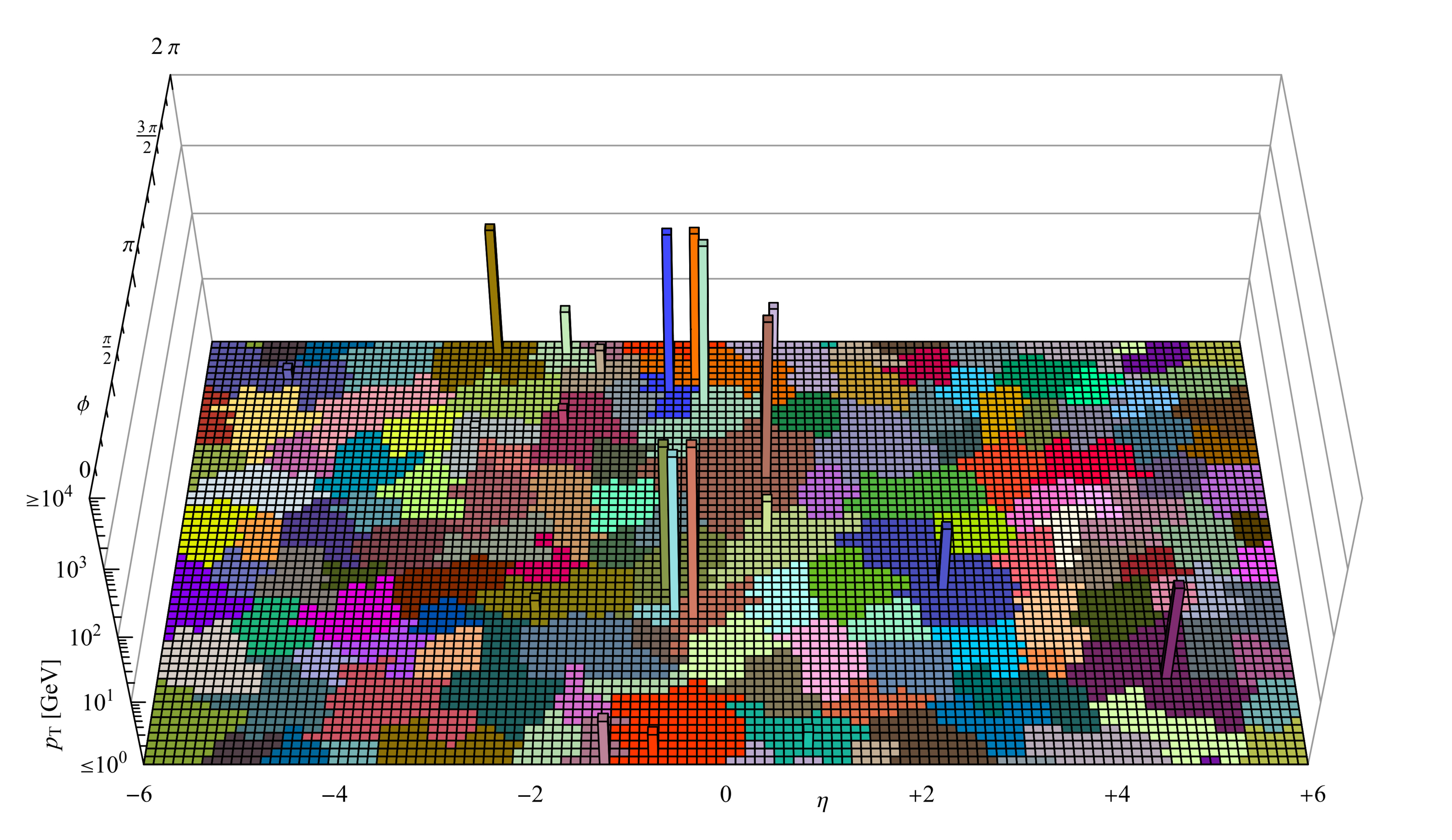} \\
\includegraphics[width=0.8\textwidth]{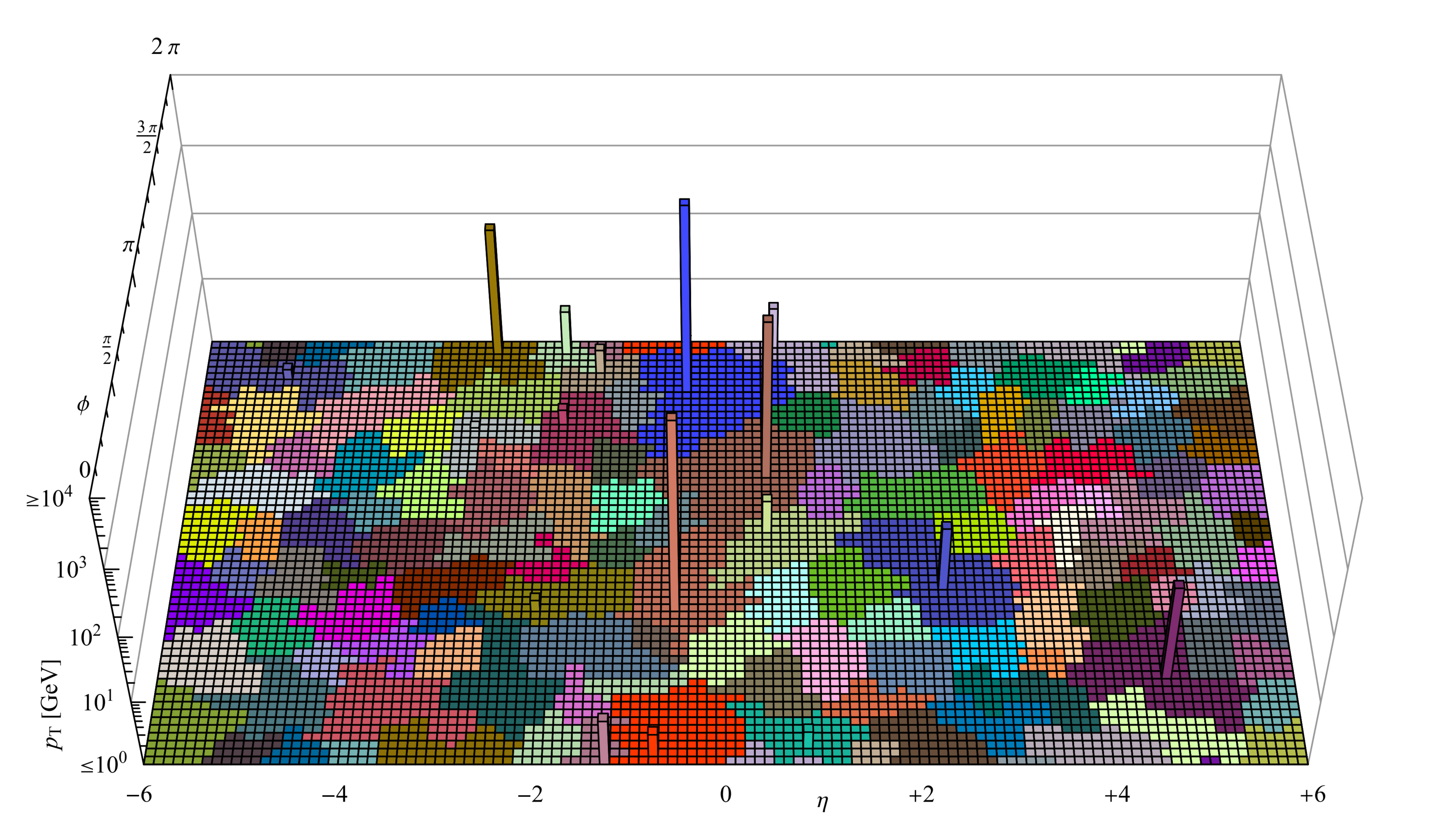}
\caption{\footnotesize
Frames representing sequential clustering of the FIG.~\ref{fig:filmframesEVT} event
using the $\kt$ algorithm.
Upper: Priority is given to the softest radiation, resulting in the
growth of dispersed catchments having a correlation length that increases in time.
Mutually hard substructures are preserved until the last stages of clustering.
Lower: The final state is characterized by irregular jet shapes with unpredictable areas.}
\label{fig:filmframesC}
\end{figure*}

\begin{figure*}[ht!]
\centering
\includegraphics[width=0.8\textwidth]{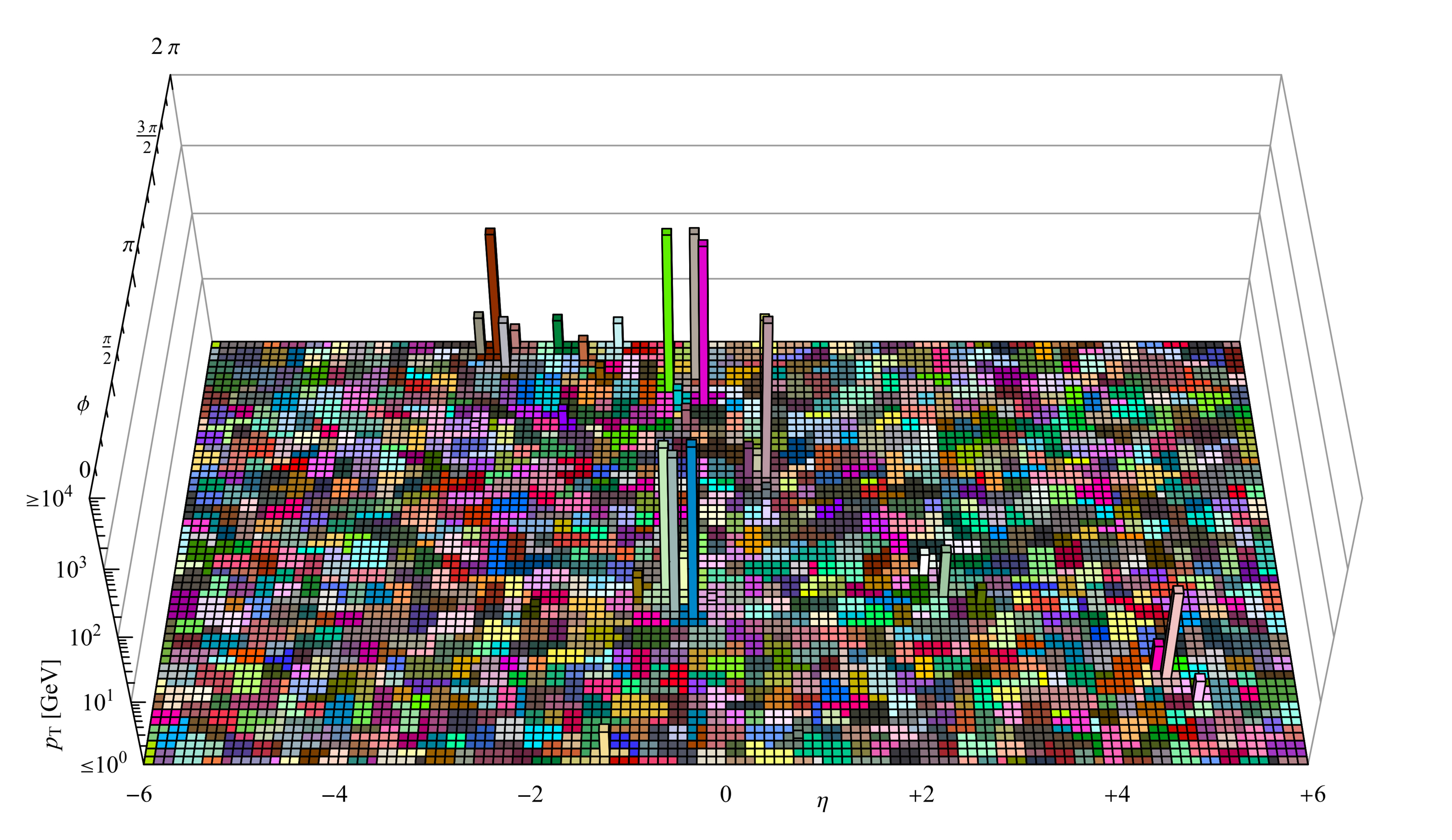} \\
\includegraphics[width=0.8\textwidth]{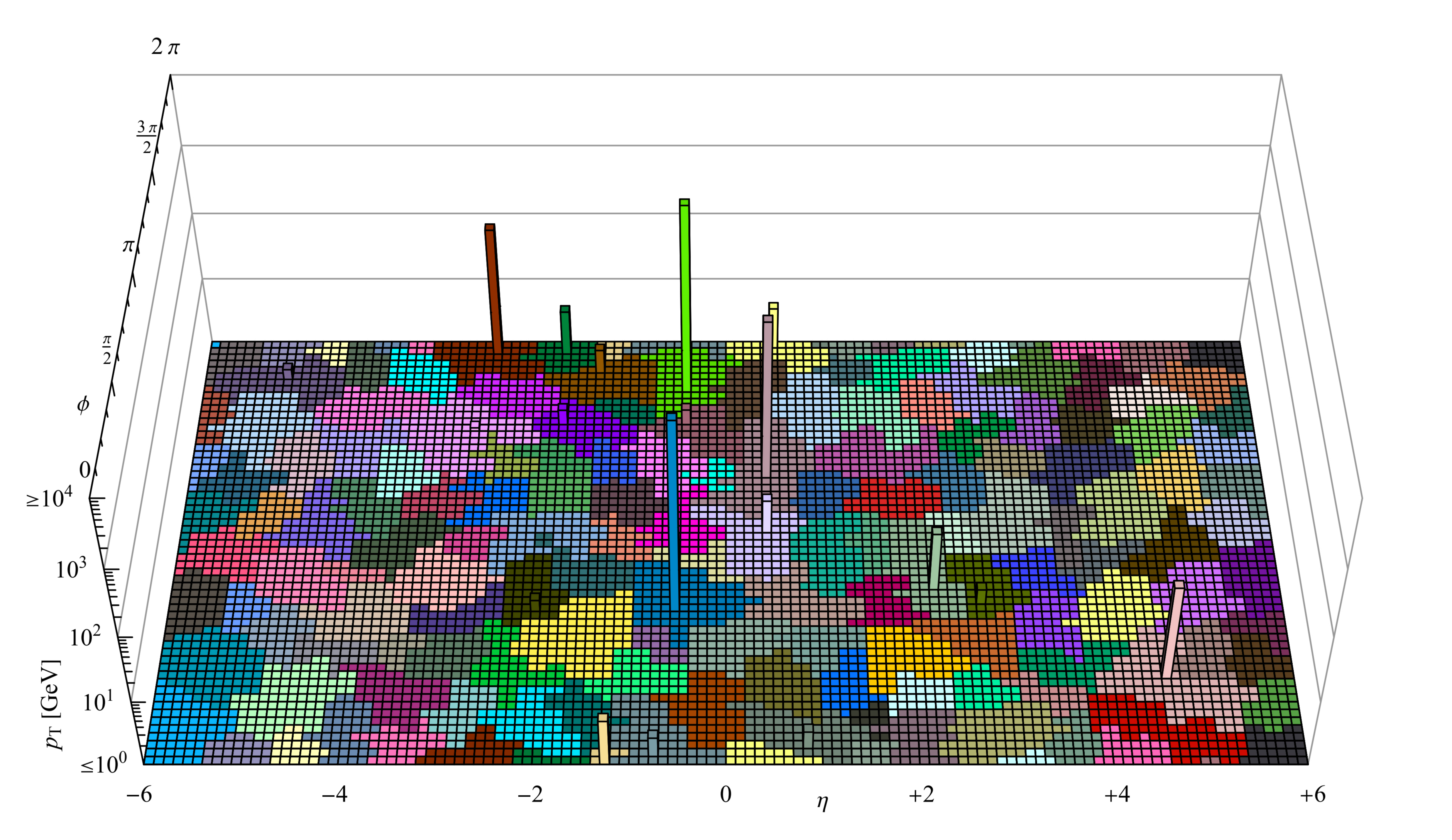}
\caption{\footnotesize
Frames representing sequential clustering of the FIG.~\ref{fig:filmframesEVT} event
using the Cambridge-Aachen algorithm.
Upper: Priority is given only the angular proximity, without reference to
the momentum scale.  Substructure is preserved only until the grain size
eclipses its angular scale.
Lower: The final state is characterized by irregular jet shapes with unpredictable areas.}
\label{fig:filmframesD}
\end{figure*}

\begin{figure*}[ht!]
\centering
\includegraphics[width=0.8\textwidth]{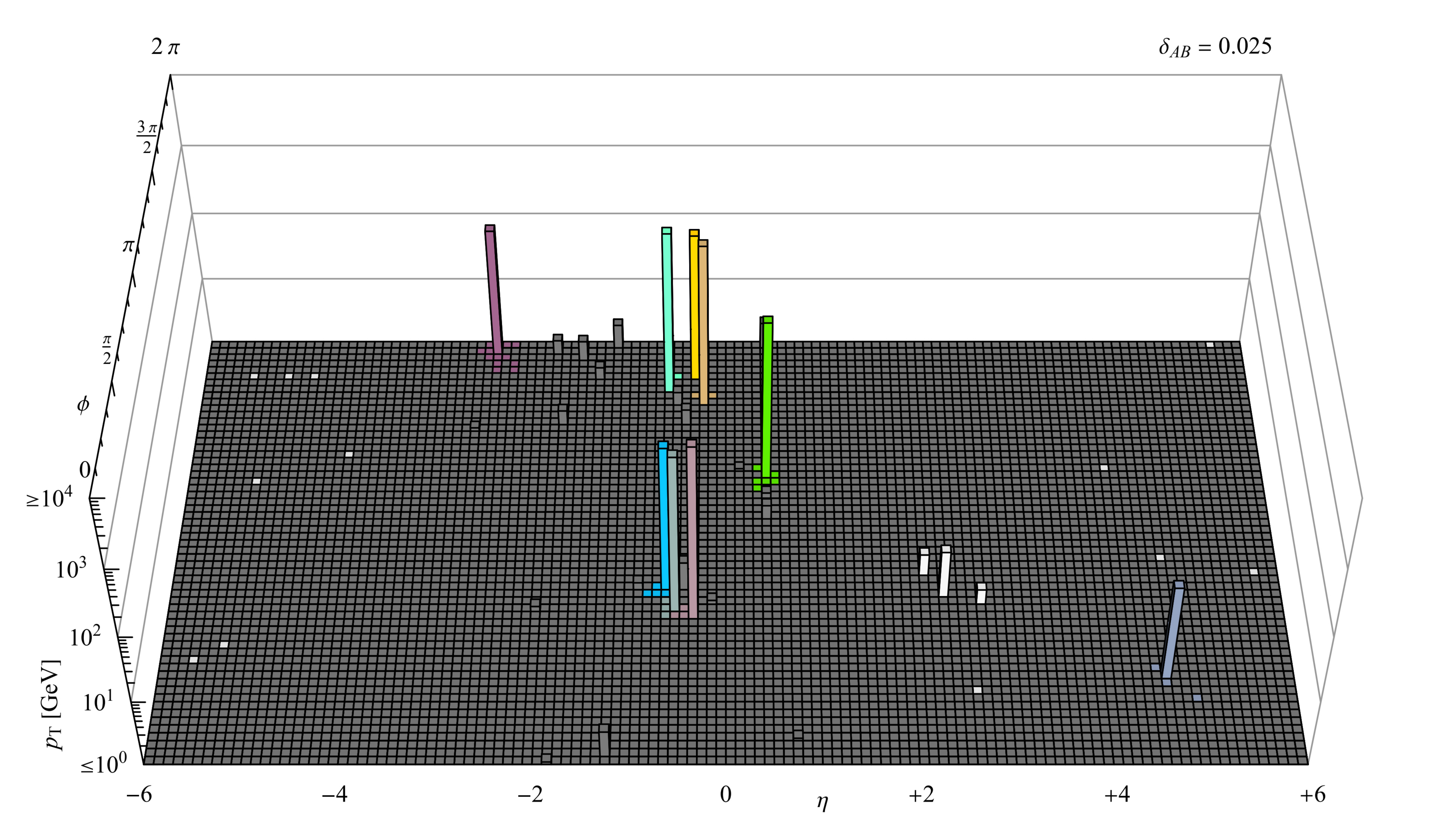} \\ 
\includegraphics[width=0.8\textwidth]{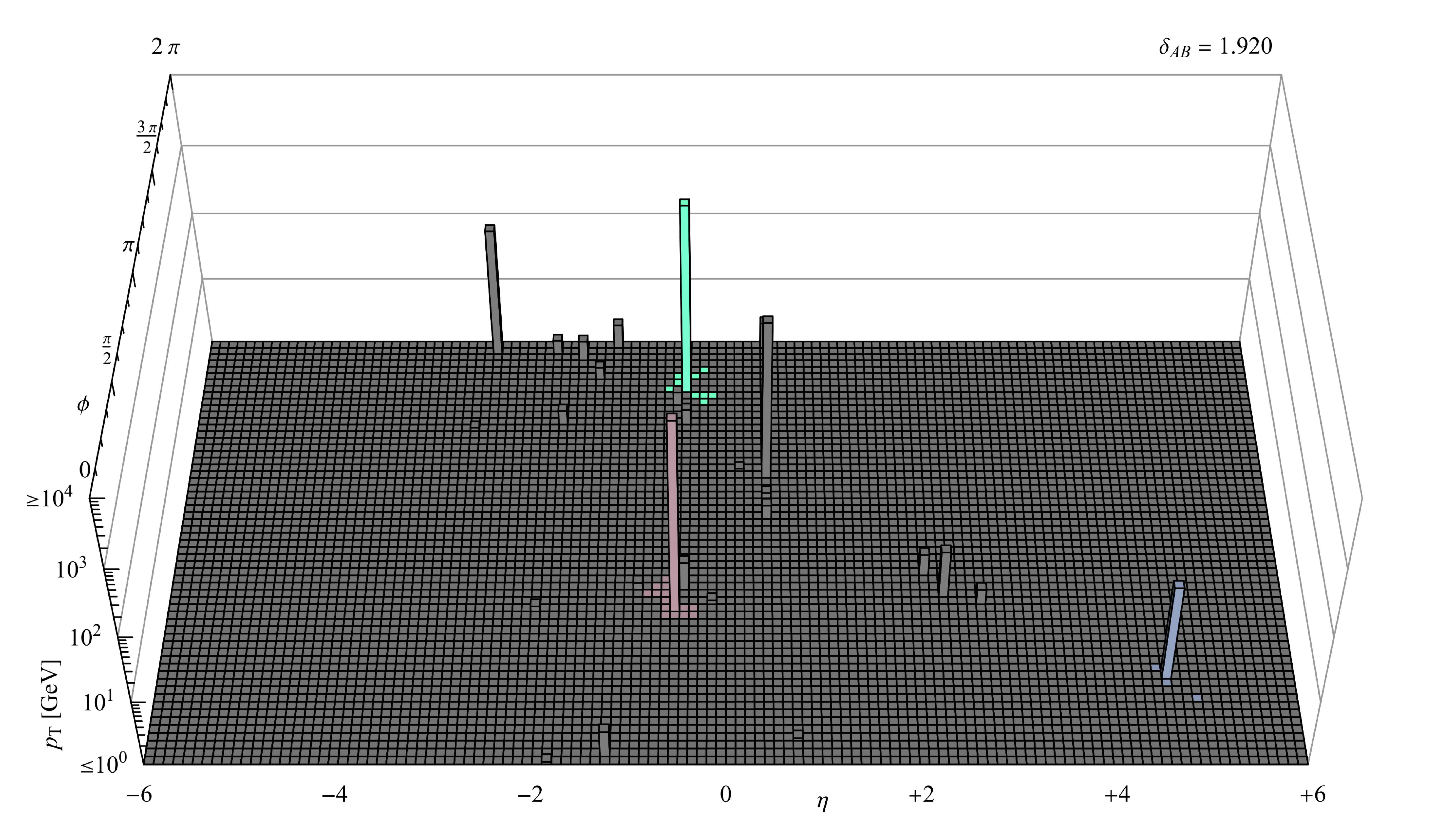}
\caption{\footnotesize
Frames representing sequential clustering of the FIG.~\ref{fig:filmframesEVT} event
using the SIFT algorithm with the application of filtering and isolation criteria.
Upper: Initial activity is dominated by the rejection of soft-wide radiation that
is paired with a hard prong by the measure but fails the filtering criterion.
Mutually hard substructures are resolved without contamination from stray radiation.
Lower: The isolation criterion triggers halting before distinct
objects associated with the initial pair production would be merged.}
\label{fig:filmframesE}
\end{figure*}

\section{The $\boldsymbol{N}$-Subjet Tree\label{sct:njettree}}

This section concludes development of the SIFT algorithm 
by formalizing the concept of an $N$-subjet tree.
This data structure records the clustering history of each
isolated final-state object from $N$ prongs down to $1$,
along with the associated value of the measure at each merger.

Final-state jets defined according to the global halting
condition outlined in Section~\ref{sct:filtering} may
still bundle multiple hard structured prongs,
since isolation requires a minimal separation of
$(\Delta\widetilde{R}_{AB} \ge \sqrt{2})$.
Within each quarantined partition,
SIFT reverts to its natural form,
as an exclusive clustering algorithm.
So, for example, it might be that reconstruction
of a doubly-hadronic $t\bar{t}$ event
would isolate the pair of top-quark remnants,
but merge each bottom with products from the associated $W$.
This is a favorable outcome, amounting to
the identification of variable large-radius jets.
But, the question of how to identify and recover the optimal
partition of each such object into $N$ subjets remains.

One could consider formulating a local halting condition
that would block the further assimilation of hard prongs
within each large-radius jet once some threshold were met. 
However, such objects are only defined in our prescription
\emph{after} having merged to exhaustion. 
More precisely, a number of candidate large-radius jets
may accumulate objects in parallel as clustering progresses,
and each will have secured a unique ``$N=1$'' configuration
at the moment of its isolation.

As such, the only available course of action appears to be proceeding
with these mergers, even potentially in the presence of substructure.
However, the fact that the SIFT measure tends to preserve mutually
hard features until the final stages of clustering suggests
that suitable proxies for the partonic event axes may be generated
automatically as a product of this sequential transition through
all possible subjet counts, especially during the last few mergers.
We refer to the superposed \emph{ensemble} of projections onto ($N = \ldots,\,3,\,2,\,1$)
prongs, i.e.,~the history of residually distinct four-vectors at each level
of the clustering flow, as an $N$-subjet tree.

In fact, it seems that interrupting the final stages of clustering
would amount to a substantial information forfeiture.
Specifically, the merger of axis candidates that are not collinear or relatively
soft imprints a sharp discontinuity on the measure, which
operates in this context like a mass-drop tagger~\cite{Butterworth:2008iy}
to flag the presence of substructure.
Additionally, the described procedure generates a basis of groomed axes that
are directly suitable for the computation of observables such as $N$-subjettiness.
In this sense, the best way to establish that a pair of constituents within a
large-radius jet should be kept apart may be to go ahead and join them,
yet to remember what has been joined and at which value of $\delta_{AB}$.

In contrast to conventional methods for substructure recovery
that involve de- and re-clustering according to a variety of disjoint prescriptions,
the finding of $N$-subjet trees representing a compound scattering
event occurs in conjunction with the filtering of stray
radiation and generation of substructure observables
during a single unified operational phase.
The performance of this approach for kinematic reconstruction
and tagging hard event prongs will be comparatively assessed
in Sections~\ref{sct:reconstruction} and~\ref{sct:tagging}.

\begin{figure*}[ht!]
\centering
\includegraphics[width=0.45\textwidth]{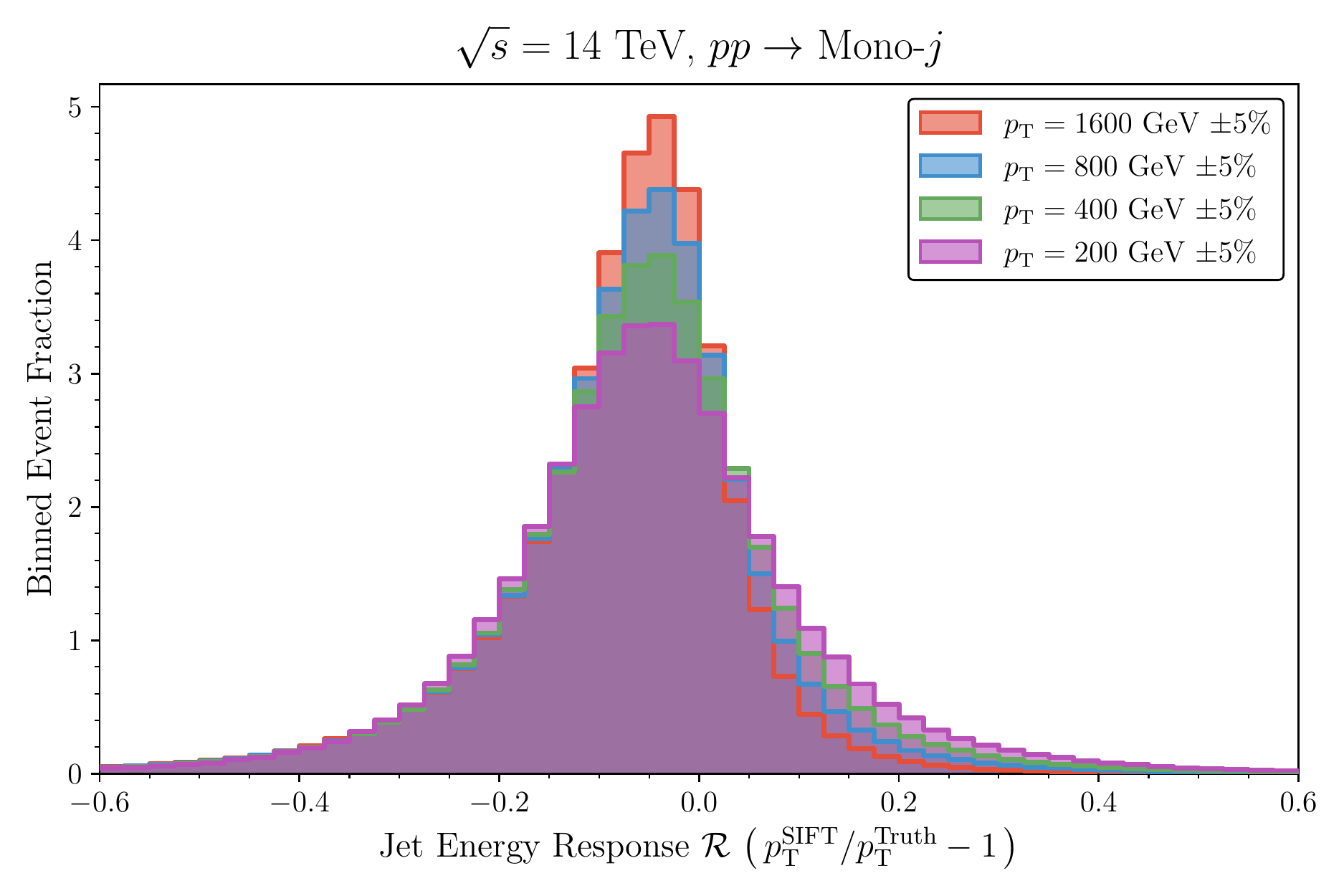} \hspace{12pt}
\includegraphics[width=0.45\textwidth]{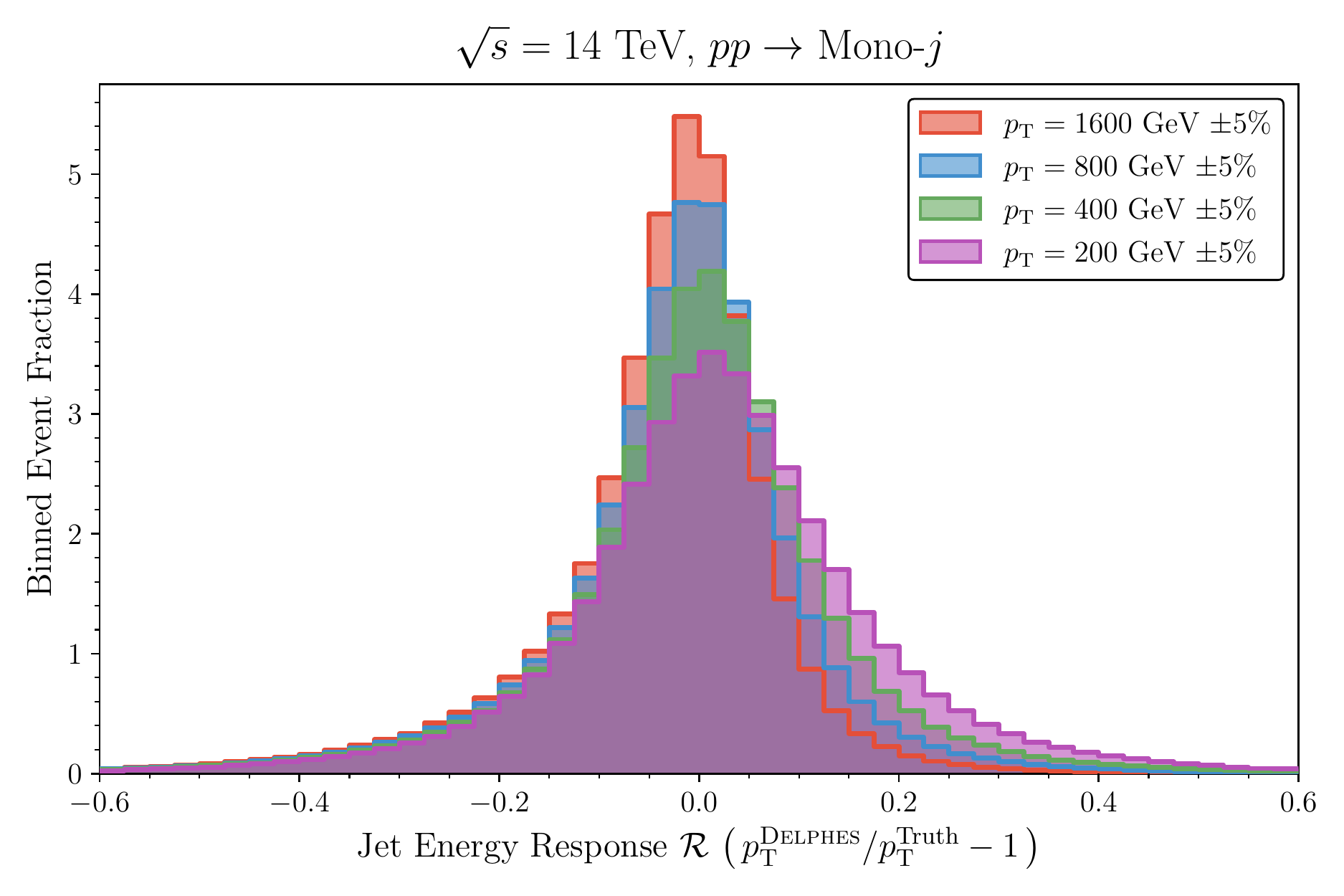} \\
\includegraphics[width=0.45\textwidth]{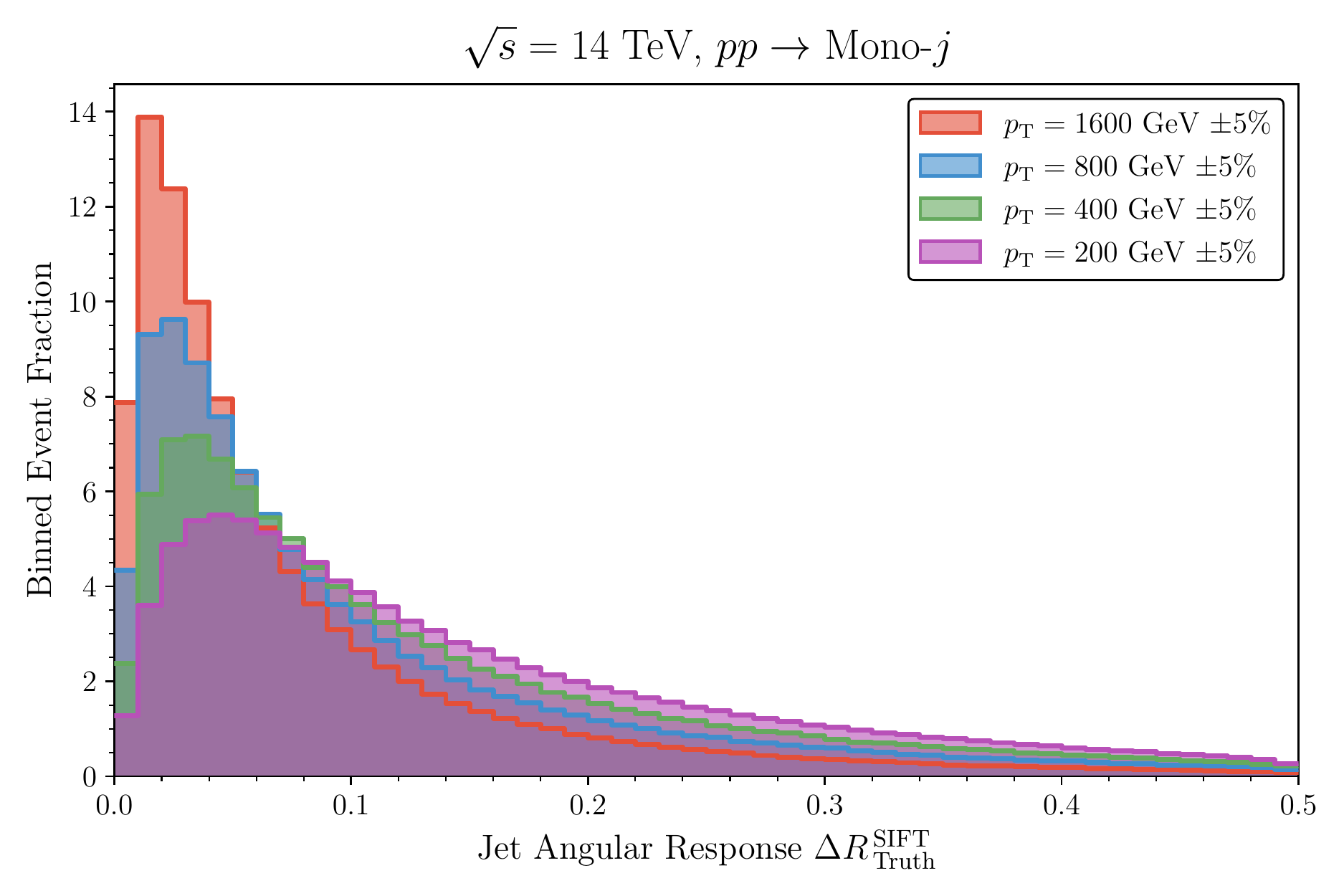} \hspace{12pt}
\includegraphics[width=0.45\textwidth]{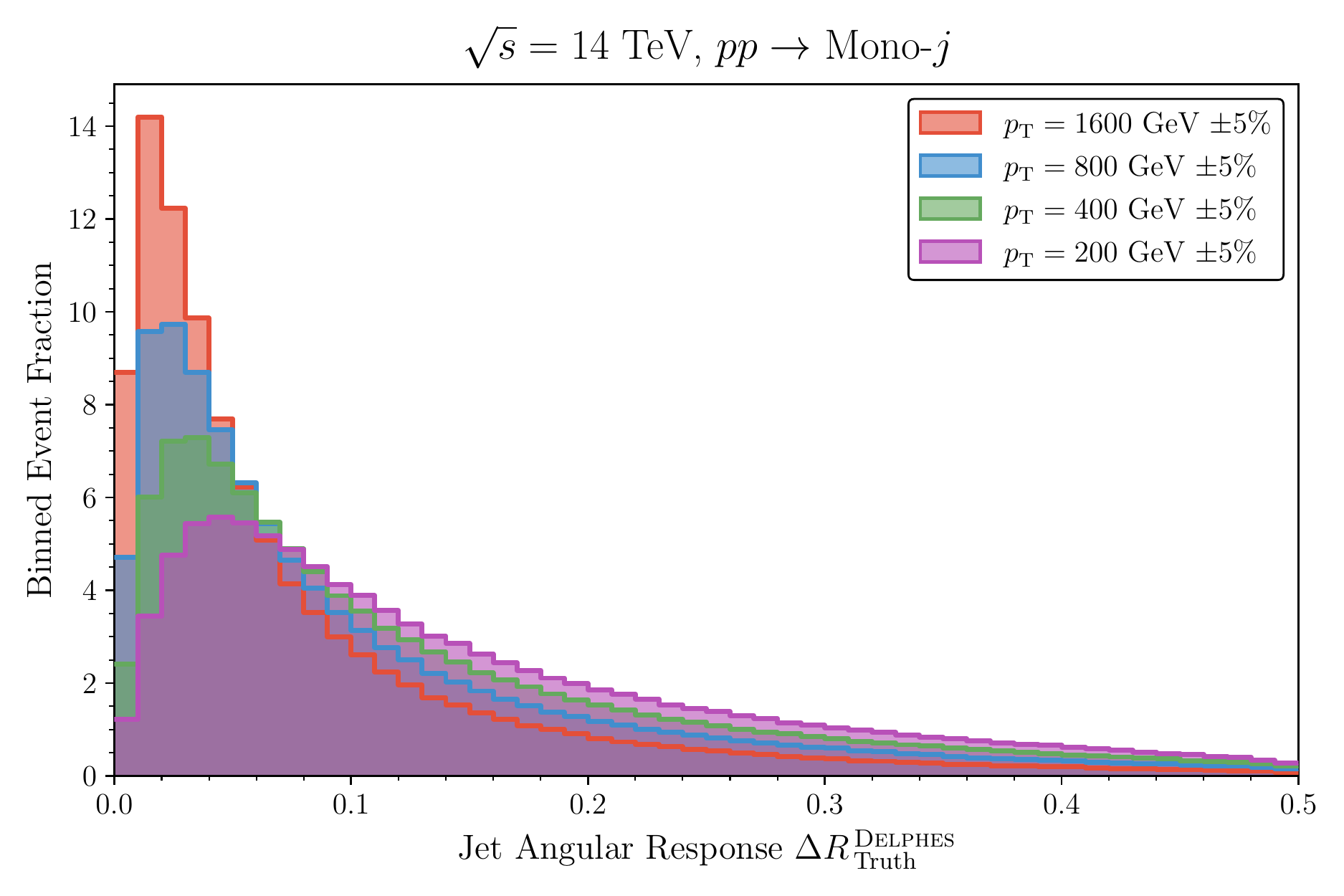}
\caption{\footnotesize
Top: Distribution $\mathcal{R}^A_B$ of reconstructed jet energy responses with detector effects relative to the partonic truth level at various transverse boosts.
Bottom: Distribution $\Delta R^{\,A}_{\,B}$ of reconstructed jet angular responses with detector effects relative to the partonic truth level at various transverse boosts.
Lefthand panels represent the leading filtered and isolated SIFT jet, while righthand panels represent the leading ($R_0 = 1$)
large-radius Soft Drop jet reported by {\sc Delphes}.  No calibration of jet energy scales is attempted for either category.}
\label{fig:JER_JAR}
\end{figure*}

\begin{table*}[htb!]
\vspace{18pt}
\bgroup
\def\arraystretch{1.25}
  \begin{center}
    \begin{tabular}{C{0.1025\textwidth}|C{0.1025\textwidth}|C{0.1025\textwidth}|C{0.1025\textwidth}|%
    C{0.1025\textwidth}|C{0.1025\textwidth}|C{0.1025\textwidth}|C{0.1025\textwidth}|C{0.1025\textwidth}}
$\pt^{{\rm GeV}\pm 5\%}$ &
$\langle \, \mathcal{R}^{\rm \,SIFT}_{\rm \,Truth} \, \rangle$ & $\sigma^{\rm SIFT}_{\mathcal{R}}$ &
$\langle \, \mathcal{R}^{\rm \,\textsc{Delphes}}_{\rm \,Truth} \, \rangle$ & $\sigma^{\rm \textsc{Delphes}}_{\mathcal{R}}$ &
$\langle \, \Delta R^{\rm \,SIFT}_{\rm \,Truth} \, \rangle$ & $\sigma^{\rm SIFT}_{\!\Delta R}$ &
$\langle \, \Delta R^{\rm \,\textsc{Delphes}}_{\rm \,Truth} \, \rangle$ & $\sigma^{\rm \textsc{Delphes}}_{\!\Delta R}$ \\ 
      \hline
100 & $-0.009$ & 0.17 & $+0.087$ & 0.17 & 0.17 & 0.12 & 0.17 & 0.12 \\
200 & $-0.046$ & 0.16 & $+0.026$ & 0.16 & 0.15 & 0.12 & 0.15 & 0.12 \\
400 & $-0.059$ & 0.15 & $-0.002$ & 0.15 & 0.13 & 0.11 & 0.13 & 0.11 \\
800 & $-0.071$ & 0.14 & $-0.024$ & 0.14 & 0.10 & 0.10 & 0.10 & 0.10 \\
1600 & $-0.081$ & 0.13 & $-0.042$ & 0.13 & 0.08 & 0.09 & 0.08 & 0.09 \\
3200 & $-0.089$ & 0.12 & $-0.058$ & 0.12 & 0.05 & 0.06 & 0.05 & 0.06 \\
    \end{tabular}
\caption{\footnotesize
Detector-level jet energy responses $\mathcal{R}^A_B$ and angular responses $\Delta R^A_B$
with associated resolutions $\sigma_{\mathcal{R}}$ and $\sigma_{\!\Delta R}$.}
    \label{tab:JER_JAR}
  \end{center}
\egroup
\end{table*}

\begin{figure*}[ht!]
\centering
\includegraphics[width=0.45\textwidth]{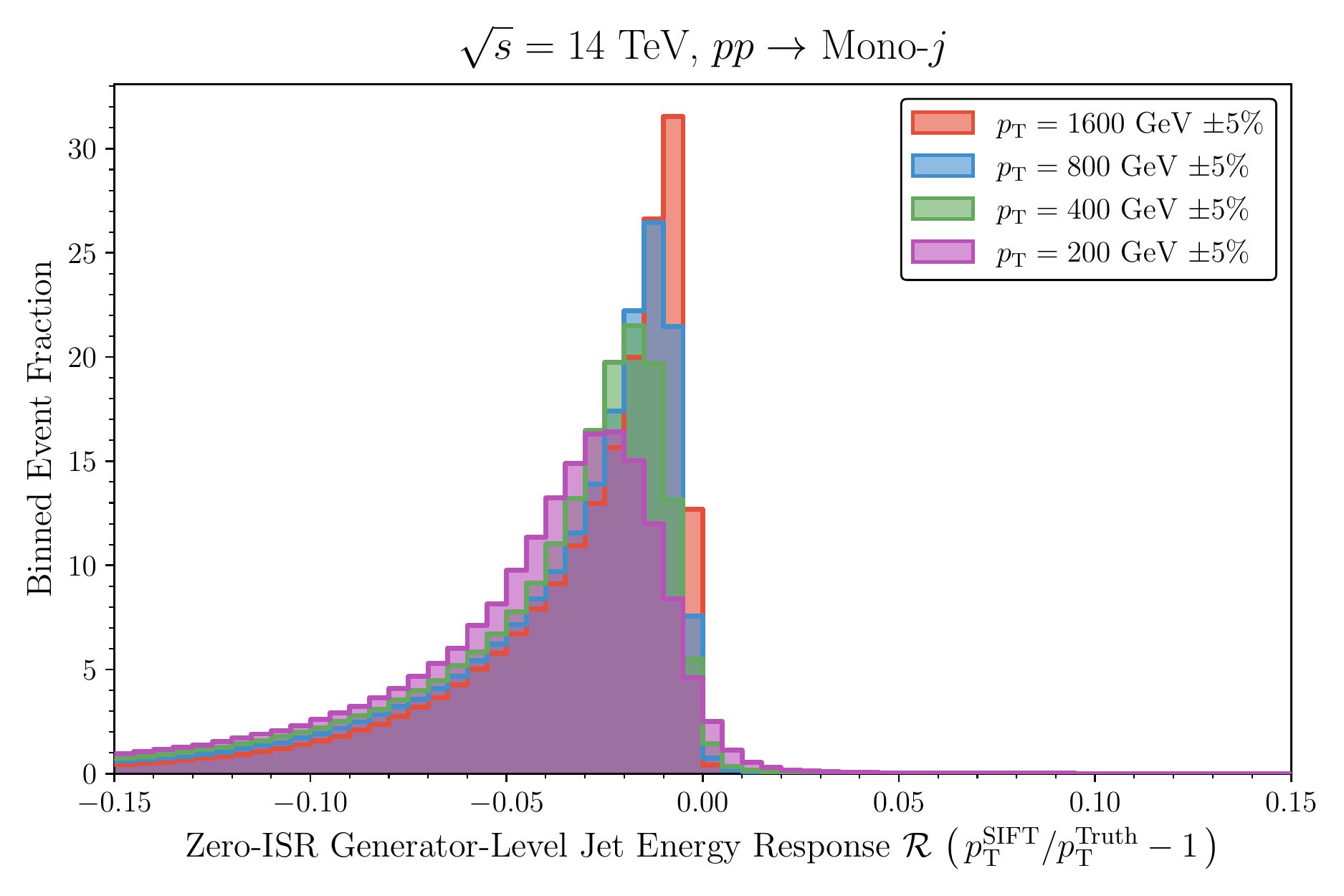} \hspace{12pt}
\includegraphics[width=0.45\textwidth]{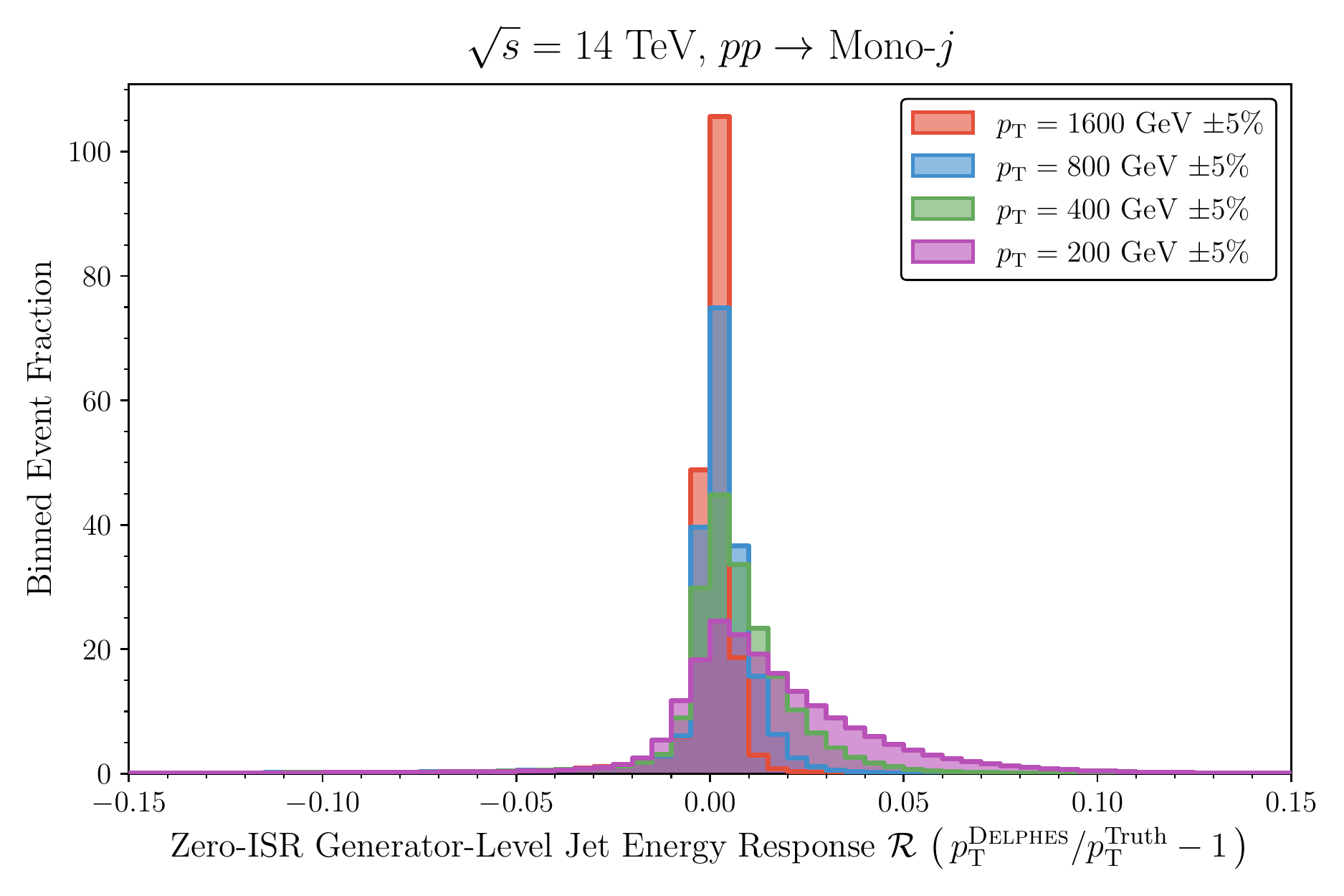} \\
\includegraphics[width=0.45\textwidth]{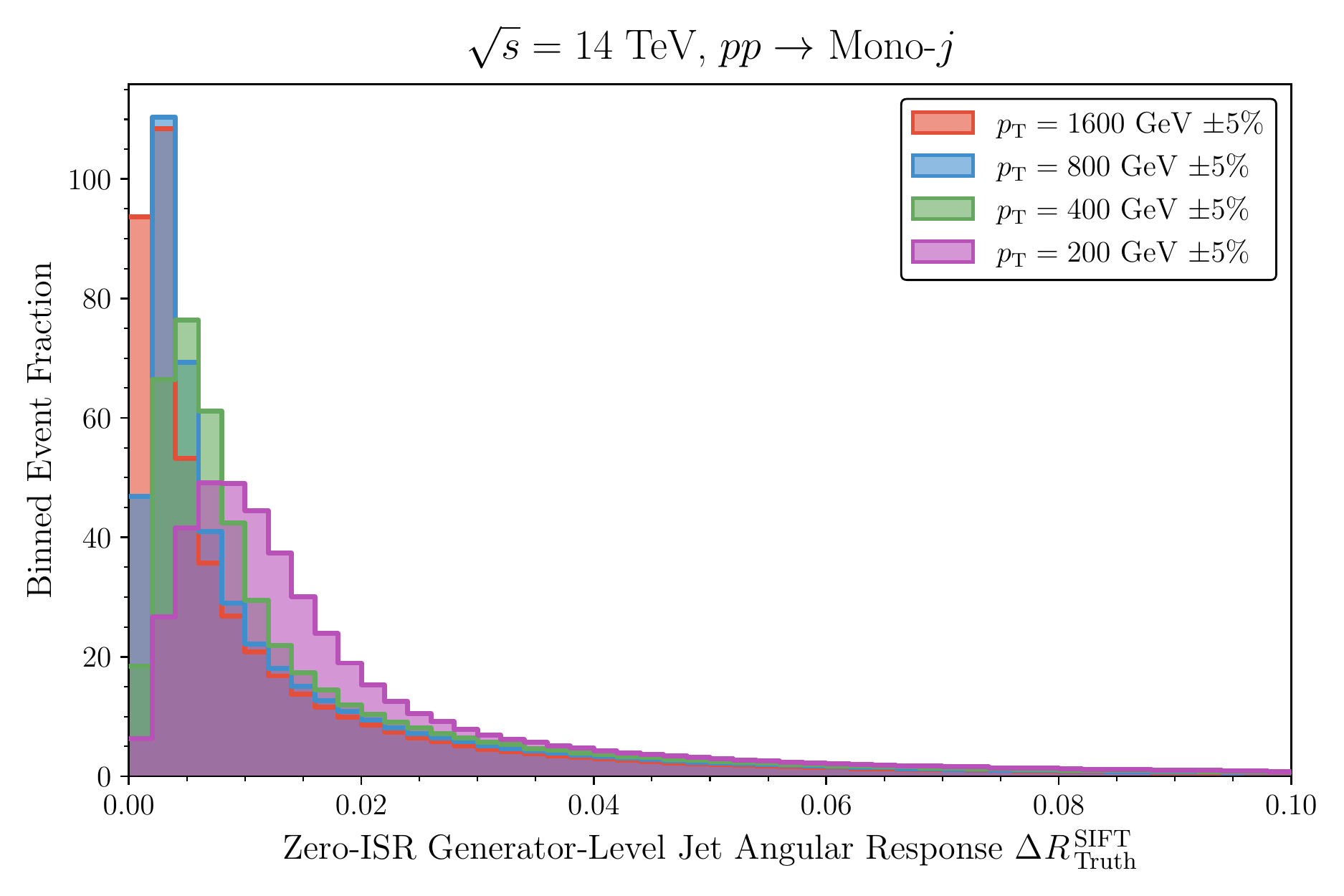} \hspace{12pt}
\includegraphics[width=0.45\textwidth]{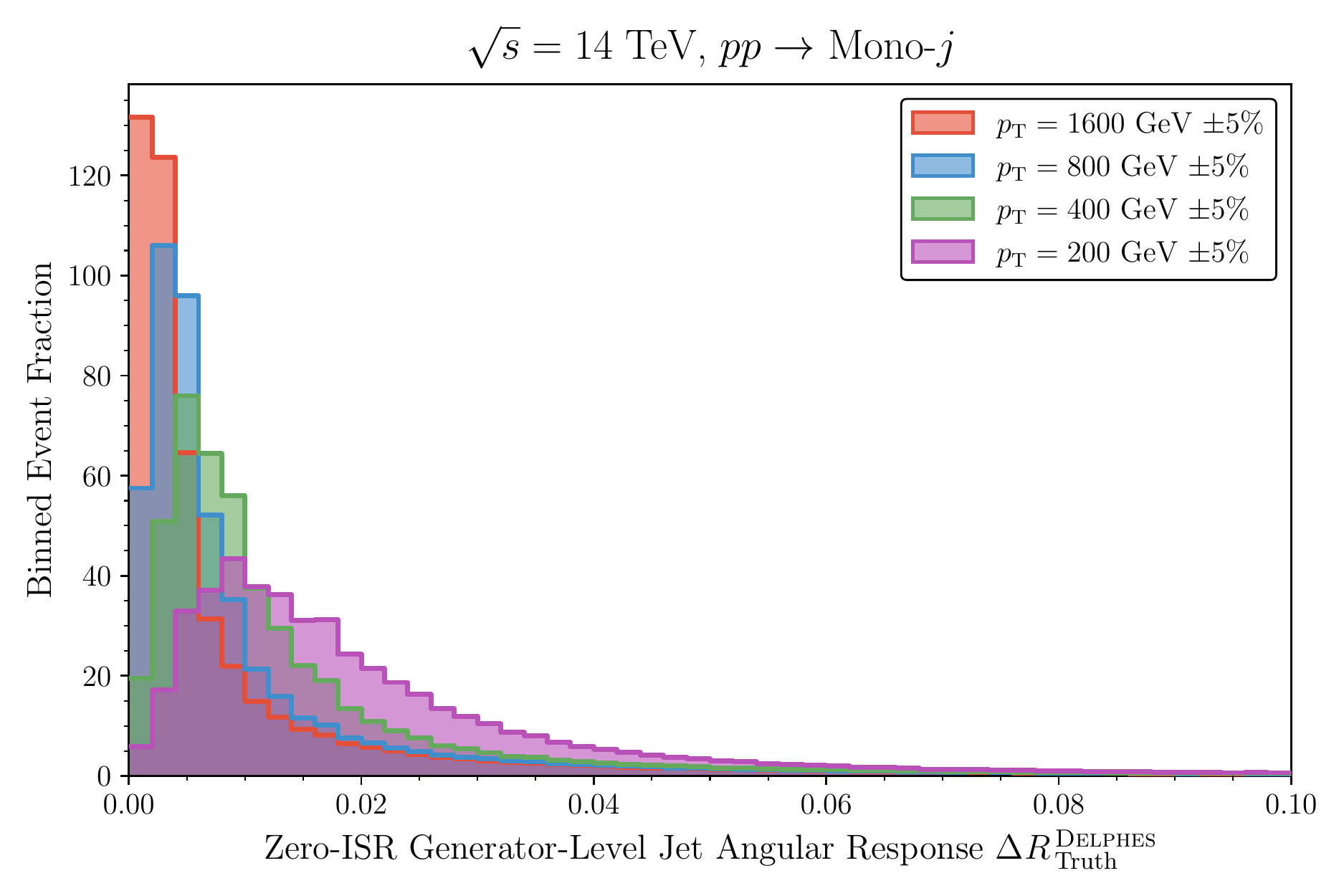}
\caption{\footnotesize
Top: Distribution $\mathcal{R}^A_B$ of reconstructed jet energy responses, at generator level and without initial-state radiation, relative to the partonic truth level at various transverse boosts.
Bottom: Distribution $\Delta R^{\,A}_{\,B}$ of reconstructed jet angular responses, at generator level and without initial-state radiation, relative to the partonic truth level at various transverse boosts.
Lefthand panels represent the leading filtered and isolated SIFT jet, while righthand panels represent the leading ($R_0 = 1$)
large-radius Soft Drop jet reported by {\sc Delphes}.  No calibration of jet energy scales is attempted for either category.}
\label{fig:JER_JAR_ZERO-ISR_GEN-JET}
\end{figure*}

\begin{table*}[htb!]
\vspace{18pt}
\bgroup
\def\arraystretch{1.25}
  \begin{center}
    \begin{tabular}{C{0.1025\textwidth}|C{0.1025\textwidth}|C{0.1025\textwidth}|C{0.1025\textwidth}|%
    C{0.1025\textwidth}|C{0.1025\textwidth}|C{0.1025\textwidth}|C{0.1025\textwidth}|C{0.1025\textwidth}}
$\pt^{{\rm GeV}\pm 5\%}$ &
$\langle \, \mathcal{R}^{\rm \,SIFT}_{\rm \,Truth} \, \rangle$ & $\sigma^{\rm SIFT}_{\mathcal{R}}$ &
$\langle \, \mathcal{R}^{\rm \,\textsc{Delphes}}_{\rm \,Truth} \, \rangle$ & $\sigma^{\rm \textsc{Delphes}}_{\mathcal{R}}$ &
$\langle \, \Delta R^{\rm \,SIFT}_{\rm \,Truth} \, \rangle$ & $\sigma^{\rm SIFT}_{\!\Delta R}$ &
$\langle \, \Delta R^{\rm \,\textsc{Delphes}}_{\rm \,Truth} \, \rangle$ & $\sigma^{\rm \textsc{Delphes}}_{\!\Delta R}$ \\ 
      \hline
100 & $-0.059$ & 0.054 & $+0.034$ & 0.052 & 0.038 & 0.052 & 0.043 & 0.049 \\
200 & $-0.052$ & 0.049 & $+0.013$ & 0.040 & 0.030 & 0.050 & 0.030 & 0.050 \\
400 & $-0.045$ & 0.045 & $+0.004$ & 0.033 & 0.025 & 0.047 & 0.022 & 0.047 \\
800 & $-0.040$ & 0.042 & $-0.001$ & 0.031 & 0.022 & 0.045 & 0.018 & 0.046 \\
1600 & $-0.035$ & 0.039 & $-0.003$ & 0.029 & 0.018 & 0.040 & 0.015 & 0.042 \\
3200 & $-0.031$ & 0.035 & $-0.004$ & 0.025 & 0.015 & 0.032 & 0.011 & 0.033 \\
    \end{tabular}
\caption{\footnotesize
Zero-ISR generator-level jet energy responses $\mathcal{R}^A_B$ and angular responses $\Delta R^A_B$
with associated resolutions $\sigma_{\mathcal{R}}$ and $\sigma_{\!\Delta R}$.}
    \label{tab:JER_JAR_ZERO-ISR_GEN-JET}
  \end{center}
\egroup
\end{table*}

\begin{figure*}[ht!]
\centering
\includegraphics[width=.45\textwidth]{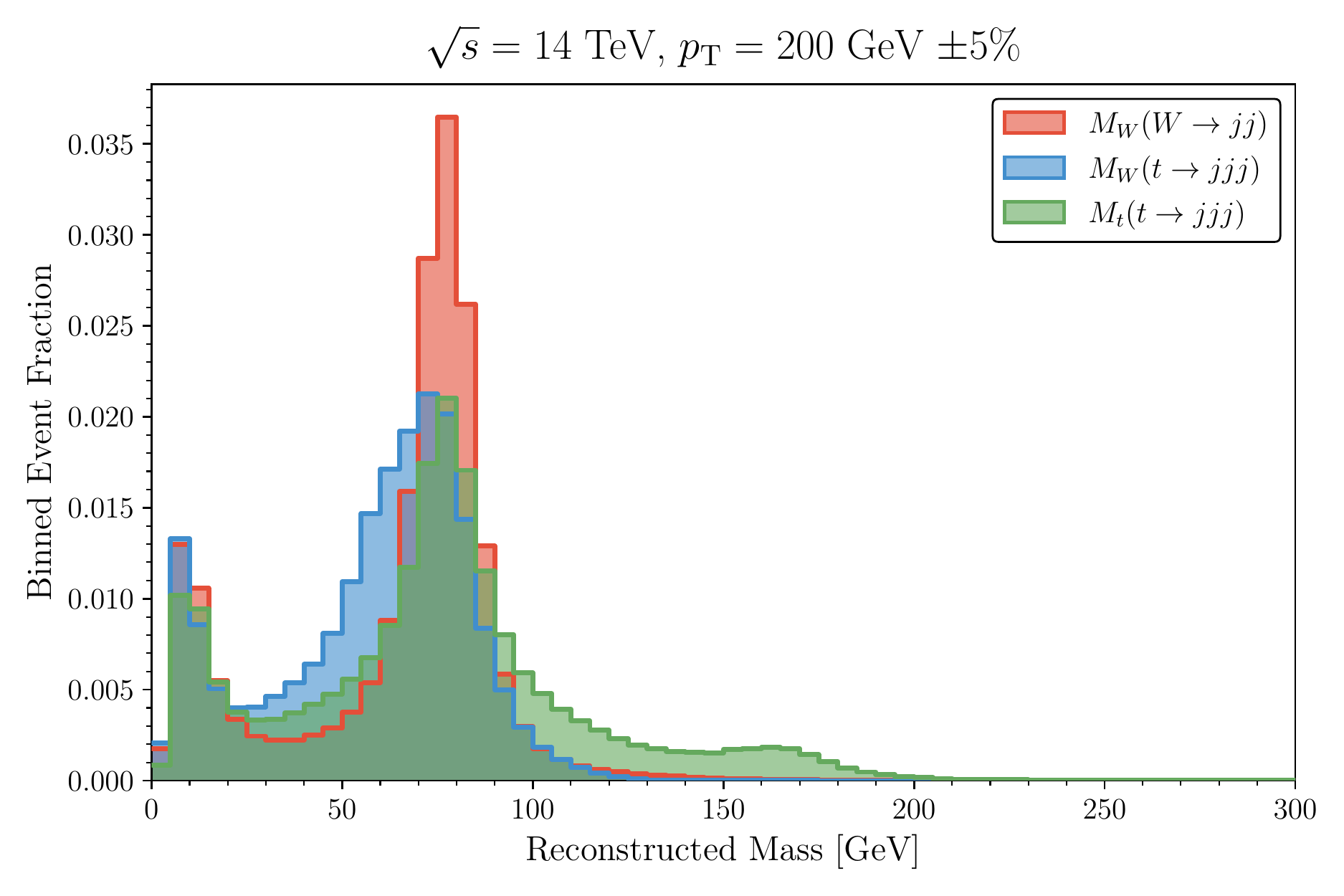} \hspace{12pt}
\includegraphics[width=.45\textwidth]{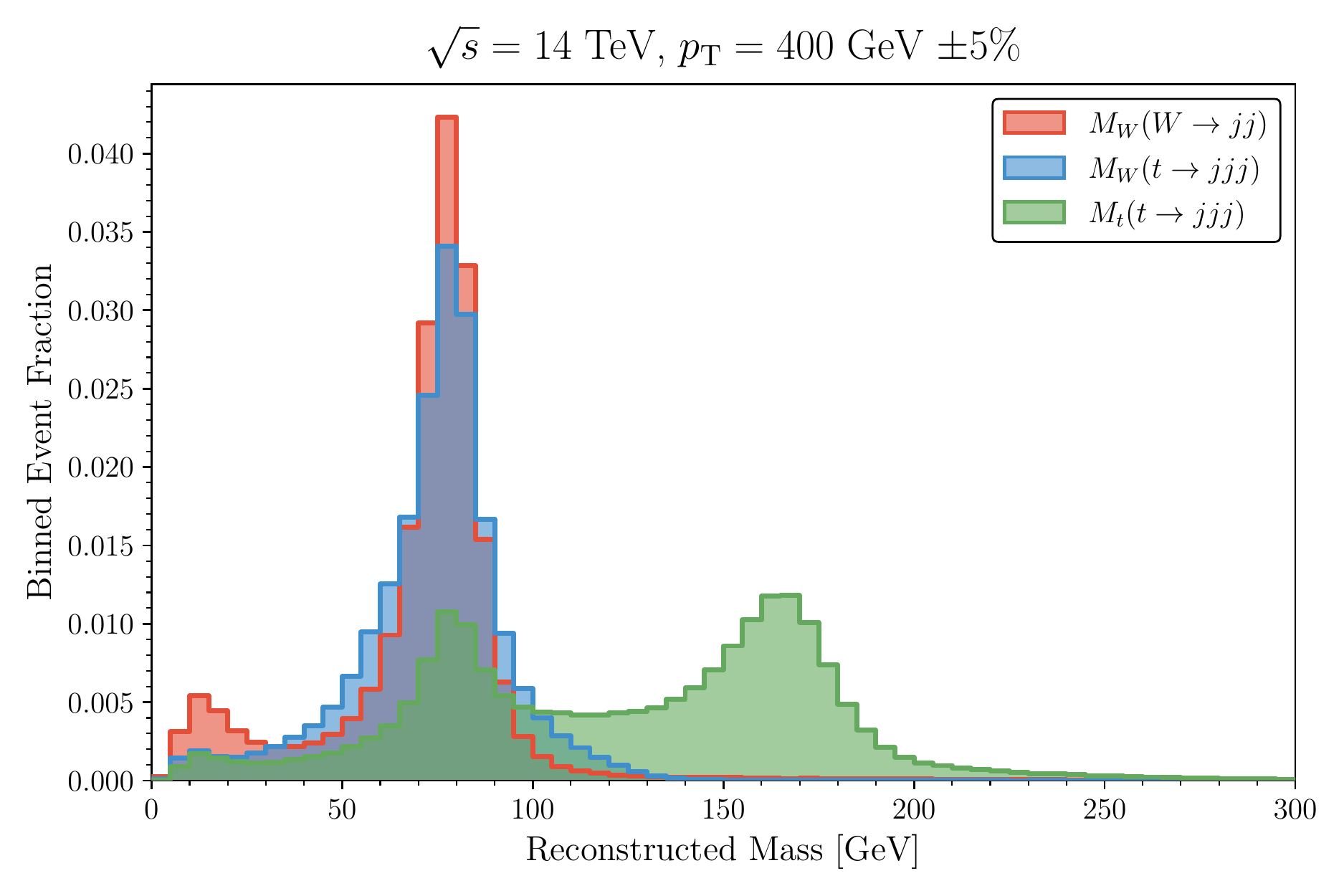} \\
\includegraphics[width=.45\textwidth]{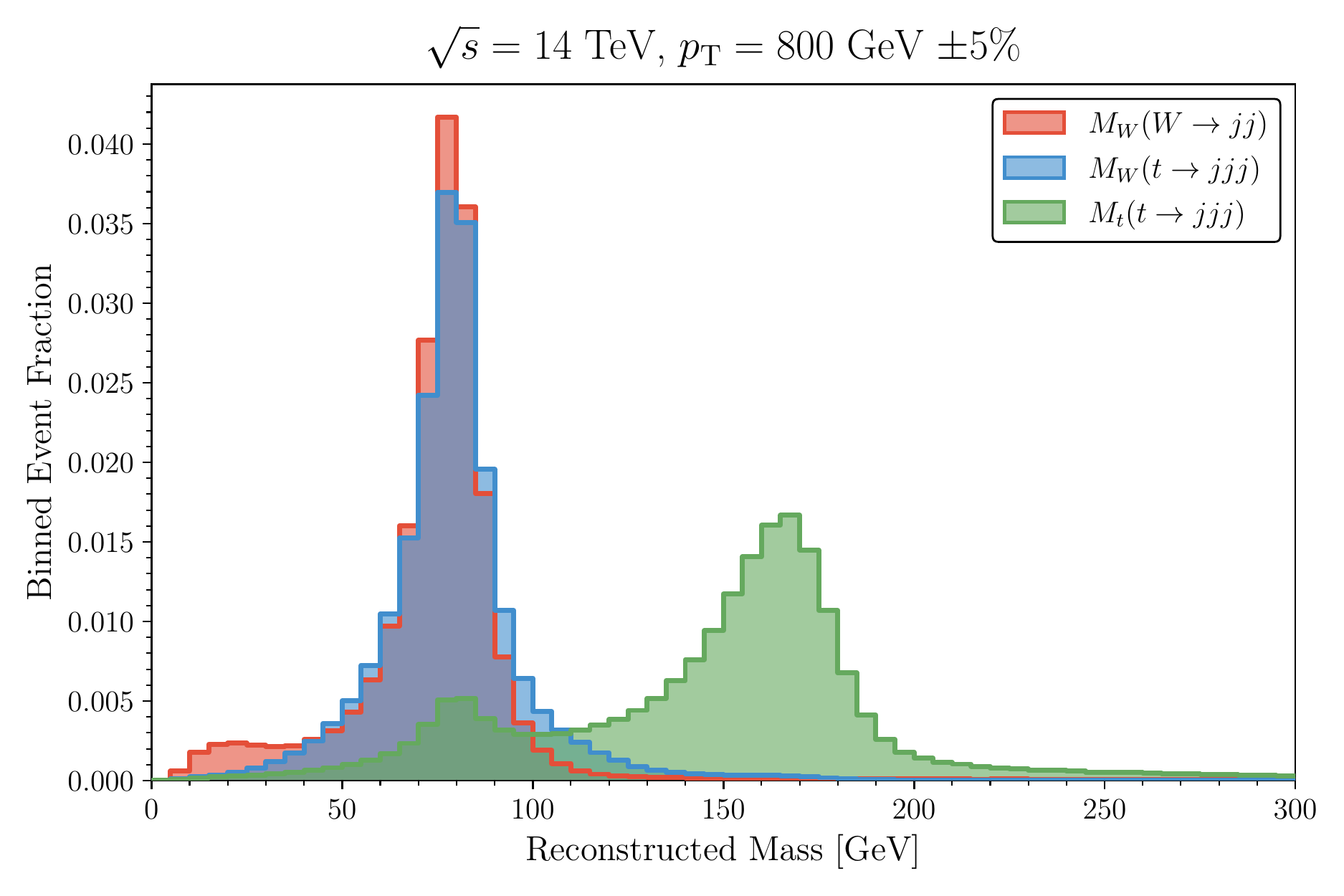} \hspace{12pt}
\includegraphics[width=.45\textwidth]{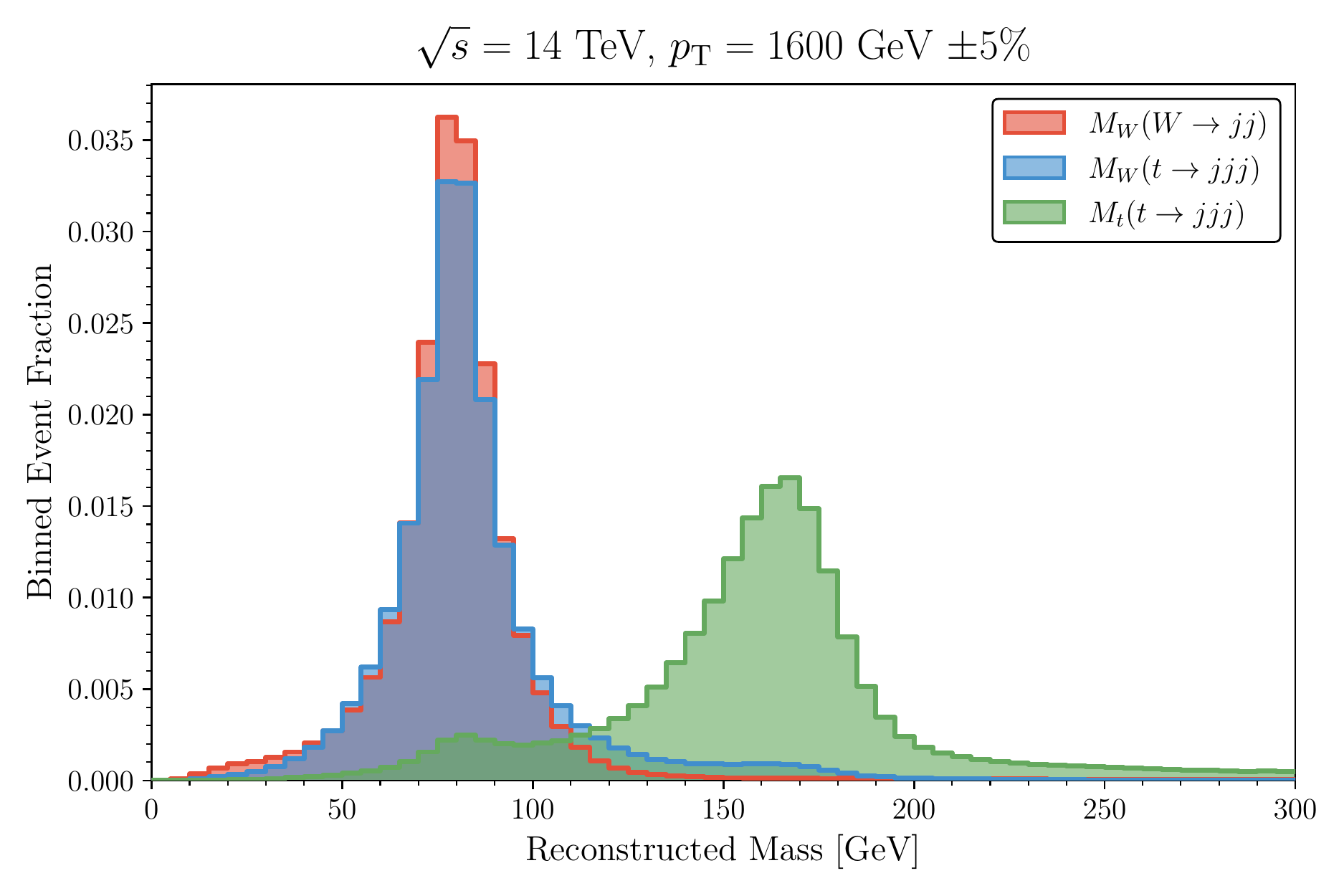}
\caption{\footnotesize
Distribution of $W$-boson and top quark masses for di, and tri-jet samples reconstructed with SIFT at various transverse boosts.}
\label{fig:mass}
\end{figure*}

\section{Comparison of Algorithms\label{sct:comparison}}

This section provides a visual comparison of merging priorities and final states
for the SIFT and $\kt$-family algorithms.  The images presented here are
still frames extracted from full-motion video simulations of each clustering sequence.
These films are provided as ancillary files with the source package
for this paper on the arXiv and may also be viewed on YouTube~\cite{youtube}.
Frames are generated for every 25th clustering action, as well as each of the initial and final 25 actions.
The Mathematica notebook used to generate these films is described in Appendix~\ref{sct:software},
and maintained with the AEACuS~\cite{aeacus} package on GitHub.

The visualized event\footnote{
This event is expected to be reasonably representative,
being the first member of its Monte Carlo sample.}
comes from a simulation in {\sc MadGraph}/{\sc MadEvent}~\cite{Alwall:2014hca}
of top quark pair production at the 14~TeV LHC with fully hadronic decays.
The scalar sum $H_{\rm T}$ of transverse momentum is approximately 1.6~TeV at the partonic level.
This large boost results in narrow collimation of the three hard prongs (quarks)
on either side of the event, as depicted in the the upper frame of FIG.~\ref{fig:filmframesEVT}.
In-plane axes represent the pseudo-rapidity $\eta$ and azimuth $\phi$, with a cell width
($\Delta R \simeq 0.1$) approximating the resolution of a modern hadronic calorimeter.
The height of each block is proportional to the log-transverse momentum $\log_{10} (\,\pt\,/{\rm [GeV]})$ it contains.

The event is passed through through {\sc Pythia8}~\cite{Sjostrand:2014zea} for showering and hadronization,
and through {\sc Delphes}~\cite{deFavereau:2013fsa} for fast detector simulation.
Detector effects are bypassed in the current context, which starts with unclustered generator-level
({\sc Pythia8}) jets and non-isolated photons/leptons extracted from the {\sc Delphes} event record by {\sc AEACuS},
but they will be included for most of the analysis in Sections~\ref{sct:reconstruction} and~\ref{sct:tagging}.
This initial state is depicted in the lower frame of FIG.~\ref{fig:filmframesEVT},
which exhibits two dense clusters of radiation that are clearly associated with
the partonic event, as well as several offset deposits having less immediate origins
in the underlying event or initial state.
Prior to clustering, a sheet of ultra-soft ghost radiation is distributed across the angular
field in order to highlight differences between the catchment area \cite{Cacciari:2008gn} of each algorithm.

We begin with an example of exclusive ($N_{\rm exc}=1$) clustering ordered by the SIFT measure from Eq.~\ref{eq:siftmeasure},
but without application of the filtering and isolation criteria described in Section~\ref{sct:filtering}.
Film~A clearly exhibits both of the previously identified pathologies,
opening with a sweep of soft-wide radiation by harder partners
(as visualized with regions of matching coloration) and closing
with the contraction of hard-wide structures into a single surviving object.
However, the described success is manifest in between, vis-\`a-vis mutual preservation
of narrowly bundled hard prongs until the end stages of clustering.
In particular, the upper frame of FIG.~\ref{fig:filmframesA} features a pair of triplets at ($\delta_{AB} \simeq 0.04$)
that fairly approximate their collinear antecedents despite bearing wide catchments.
Subsequently, this substructure collapses into an image of the original pair production,
as depicted at ($\delta_{AB} \simeq 0.3$) in the lower frame of FIG.~\ref{fig:filmframesA}.
Nothing further occurs until ($\delta_{AB} \gtrsim 1.0$), beyond which
residual structures begin to merge and migrate in unphysical ways.

For comparison, we process the same event using the anti-$\kt$ algorithm at ($R_0 = 0.5$).
Film~B shows how early activity is dominated by the hardest radiative seeds, which
promptly capture all available territory up to the stipulated radial boundary.
In particular, any substructure that is narrower than $R_0$ will be rapidly erased,
as illustrated in the upper frame of FIG.~\ref{fig:filmframesB}.
The subsequent stages of clustering are of lesser interest, being progressively
occupied with softer seeds gathering up yet softer unclaimed scraps.
At termination, the lower frame of FIG.~\ref{fig:filmframesB} exhibits the regular
cone shapes with uniform catchment areas that are a hallmark of anti-$\kt$.
This property is linked to the anchoring of new cones on hard prongs that 
are less vulnerable to angular drift.  It is favored by experimentalists for facilitating
calibration of jet energy scales and subtraction of soft pileup radiation.

Proceeding, we repeat the prior exercise using the $\kt$ algorithm at ($R_0 = 0.5$).
In contrast to anti-$\kt$, clustering is driven here by the softest seeds.
Film~C demonstrates the emergence of a fine grain structure in the
association pattern of objects from adjacent regions that grows in
size as the algorithm progresses.  Unlike SIFT, which preferentially
binds soft radiation to a hard partner, mutually soft objects without a strong
physical correlation are likely to pair in this case.  Since summing geometrically
adjacent partners tends increase $\pt$, merged objects become
less immediately attractive to the measure.  As a result, activity is
dispersed widely across the plane, and attention jumps rapidly 
from one location to the next.  However, the combination of mutually hard prongs
is actively deferred, which causes collimated substructures to be preserved,
as shown in the upper frame of FIG.~\ref{fig:filmframesC}.
In contrast to SIFT, this hardness criterion is absolute, rather than relative.
Ultimately, structures more adjacent than the fixed angular cutoff $R_0$ will still be absorbed.
Jet centers are likely to drift substantially,
and associated catchment shapes are thus highly irregular, as shown in the
lower frame of FIG.~\ref{fig:filmframesC}.

Similarly, we also cluster using the Cambridge Aachen algorithm at ($R_0 = 0.5$).
Pairings are driven here solely by angular proximity, and Film~D shows an associated
growth of grain size that is like that of the $\kt$ algorithm.
The banded sequencing is simply an artifact of the way we disperse
ghost jets, randomizing $\pt$ but regularizing placement on the grid.
In contrast, mutually hard substructures are not specifically protected
and will last only until the correlation length catches up to their
separation, as shown in the upper frame of FIG.~\ref{fig:filmframesD}.
As before, the angular cutoff $R_0$ limits resolution of structure.
Likewise, jet drift leads to irregular catchment shapes, as shown in the
lower frame of FIG.~\ref{fig:filmframesD}.

We conclude this section with a reapplication of the SIFT algorithm,
enabling the filtering and isolation criteria from Section~\ref{sct:filtering}.
Film~E demonstrates that the soft ghost radiation is still targeted first,
but it is now efficiently discarded (as visualized with dark grey)
rather than clustered, suggesting resiliency to soft pileup.
Hard substructures are resolved without accumulating stray radiation,
as shown in the upper frame of FIG.~\ref{fig:filmframesE}.
This helps to stabilize reconstructed jet kinematics relative to the parton-level event.
Objects decayed and showered from the pair of opposite-hemisphere
top quarks are fully isolated from each other in the final state,
as shown in the lower frame of FIG.~\ref{fig:filmframesE}.
Each such object selectively associates constituents within a
``fuzzy'' scale-dependent catchment boundary.
Nevertheless, it may still be possible to establish an ``effective''
jet radius by integrating the pileup distribution function
up to the maximal radius $(\Delta \widetilde{R}_{AB} < \sqrt{2})$.
In any case, the traditional approach
to pileup subtraction has been somewhat superseded by the
emergence of techniques for event-by-event, and per-particle
pileup estimation like {\sc PUPPI}~\cite{Bertolini:2014bba}.

\begin{figure*}[ht!]
\centering
\includegraphics[width=0.45\textwidth]{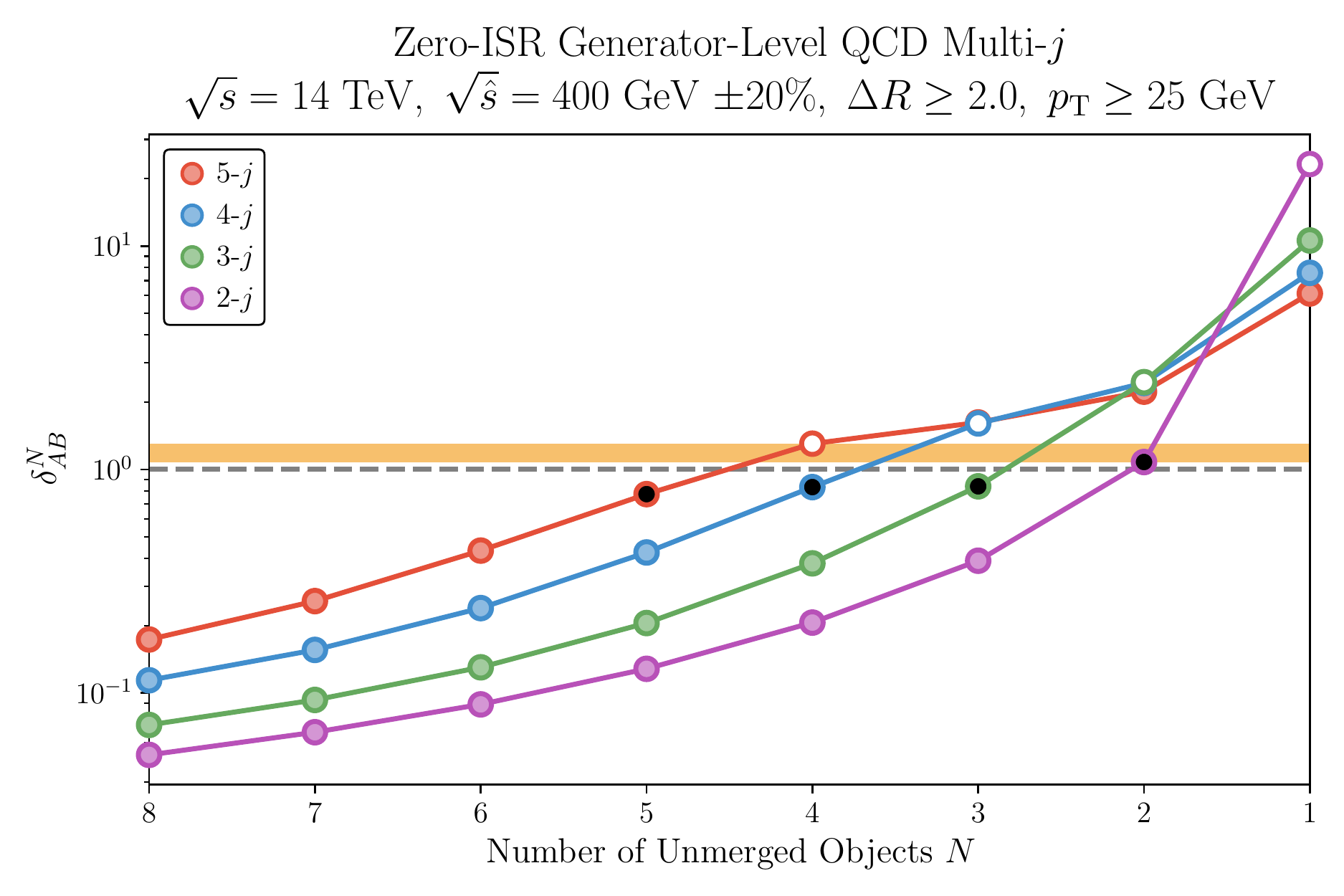} \hspace{12pt}
\includegraphics[width=0.45\textwidth]{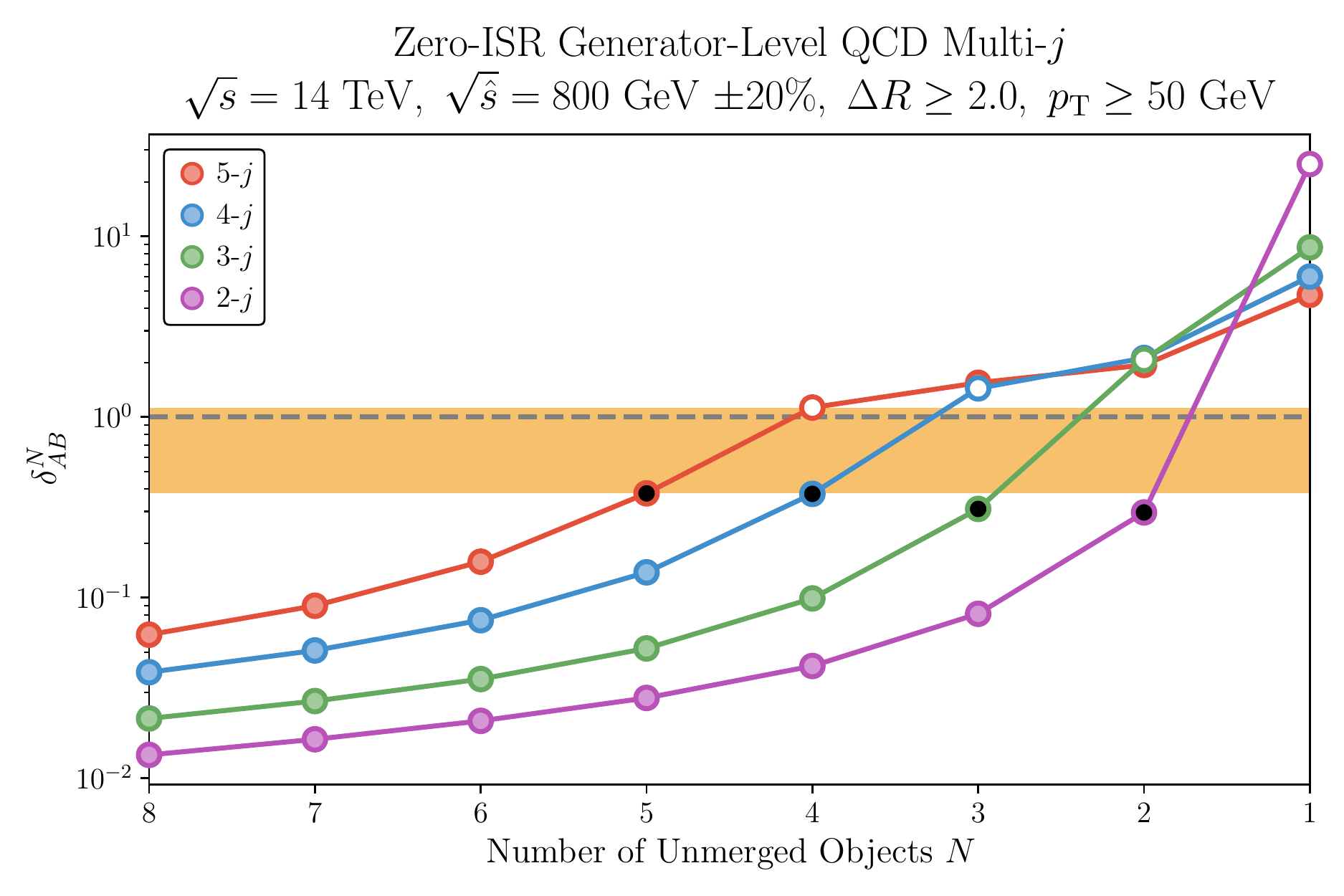} \\
\includegraphics[width=0.45\textwidth]{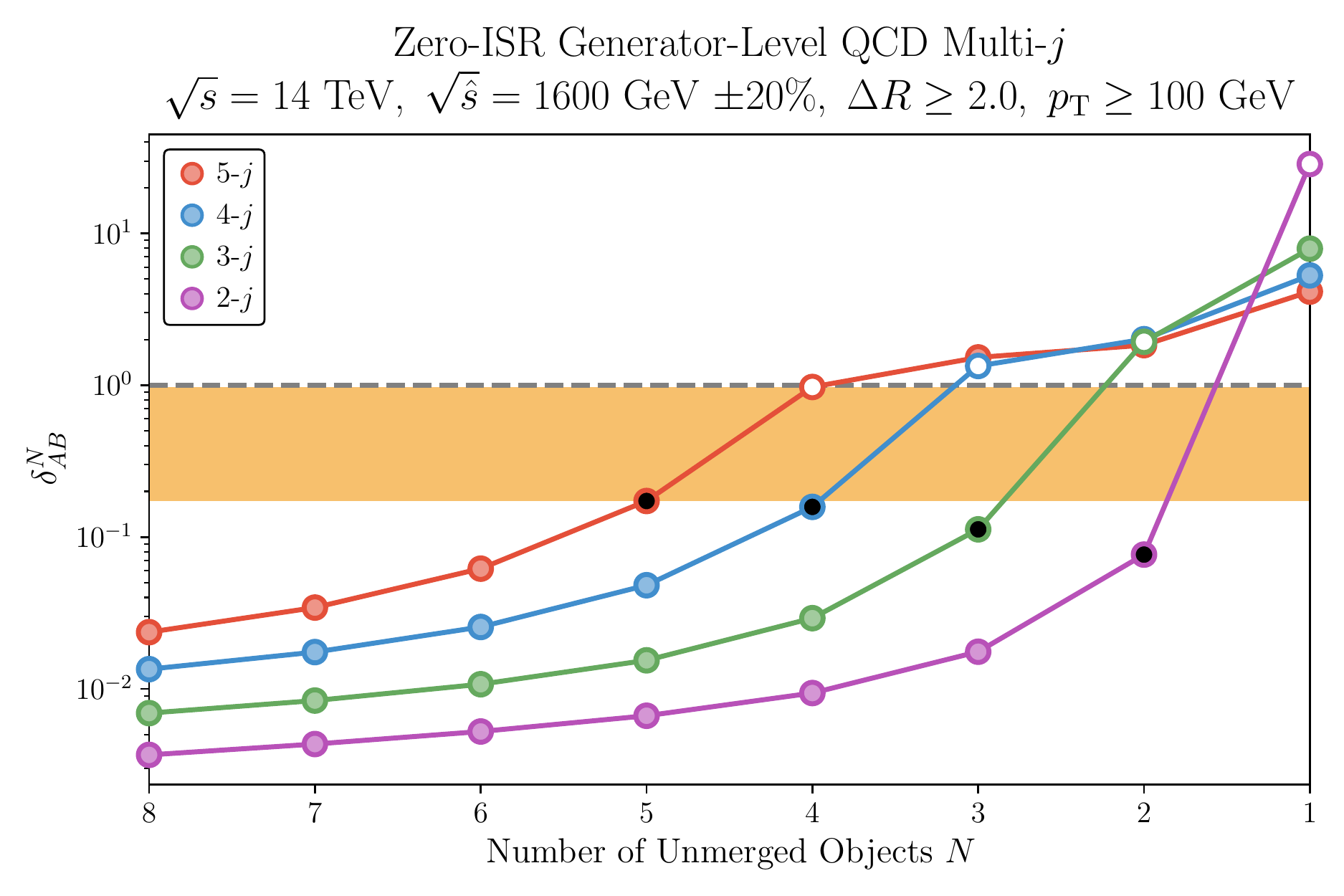} \hspace{12pt}
\includegraphics[width=0.45\textwidth]{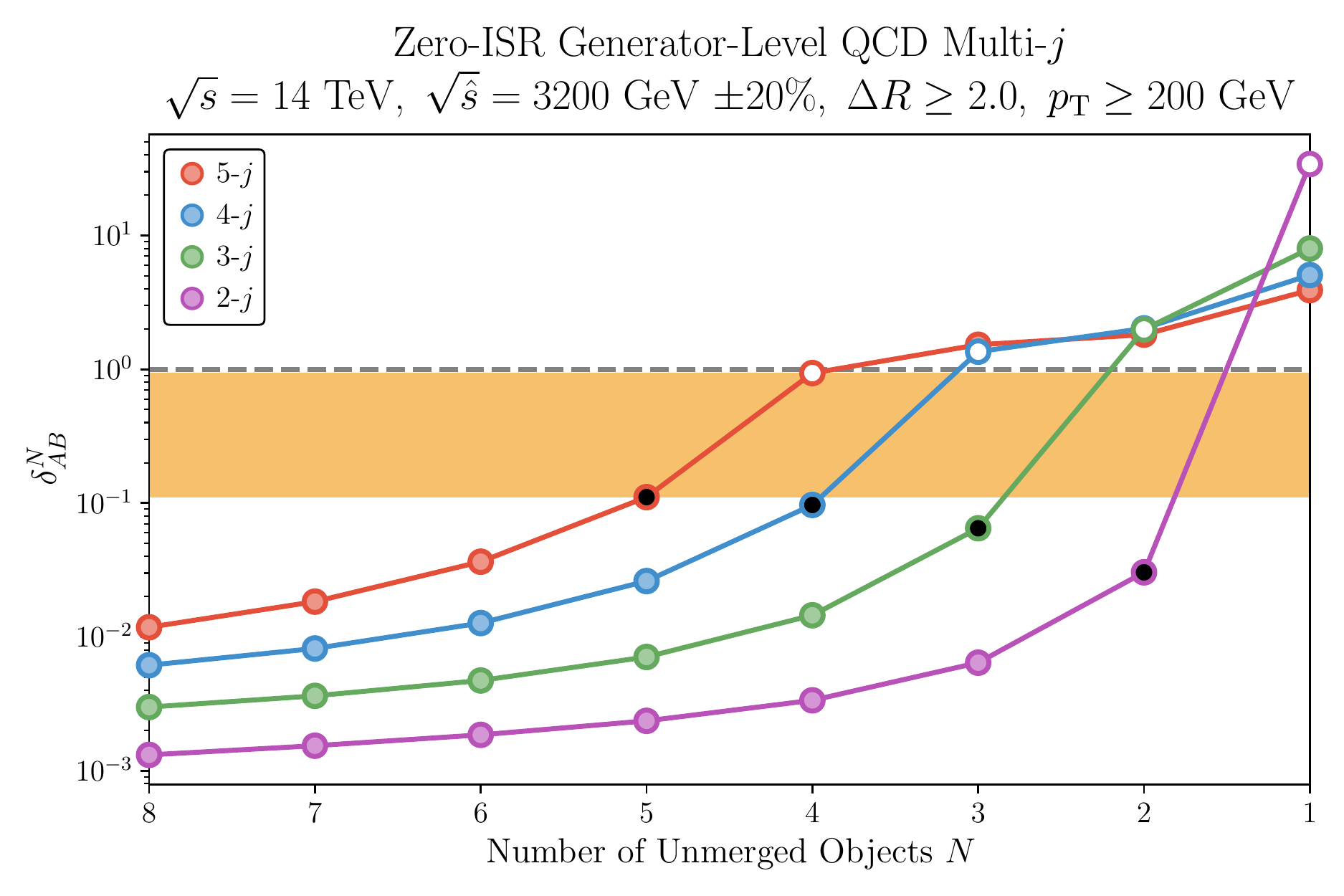}
\caption{\footnotesize
Evolution of the measure $\delta^N_{AB}$
as a function of the number $N$ of unmerged objects for QCD
multi-jet production at various values of $\sqrt{\hat{s}}$.
Partons have an angular separation ($\Delta R \geq 2.0$)
and the $\pt$ threshold is stepped in proportion to $\sqrt{\hat{s}}$.
Initial-state radiation is suppressed and analysis is at generator level.
The orange band indicates the interval where all samples have
merged to the point of their natural partonic count (white) but not beyond (black).
The grey dashed line marks the isolation threshold at ($\delta_{AB}=1$).
}
\label{fig:evo}
\end{figure*}

\begin{figure*}[ht!]
\centering
\includegraphics[width=.45\textwidth]{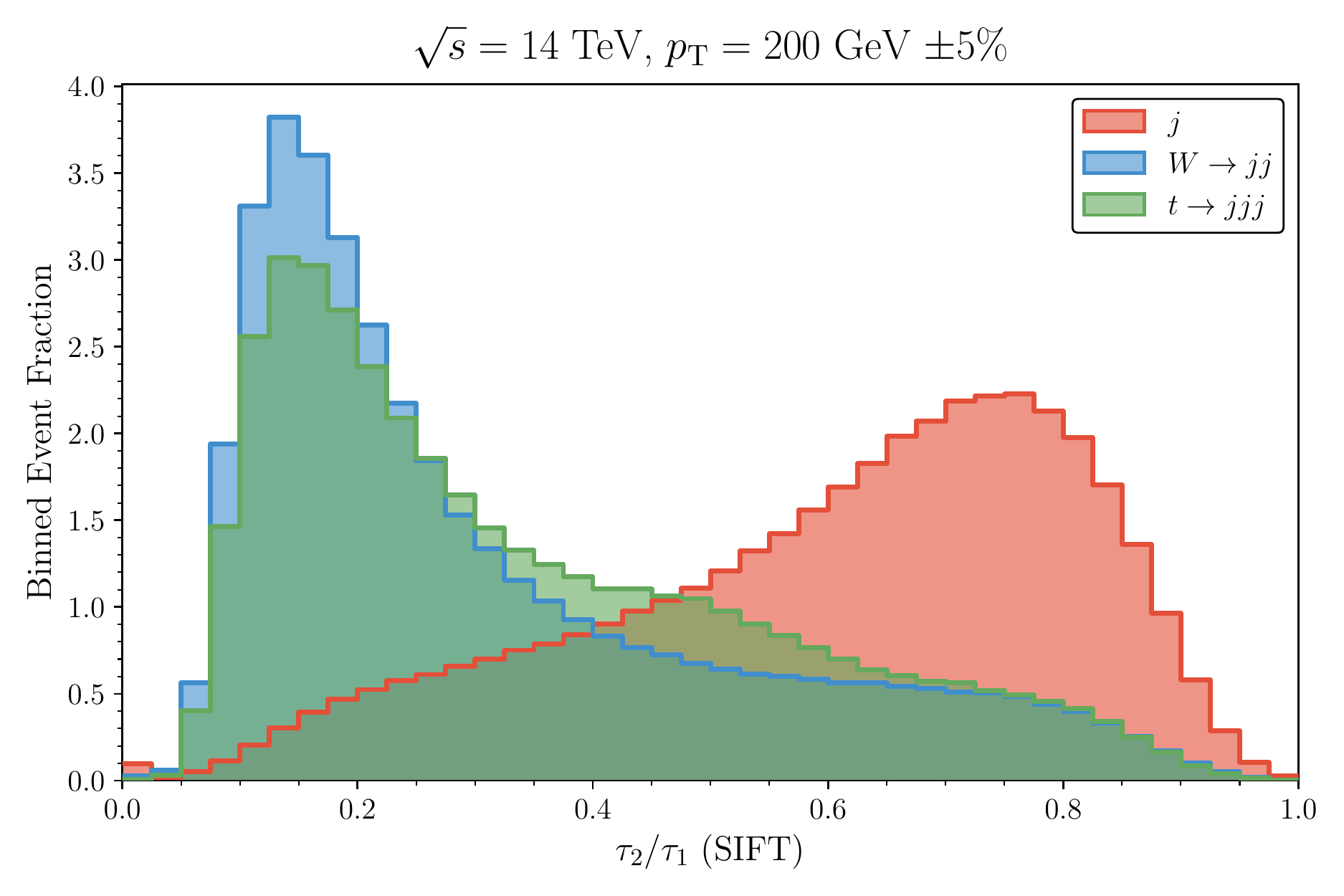} \hspace{12pt}
\includegraphics[width=.45\textwidth]{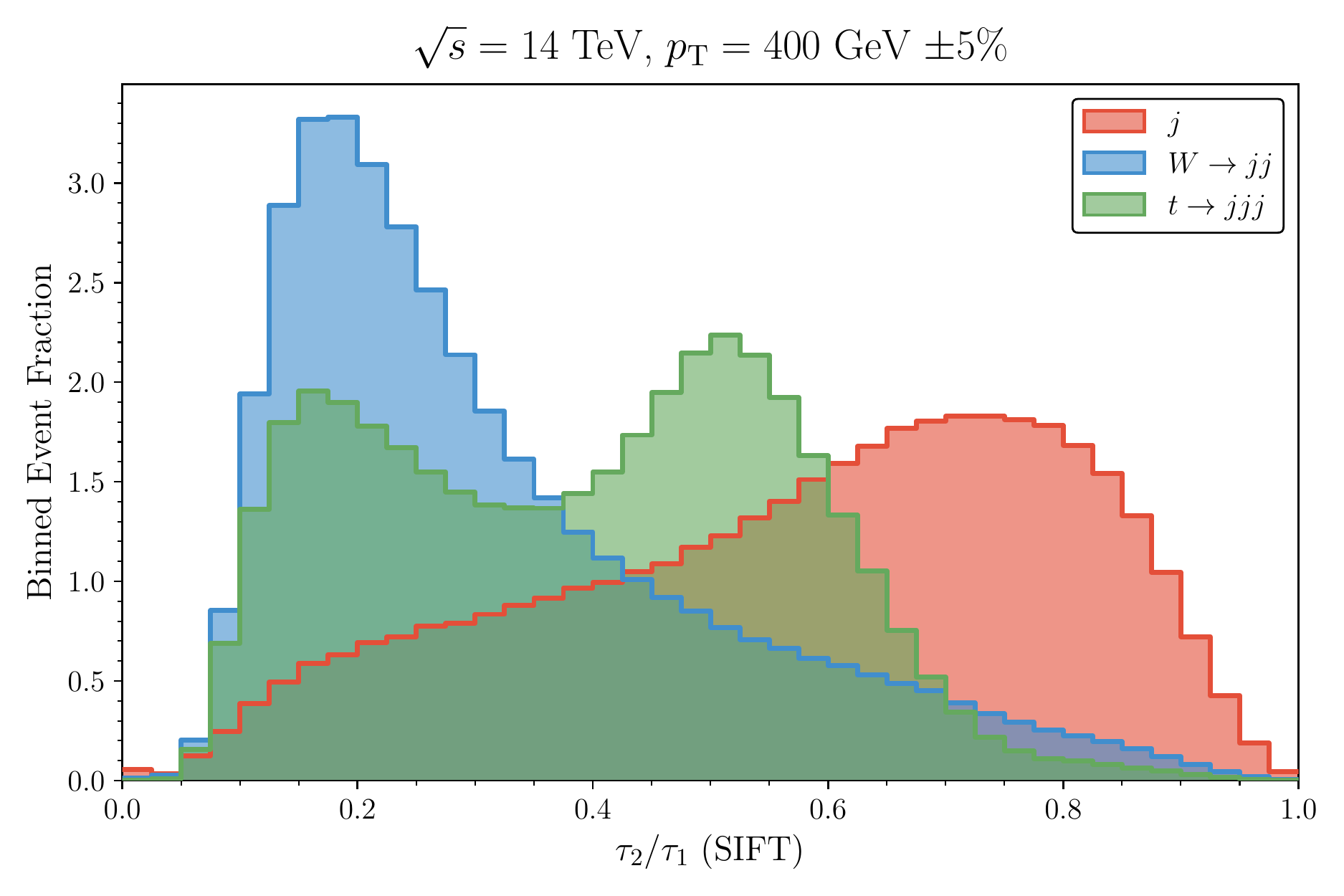} \\
\includegraphics[width=.45\textwidth]{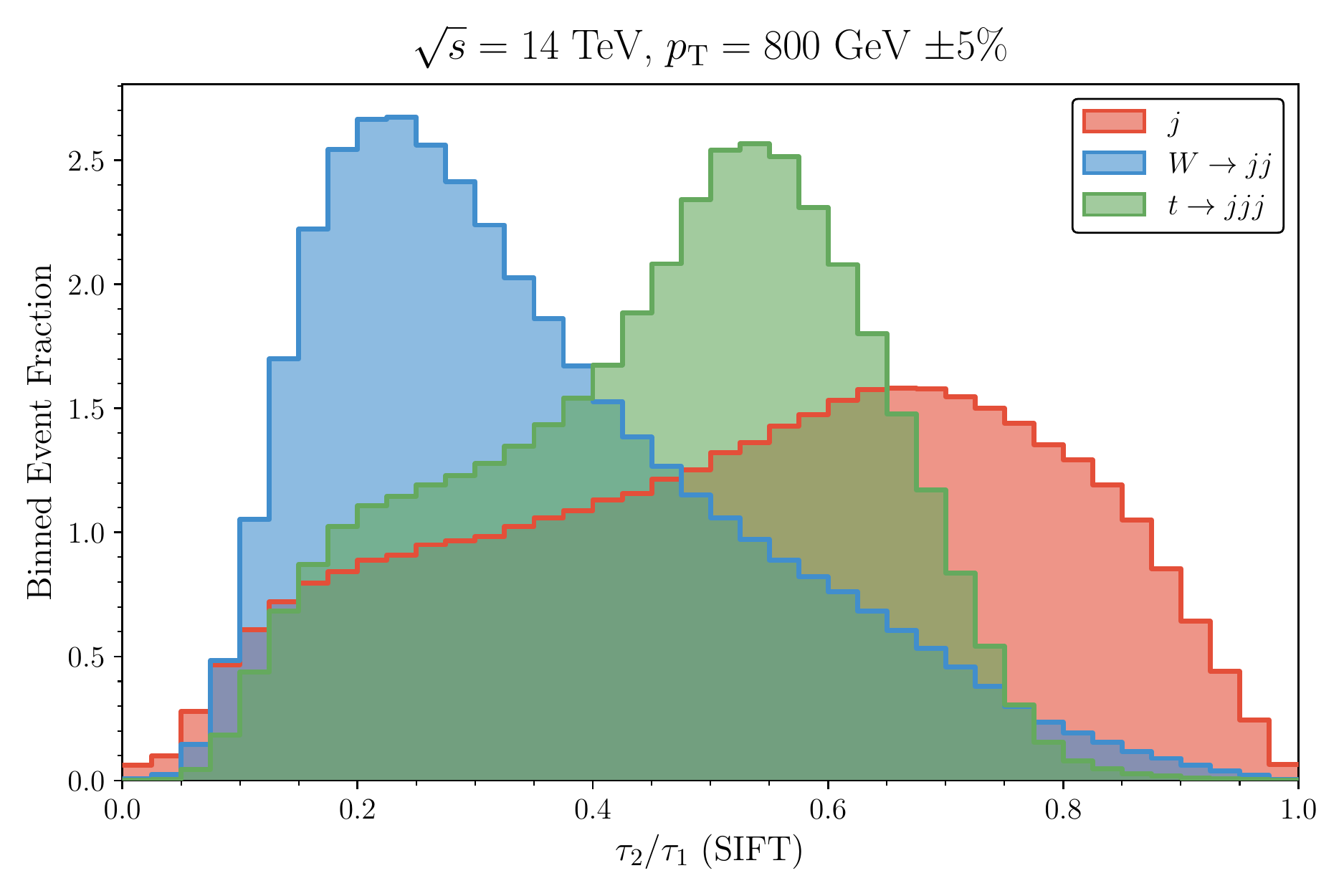} \hspace{12pt}
\includegraphics[width=.45\textwidth]{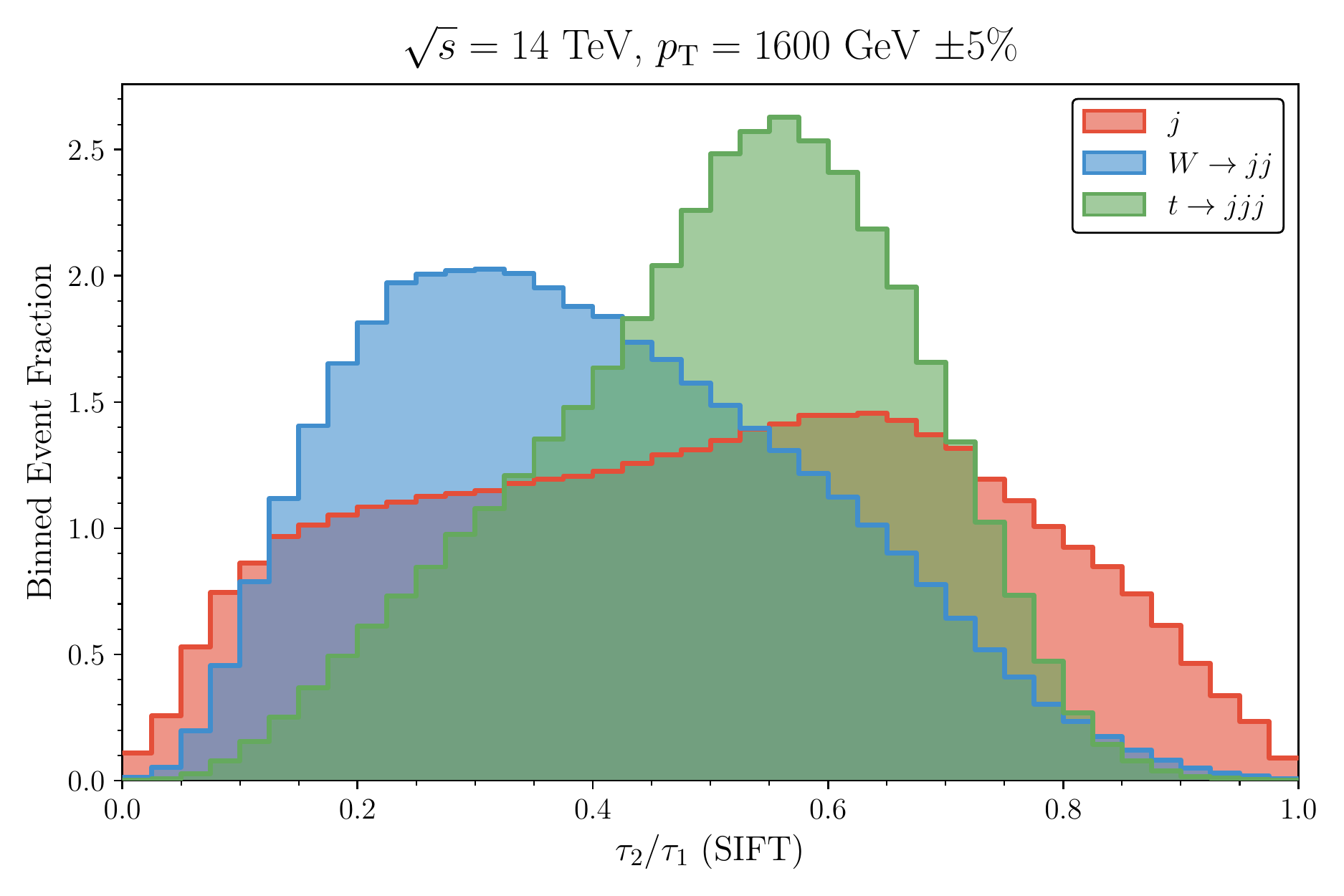}
\caption{\footnotesize
Distribution of $\tau_2/\tau_1$ computed with SIFT axes for mono-, di-, and tri-jet samples at various transverse boosts.
}
\label{fig:tau21}
\end{figure*}

\begin{figure*}[ht!]
\centering
\includegraphics[width=.45\textwidth]{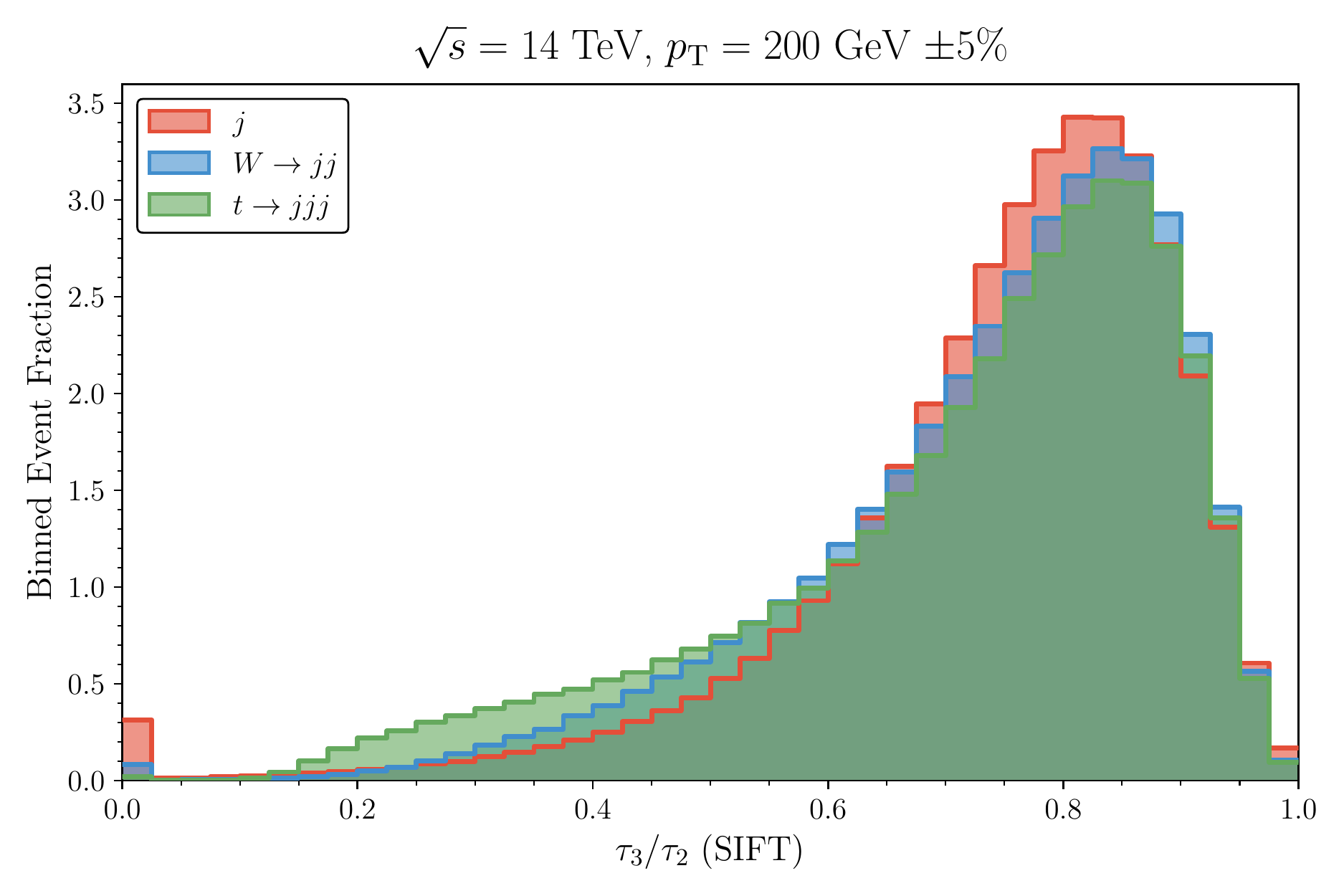} \hspace{12pt}
\includegraphics[width=.45\textwidth]{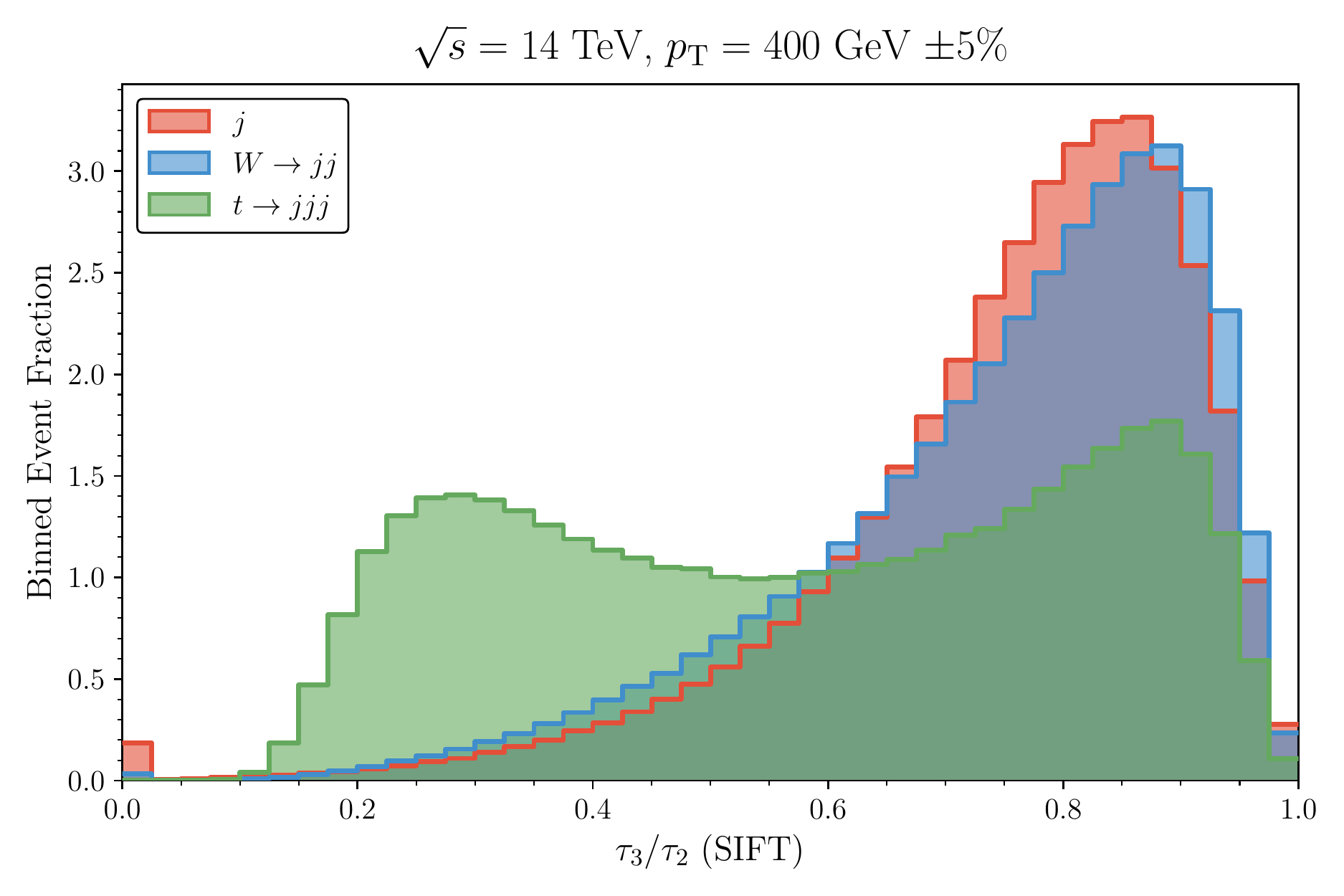} \\
\includegraphics[width=.45\textwidth]{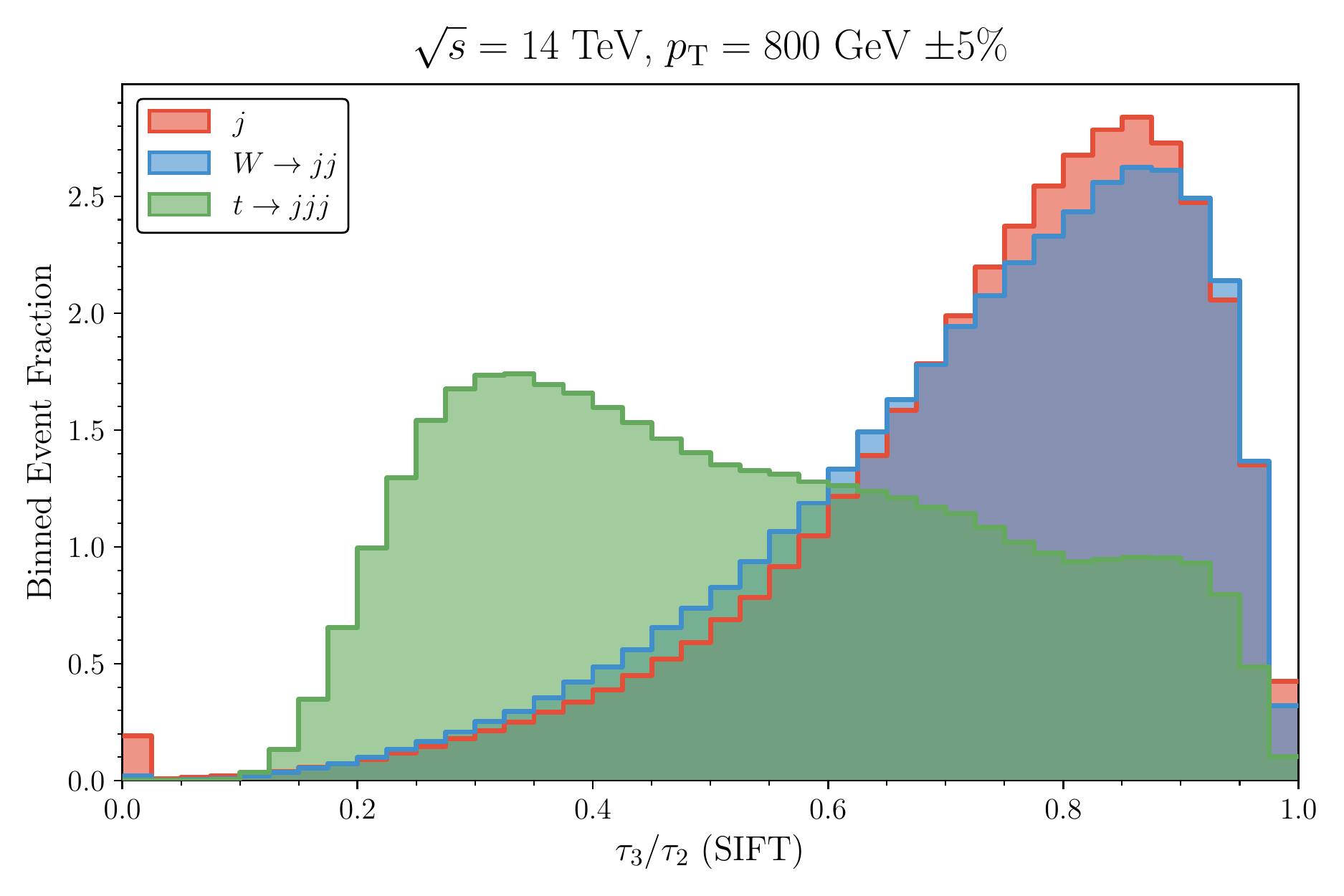} \hspace{12pt}
\includegraphics[width=.45\textwidth]{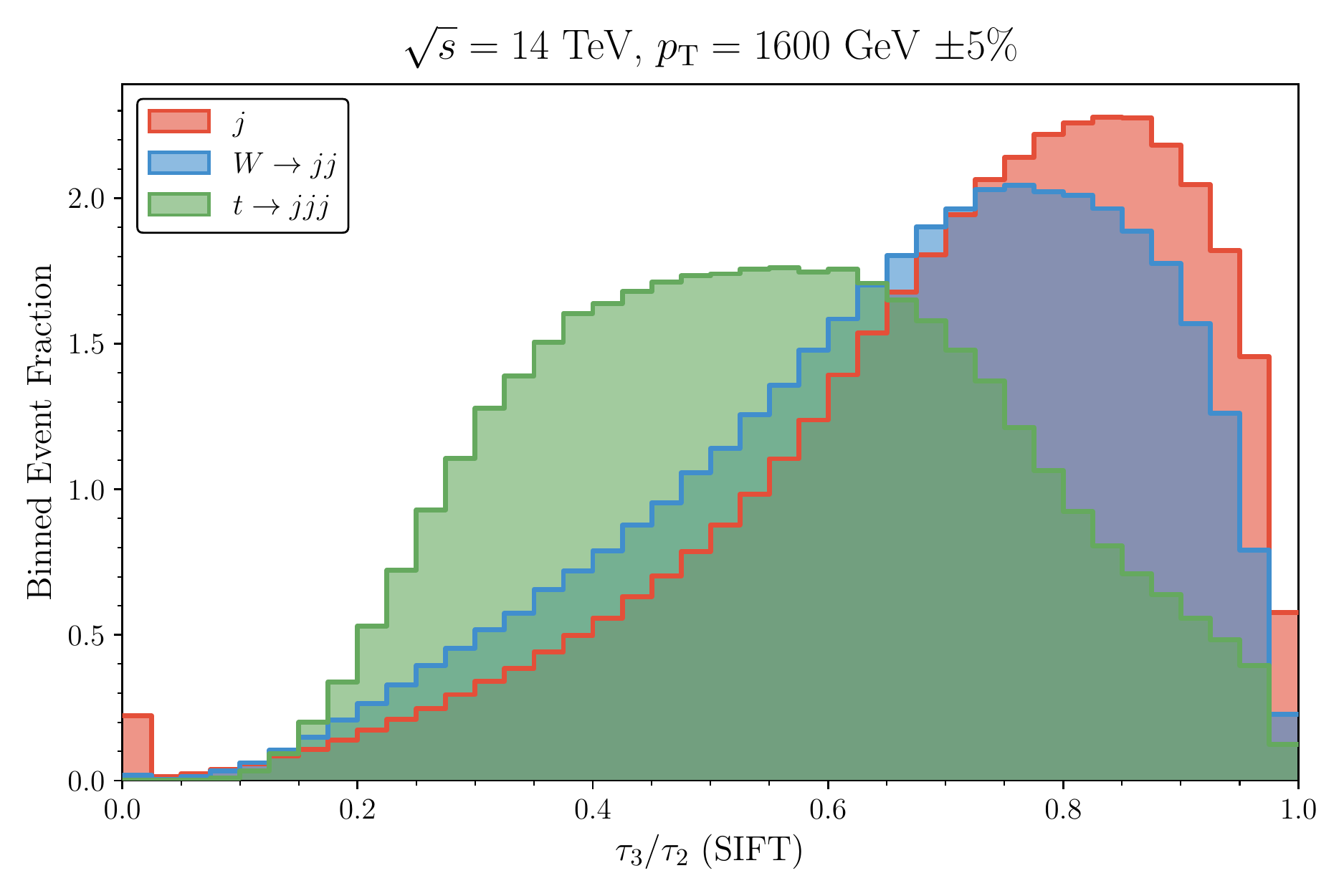}
\caption{\footnotesize
Distribution of $\tau_3/\tau_2$ computed with SIFT axes for mono-, di-, and tri-jet samples at various transverse boosts.
}
\label{fig:tau32}	
\end{figure*}

\begin{figure*}[ht!]
\centering
\includegraphics[width=.45\textwidth]{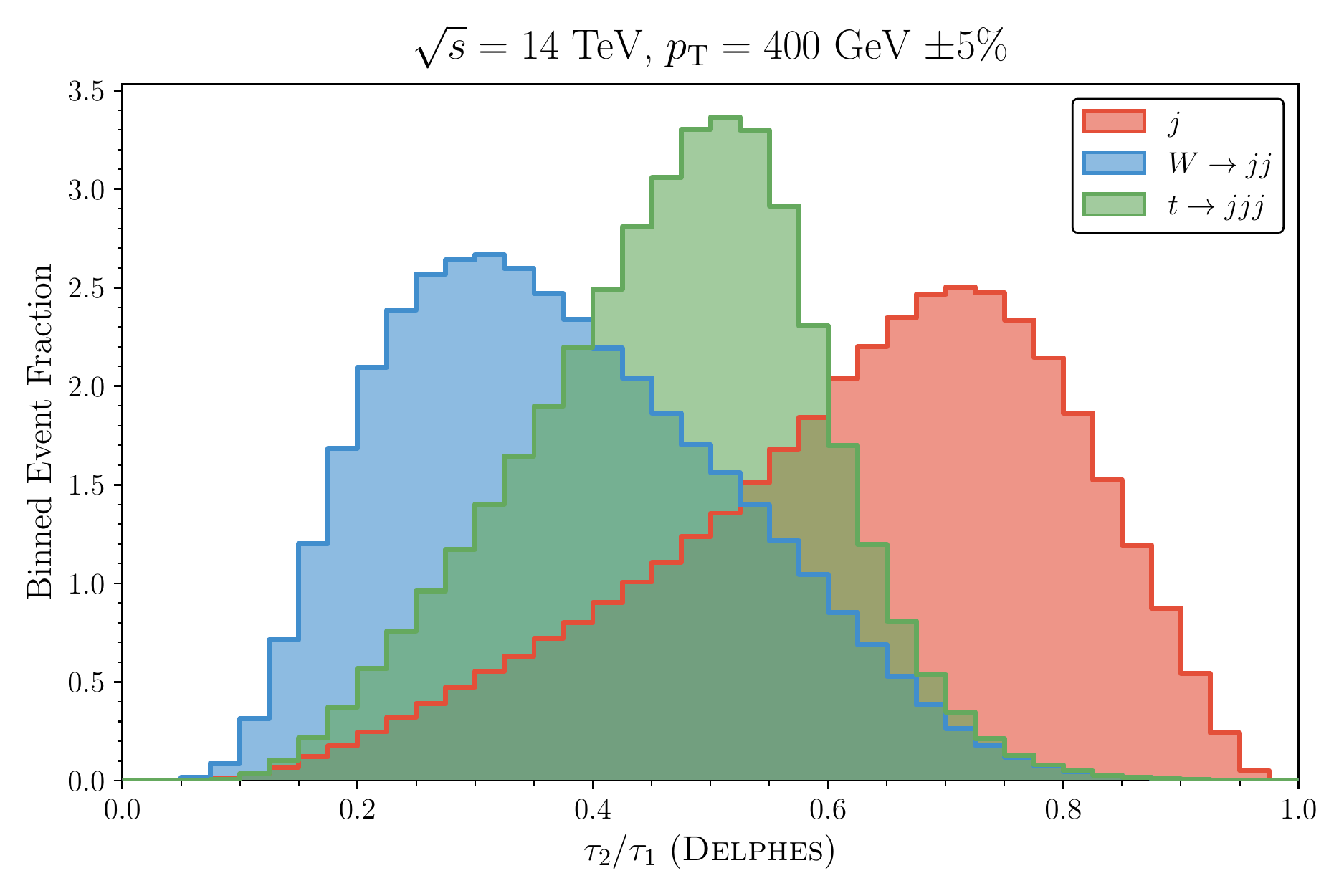} \hspace{12pt}
\includegraphics[width=.45\textwidth]{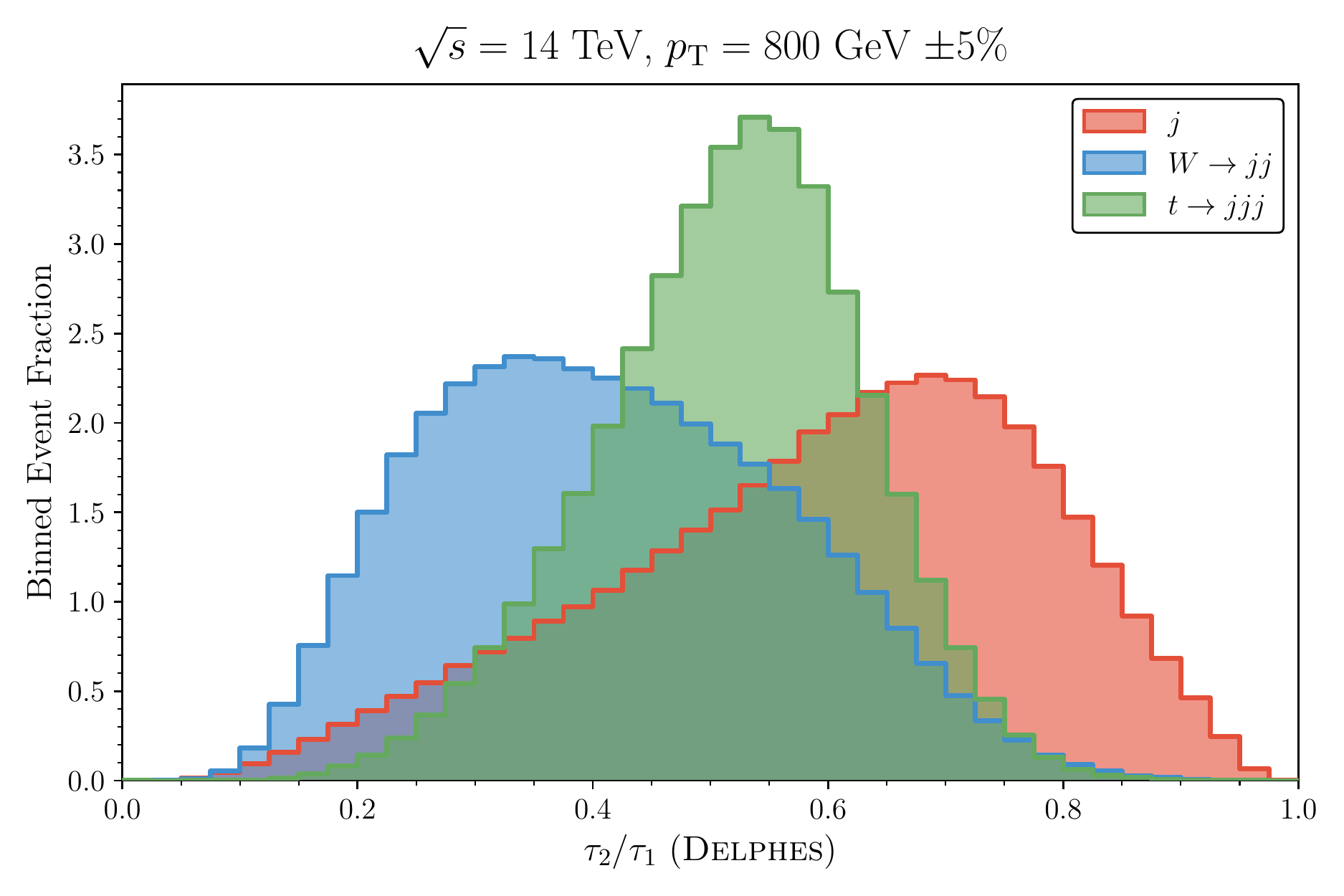} \\
\includegraphics[width=.45\textwidth]{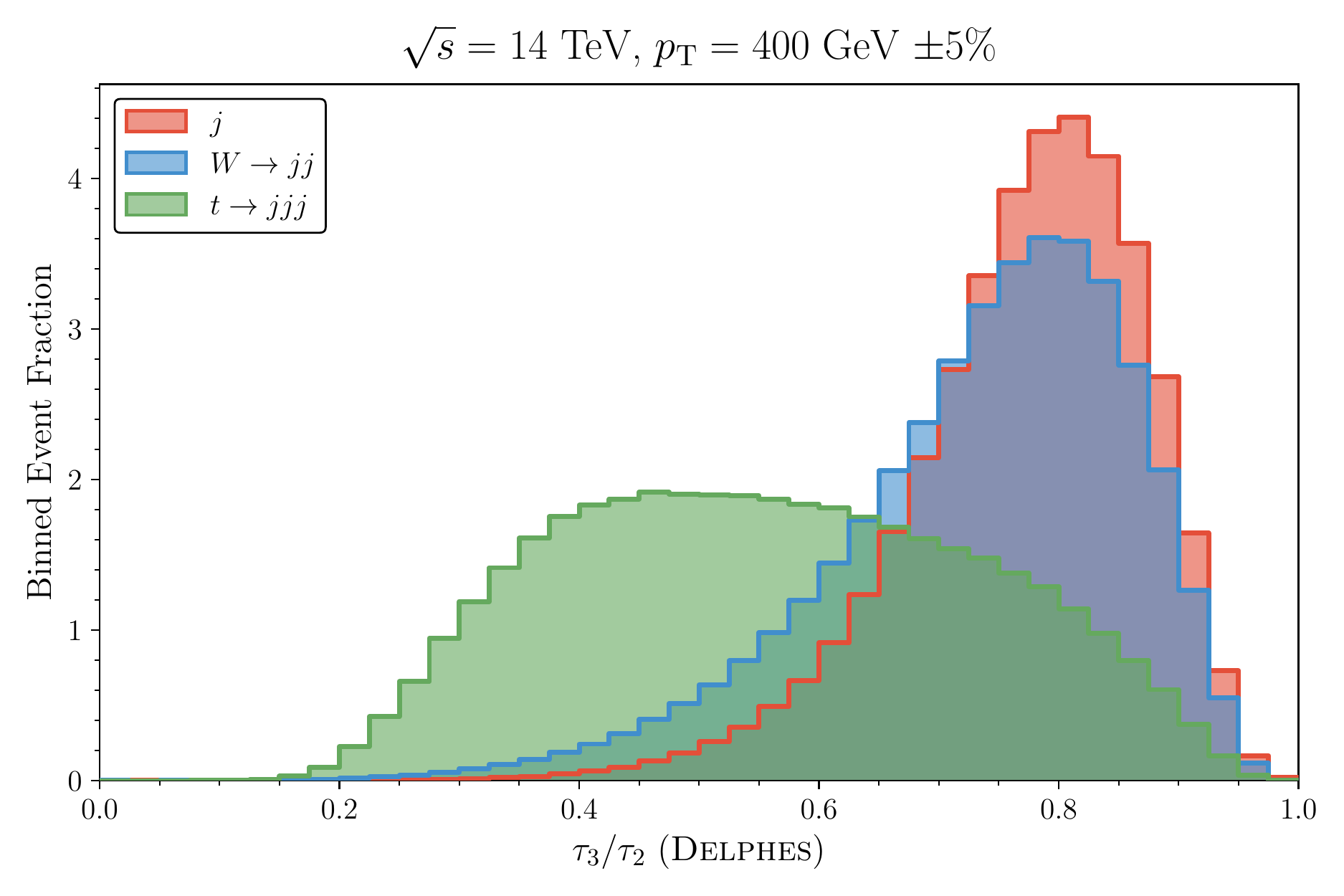} \hspace{12pt}
\includegraphics[width=.45\textwidth]{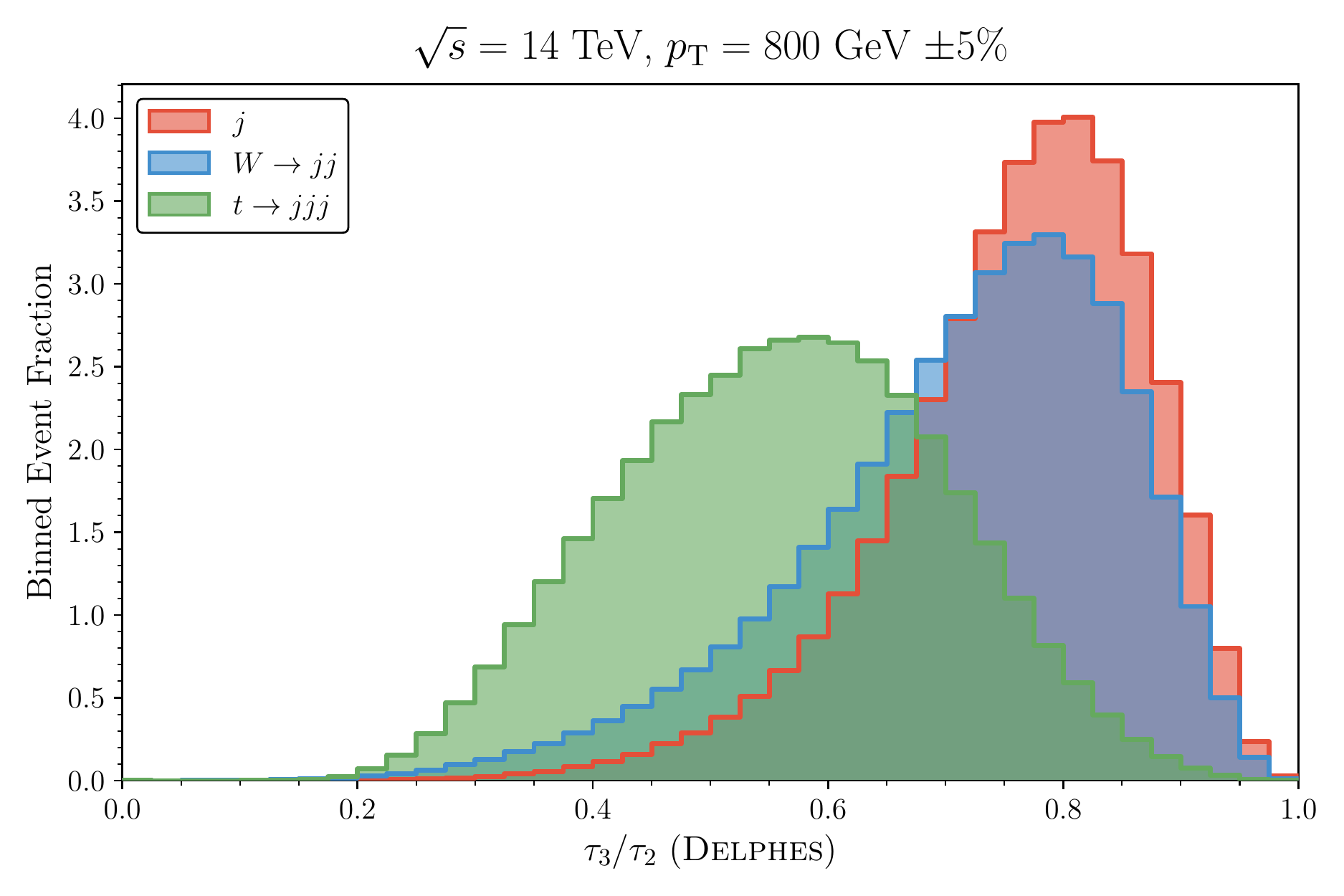}
\caption{\footnotesize
Reference distributions of $\tau_2/\tau_1$ and $\tau_3/\tau_2$ computed by {\sc Delphes} at various transverse boosts.
}
\label{fig:tau2132Delphes}	
\end{figure*}

\begin{figure*}[ht!]
\centering
\includegraphics[width=.45\textwidth]{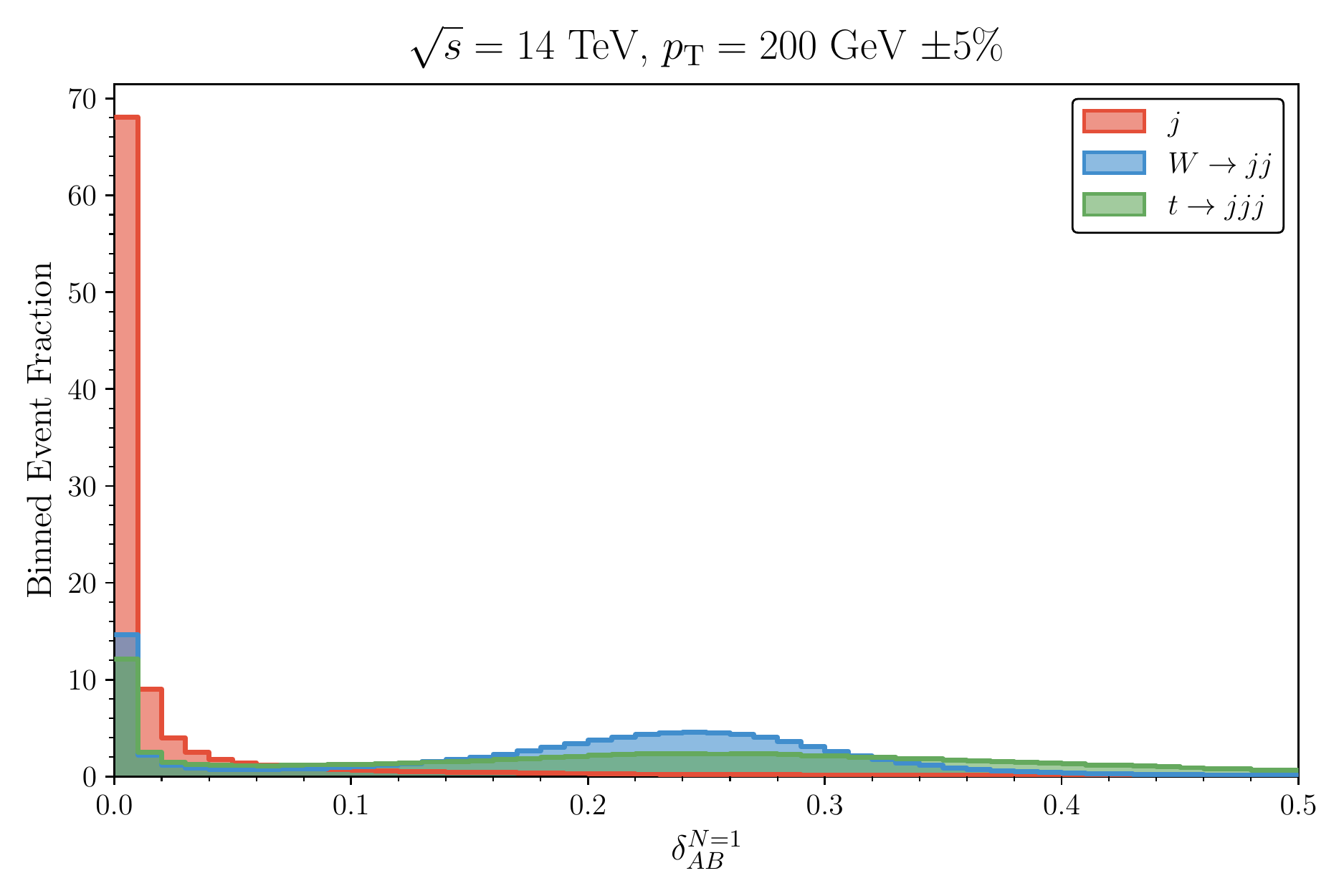} \hspace{12pt}
\includegraphics[width=.45\textwidth]{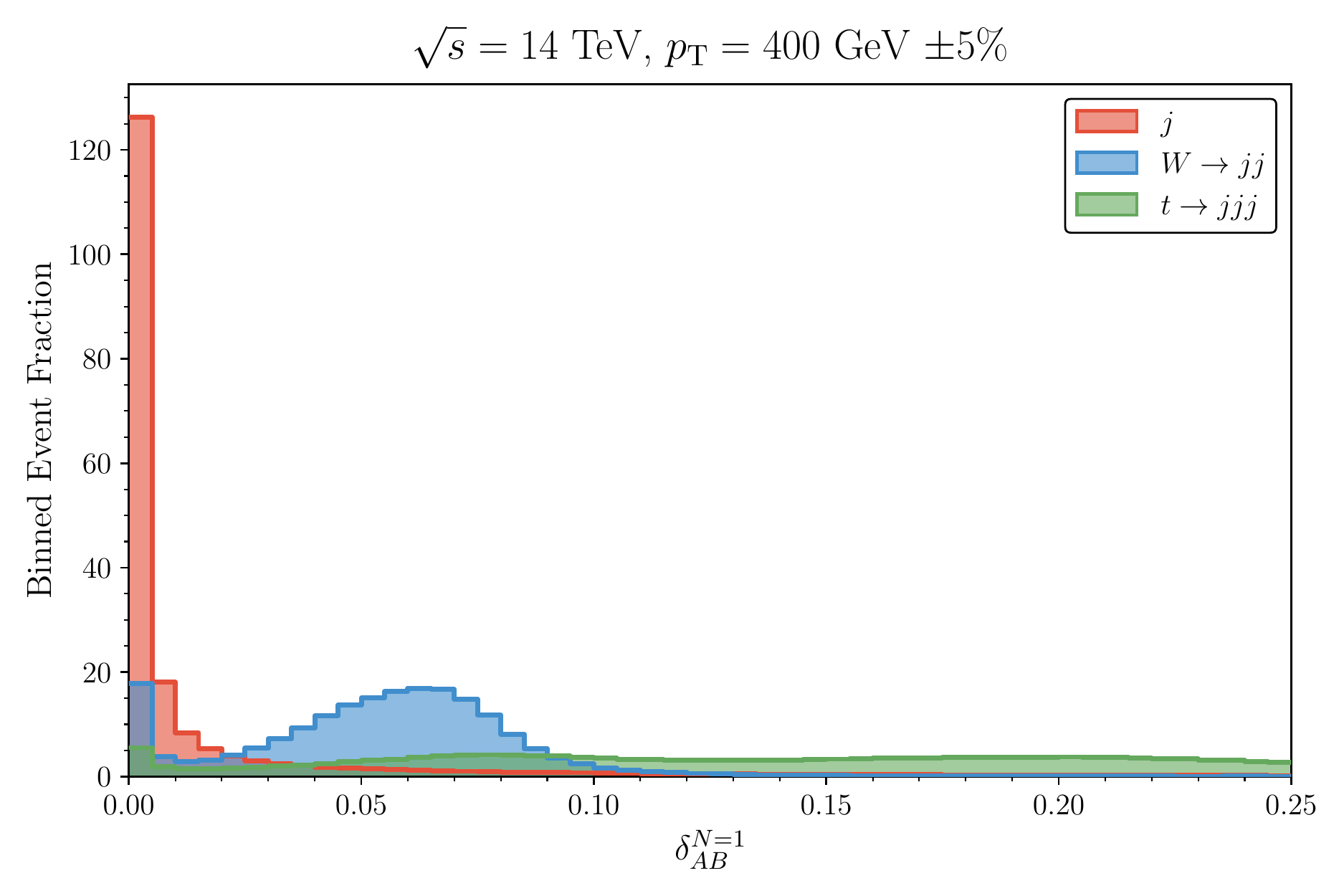} \\
\includegraphics[width=.45\textwidth]{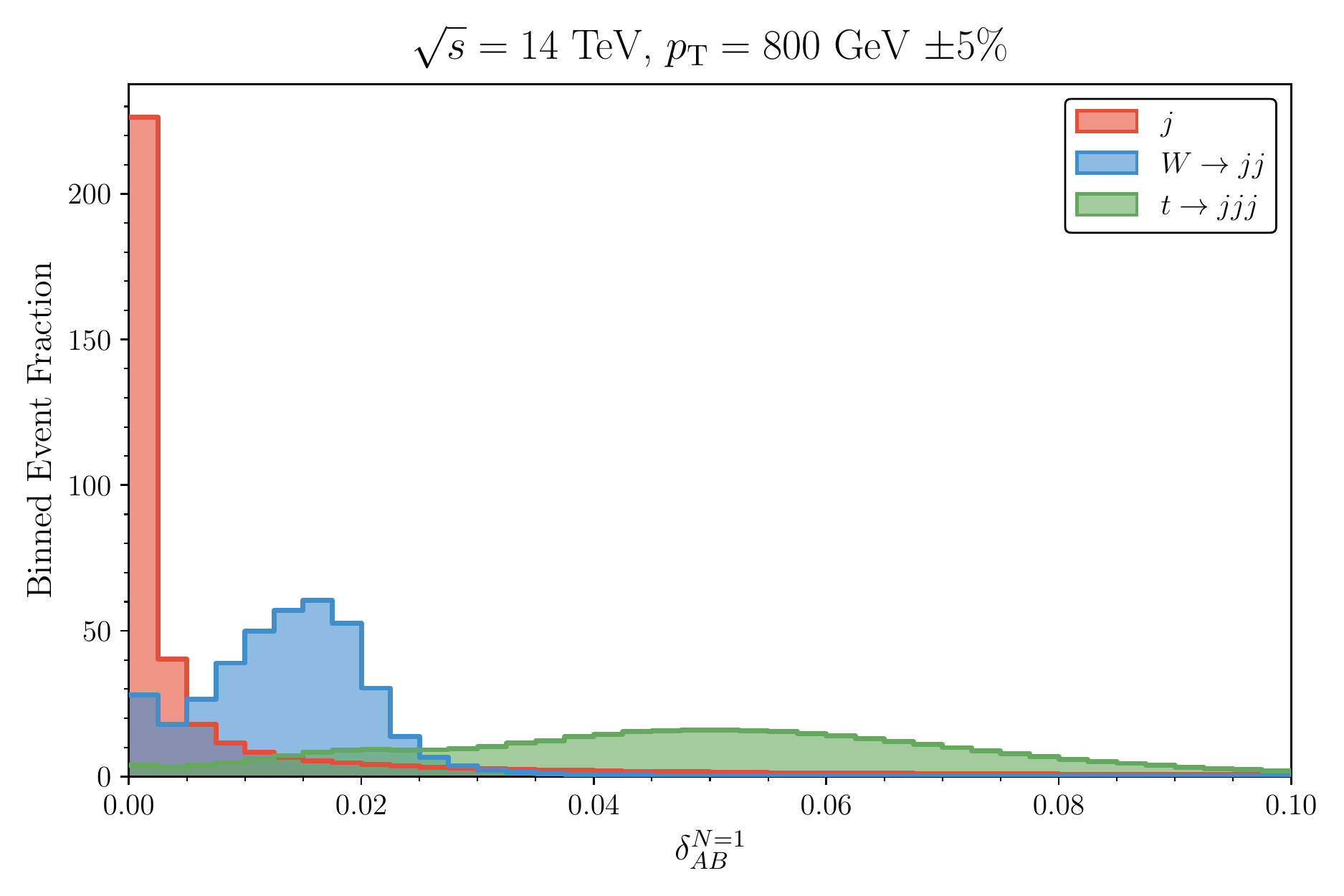} \hspace{12pt}
\includegraphics[width=.45\textwidth]{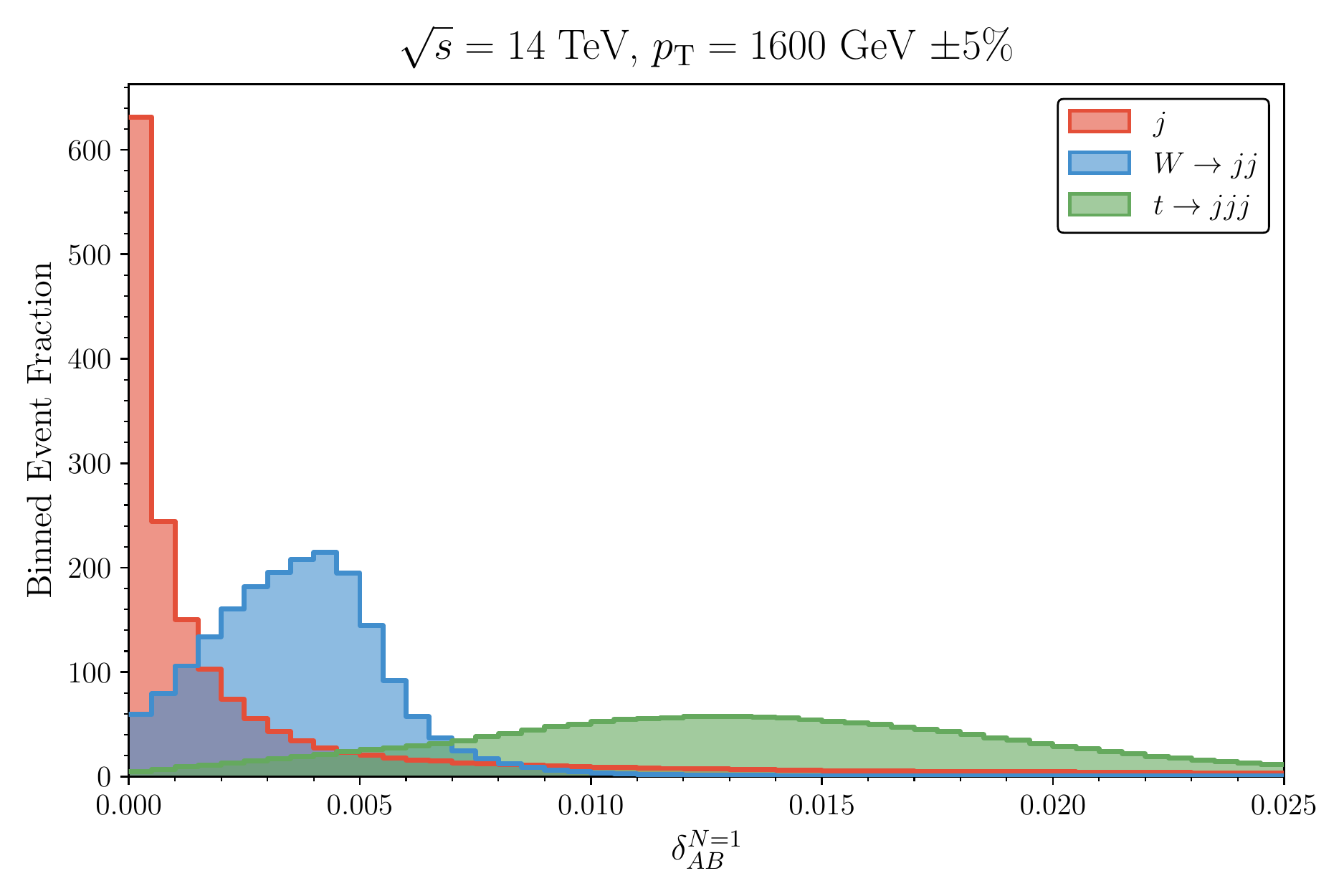}
\caption{\footnotesize
Distribution of $\delta_{AB}$ at the $N=1$ stage of clustering with SIFT for mono-, di-, and tri-jet samples at various transverse boosts.
}
\label{fig:siftdab1}
\end{figure*}

\begin{figure*}[ht!]
\centering
\includegraphics[width=.45\textwidth]{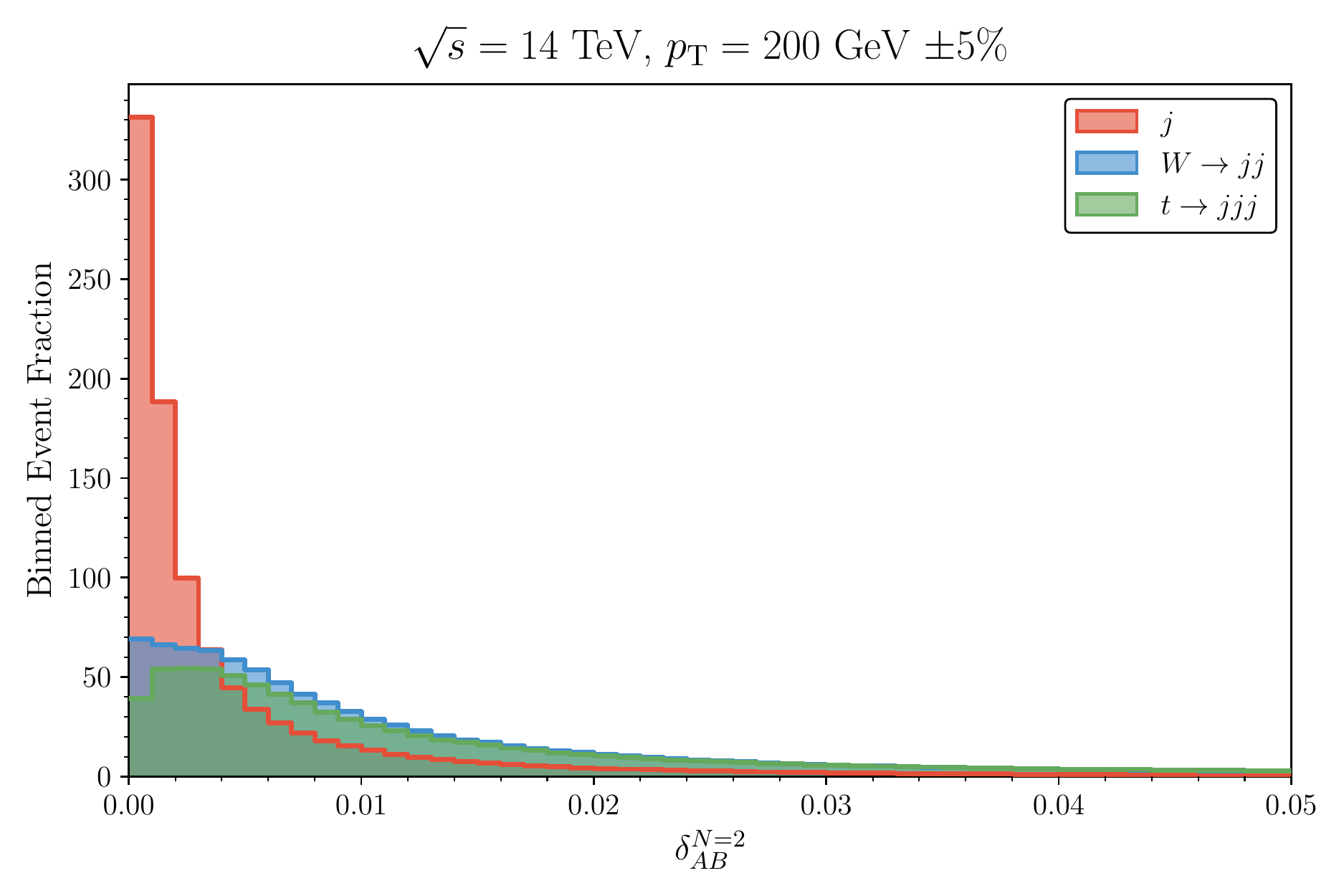} \hspace{12pt}
\includegraphics[width=.45\textwidth]{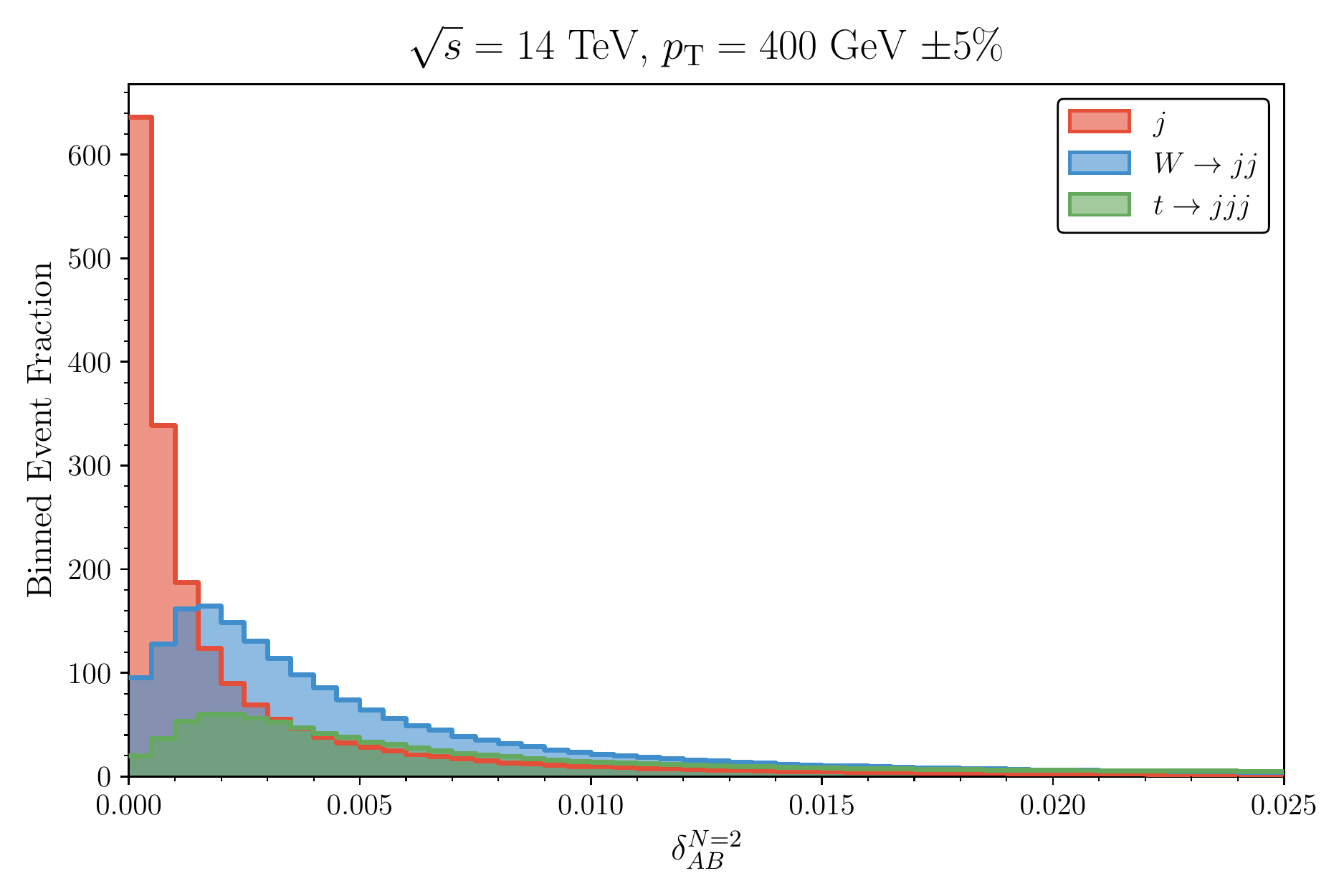} \\
\includegraphics[width=.45\textwidth]{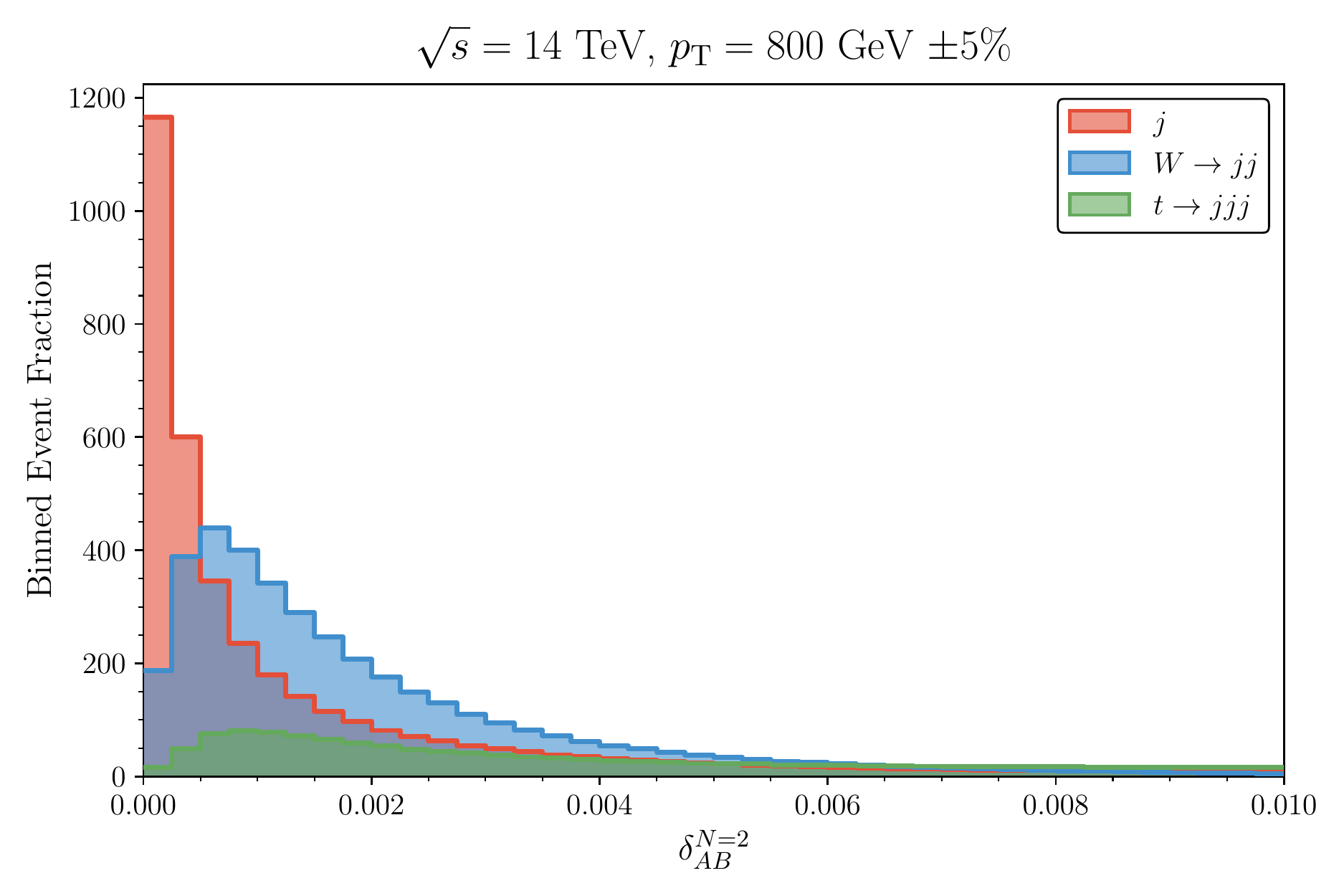} \hspace{12pt}
\includegraphics[width=.45\textwidth]{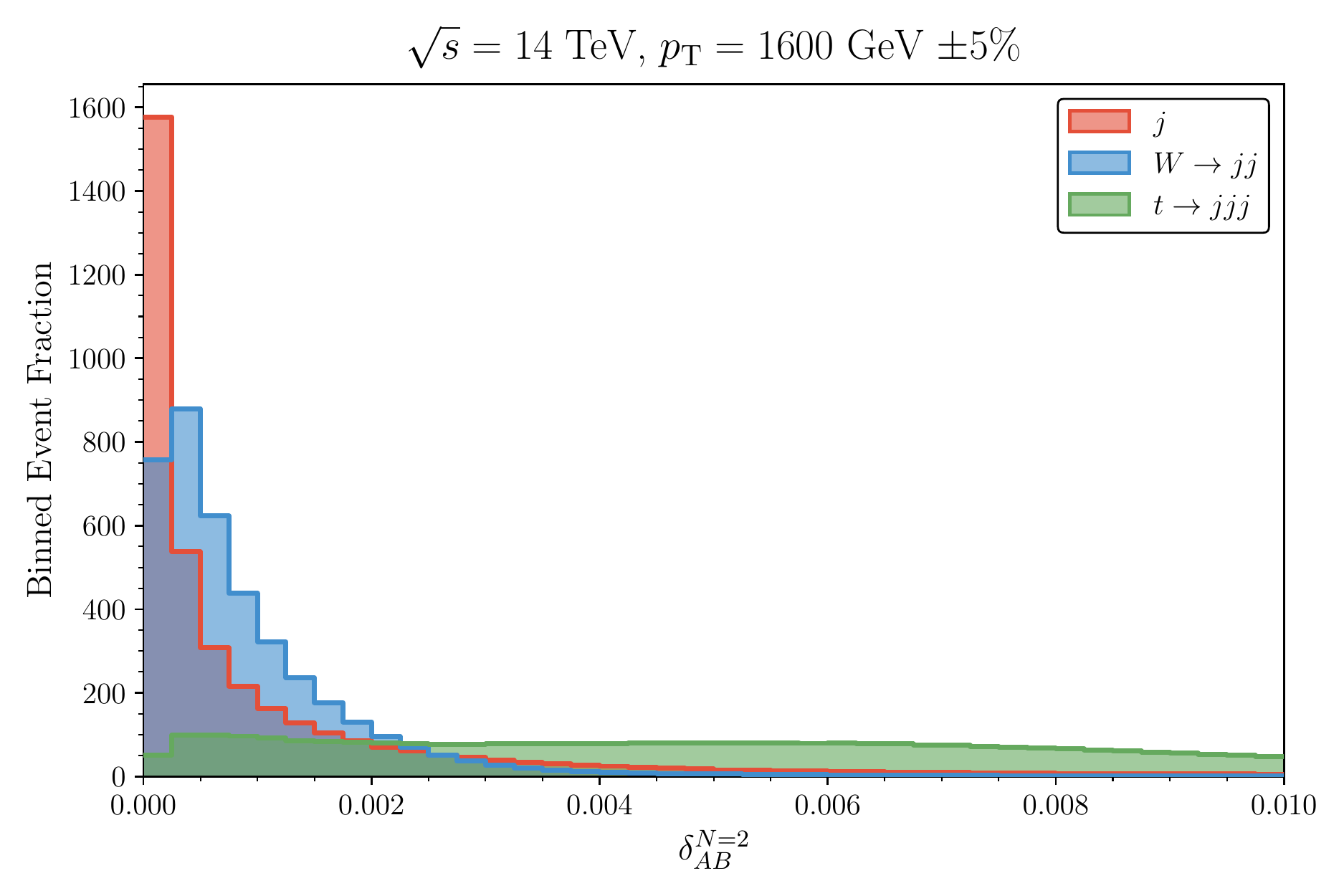}
\caption{\footnotesize
Distribution of $\delta_{AB}$ at the $N=2$ stage of clustering with SIFT for mono-, di-, and tri-jet samples at various transverse boosts.
}
\label{fig:siftdab2}
\end{figure*}

\begin{table}[ht!]
\bgroup
\def\arraystretch{1.25}
  \begin{center}
    \begin{tabular}{C{0.0875\textwidth}|C{0.0875\textwidth}|C{0.0875\textwidth}|C{0.0875\textwidth}|C{0.0875\textwidth}}
$\pt^{{\rm GeV}\pm 5\%}$ & $\tau_{\textsc{Delphes}}^{N+1/N}$ & $\tau_{\rm SIFT}^{N+1/ N}$ & $\delta_{AB}^N$ & $\delta+\tau$ \\
      \hline
100 & 0.62 & 0.68 & 0.69 & 0.70 \\
200 & 0.91 & 0.86 & 0.88 & 0.89 \\
400 & 0.89 & 0.85 & 0.91 & 0.92 \\
800 & 0.82 & 0.79 & 0.92 & 0.93 \\
1600 & 0.77 & 0.74 & 0.91 & 0.92 \\
3200 & 0.78 & 0.76 & 0.88 & 0.90 \\
    \end{tabular}
\caption{\footnotesize
Area under curve ROC scores for discrimination of resonances with hard 1- and 2-prong substructure using a BDT trained on various sets of event observables.}
    \label{tab:BDT_21}
  \end{center}
\egroup
\end{table}

\begin{table}[ht!]
\bgroup
\def\arraystretch{1.25}
  \begin{center}
    \begin{tabular}{C{0.0875\textwidth}|C{0.0875\textwidth}|C{0.0875\textwidth}|C{0.0875\textwidth}|C{0.0875\textwidth}}
$\pt^{{\rm GeV}\pm 5\%}$ & $\tau_{\textsc{Delphes}}^{N+1/N}$ & $\tau_{\rm SIFT}^{N+1/ N}$ & $\delta_{AB}^N$ & $\delta+\tau$ \\
      \hline
100 & 0.61 & 0.61 & 0.63 & 0.65 \\
200 & 0.63 & 0.60 & 0.71 & 0.72 \\
400 & 0.82 & 0.74 & 0.90 & 0.90 \\
800 & 0.85 & 0.80 & 0.94 & 0.95 \\
1600 & 0.77 & 0.77 & 0.97 & 0.97 \\
3200 & 0.77 & 0.79 & 0.98 & 0.99 \\
    \end{tabular}
\caption{\footnotesize
Area under curve ROC scores for discrimination of resonances with hard 2- and 3-prong substructure using a BDT trained on various sets of event observables.}
    \label{tab:BDT_32}
  \end{center}
\egroup
\end{table}

\begin{table}[ht!]
\bgroup
\def\arraystretch{1.25}
  \begin{center}
    \begin{tabular}{C{0.0875\textwidth}|C{0.0875\textwidth}|C{0.0875\textwidth}|C{0.0875\textwidth}|C{0.0875\textwidth}}
$\pt^{{\rm GeV}\pm 5\%}$ & $\tau_{\textsc{Delphes}}^{N+1/N}$ & $\tau_{\rm SIFT}^{N+1/ N}$ & $\delta_{AB}^N$ & $\delta+\tau$ \\
      \hline
100 & 0.70 & 0.75 & 0.77 & 0.77 \\
200 & 0.86 & 0.87 & 0.90 & 0.90 \\
400 & 0.93 & 0.91 & 0.95 & 0.96 \\
800 & 0.91 & 0.89 & 0.96 & 0.96 \\
1600 & 0.84 & 0.83 & 0.94 & 0.95 \\
3200 & 0.76 & 0.78 & 0.91 & 0.92 \\
    \end{tabular}
\caption{\footnotesize
Area under curve ROC scores for discrimination of resonances with hard 3- and 1-prong substructure using a BDT trained on various sets of event observables.}
    \label{tab:BDT_13}
  \end{center}
\egroup
\end{table}

\begin{figure*}[ht!]
\centering
\includegraphics[width=.45\textwidth]{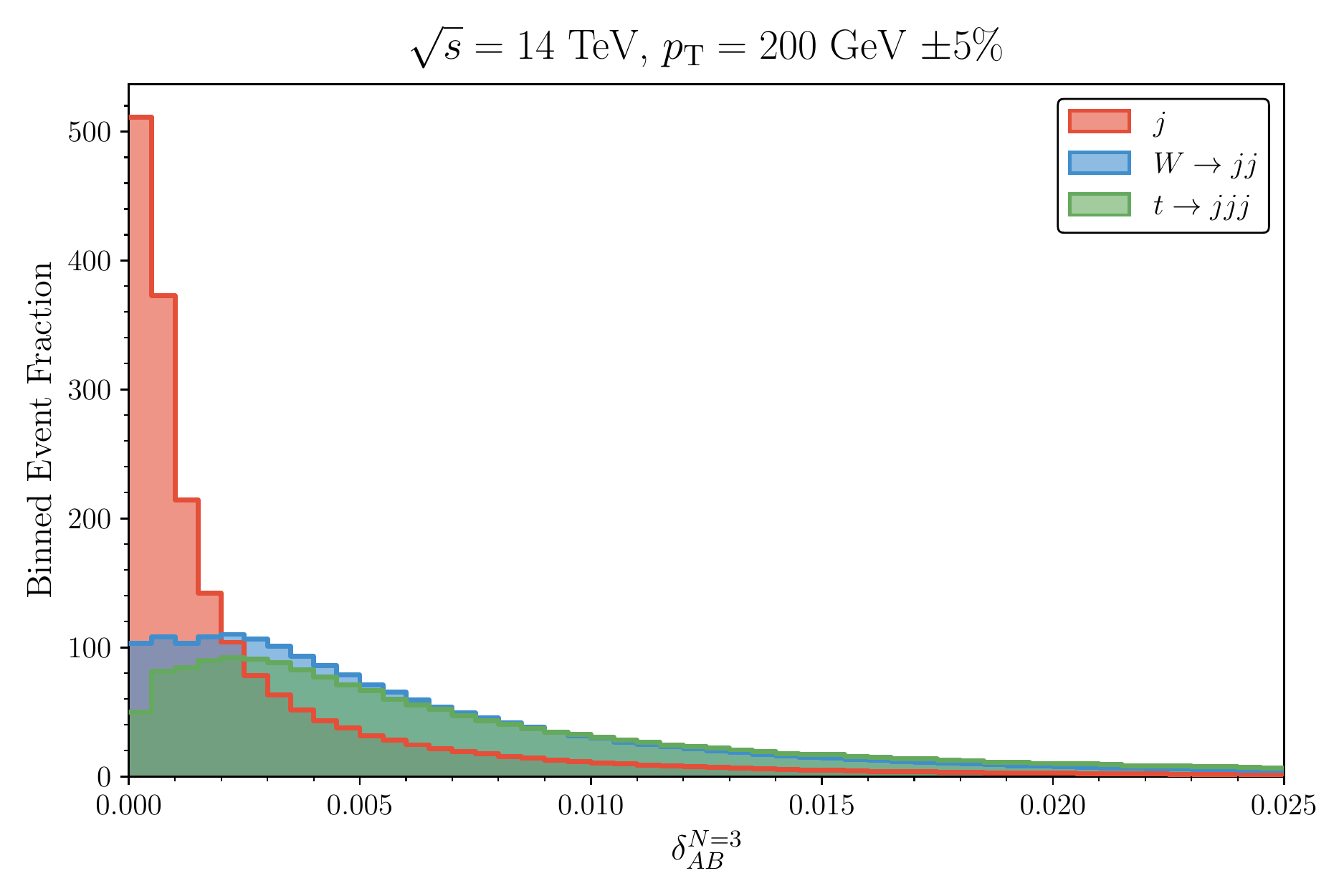} \hspace{12pt}
\includegraphics[width=.45\textwidth]{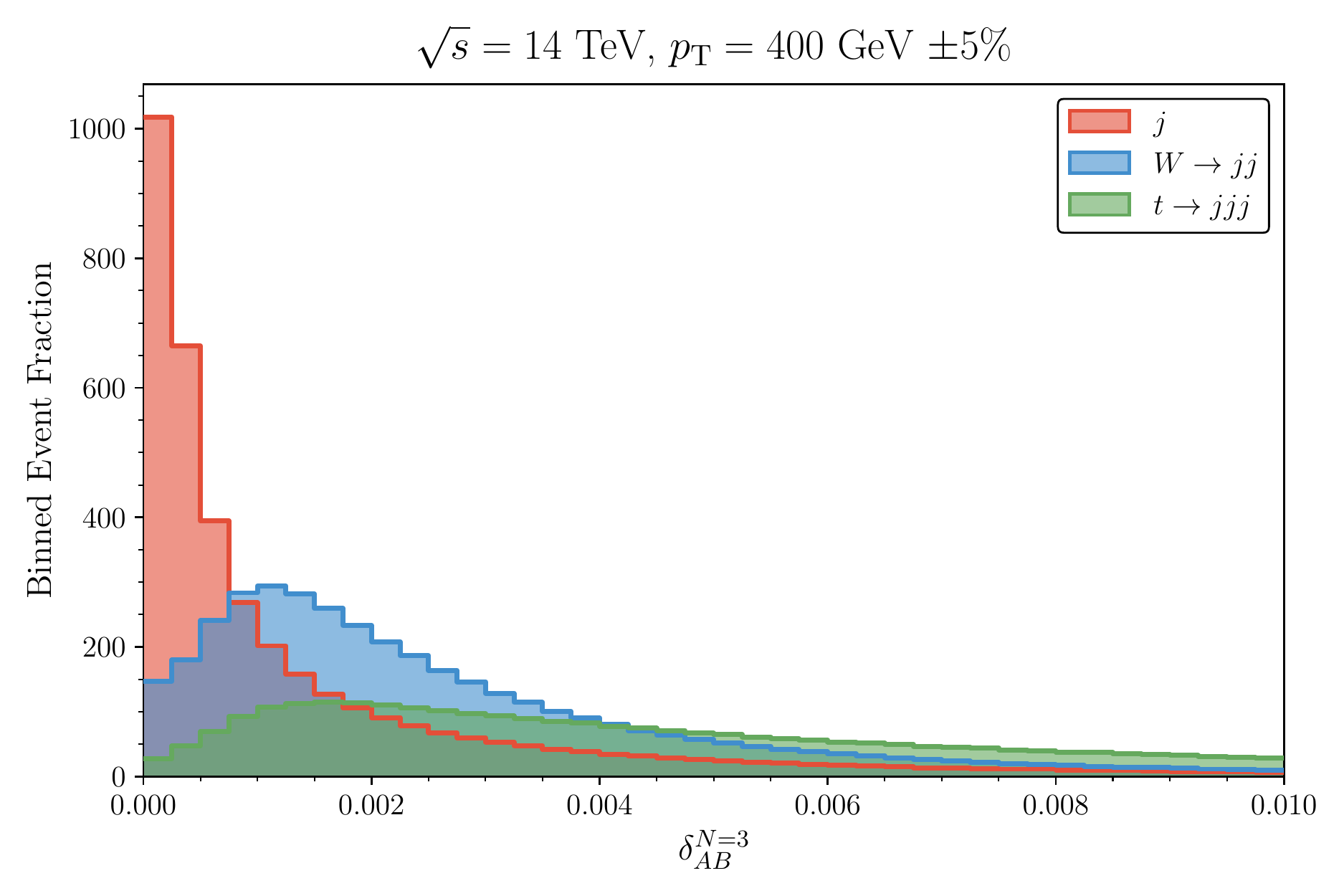} \\
\includegraphics[width=.45\textwidth]{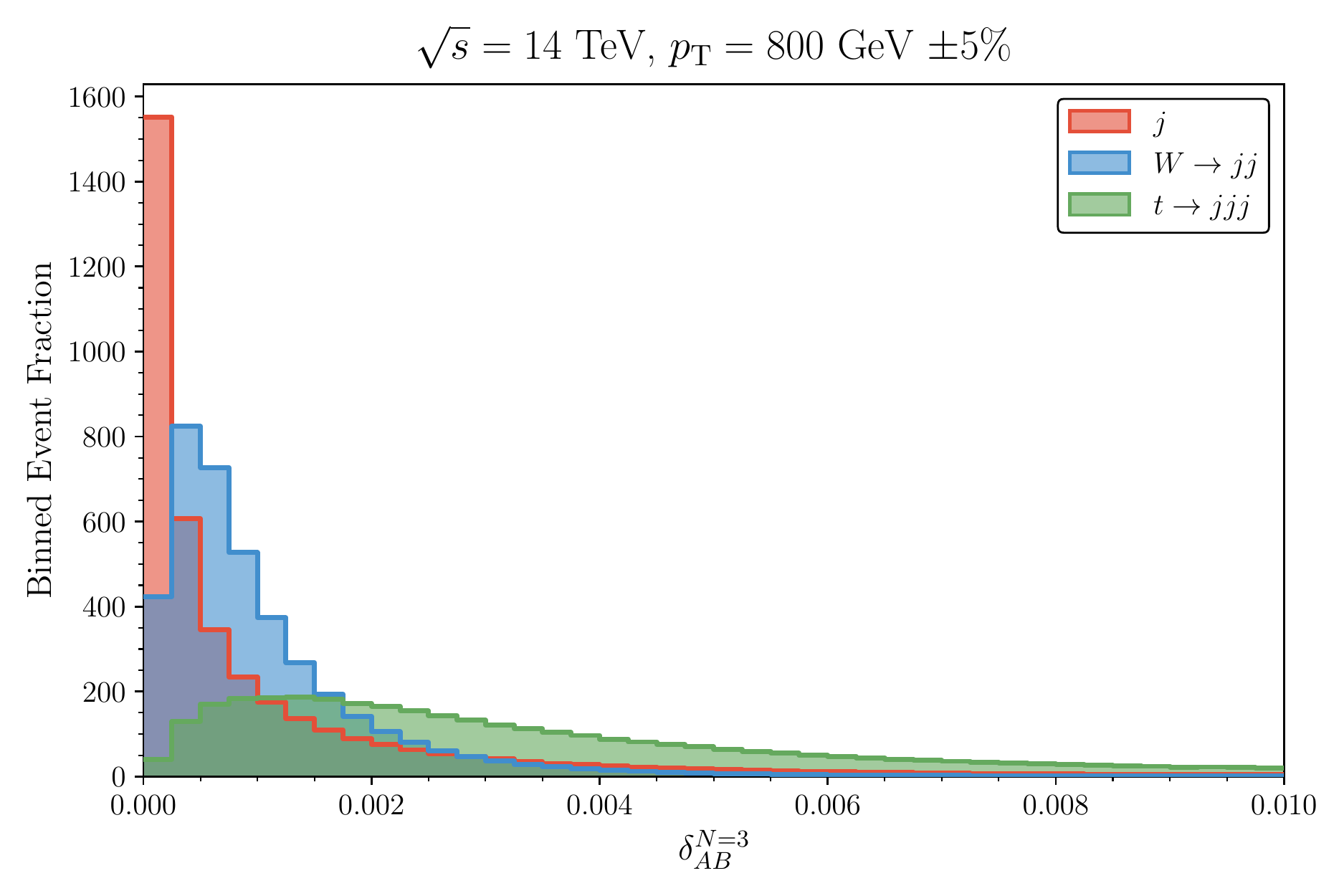} \hspace{12pt}
\includegraphics[width=.45\textwidth]{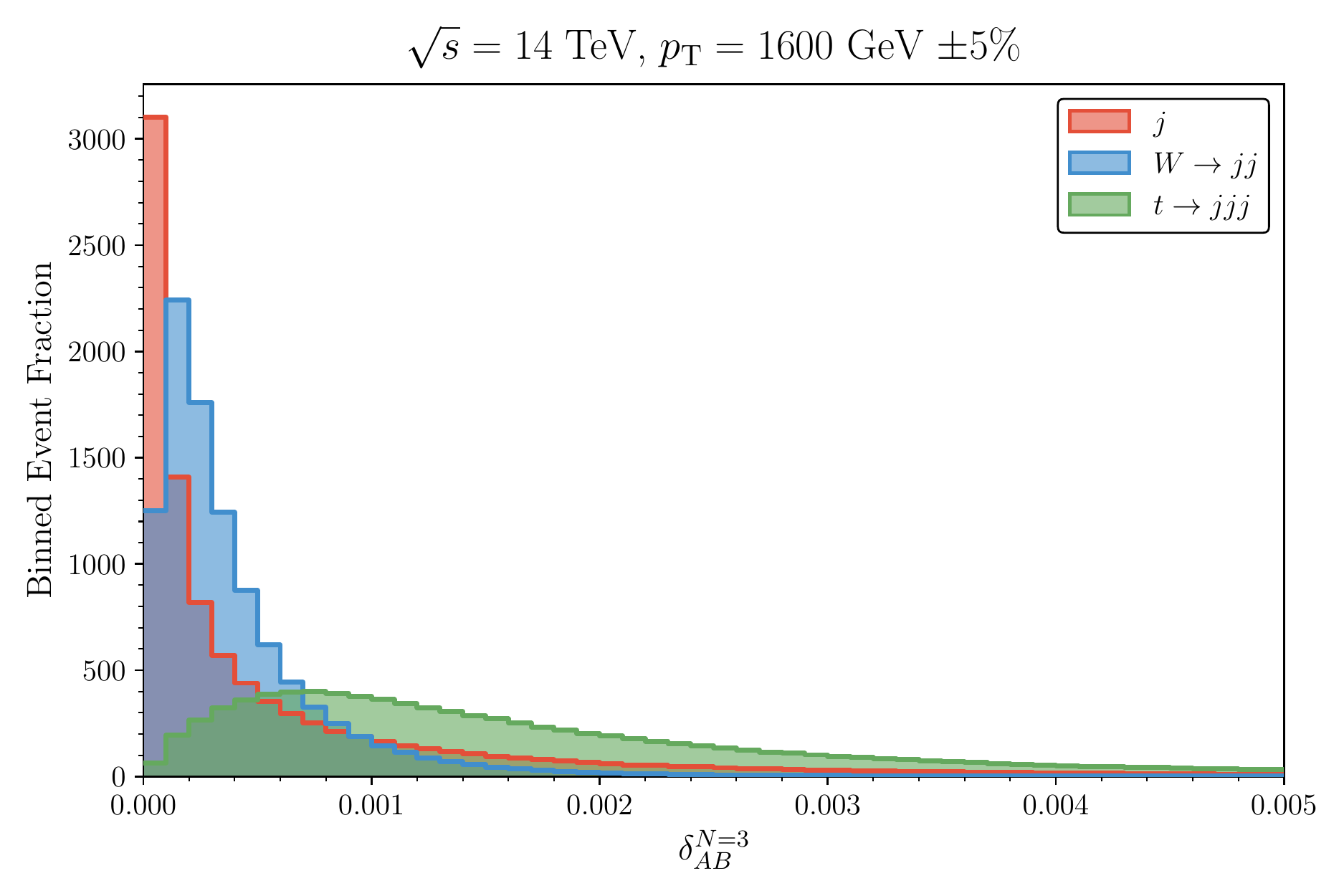}
\caption{\footnotesize
Distribution of $\delta_{AB}$ at the $N=3$ stage of clustering with SIFT for mono-, di-, and tri-jet samples at various transverse boosts.
}
\label{fig:siftdab3}
\end{figure*}

\section{Resolution and Reconstruction\label{sct:reconstruction}}

This section characterizes SIFT's angular and energetic
response functions for the resolution of hard mono-jets
and tests the reconstruction of collimated di- and tri-jet
systems associated with a massive resonance.
The best performance is achieved for large transverse boosts.

We generate Monte Carlo collider data modeling the
$\sqrt{s} = 14$~TeV LHC using {\sc MadGraph}/{\sc MadEvent},
{\sc Pythia8}, and {\sc Delphes} as before.
Clean ($N=1,2,3$) prong samples are obtained by simulating the processes
($p p \Rightarrow j Z \Rightarrow j +\nu \bar{\nu}$),
($p p \Rightarrow W^\pm Z \Rightarrow j j +\nu \bar{\nu}$), and
($p p \Rightarrow tW^- \Rightarrow j j j + \bar{\nu}\ell^-$) plus conjugate,
respectively.  In the latter case, an angular isolation cone
with ($\Delta R = 0.5$) is placed around the visible lepton.
Hard partonic objects are required to carry a minimal transverse
momentum ($\pt \ge 25$~GeV) and be inside ($\vert\eta\vert <= 3.0$).
No restrictions are placed on the angular separation of decay products. 
Jets consist of gluons and/or light first-generation quarks ($u,\,d$),
as well as $b$-quarks where required by a third-generation process.
In order to represent a wide range of event scales,
we tranche in the transverse momentum (vector sum magnitude)
of the hadronic system, considering six log-spaced intervals 
$\pt = (\,100,\,200,\,400,\,800,\,1600,\,3200\,)$~GeV $\pm 5\%$
and giving attention primarily to the inner four.

Clustering is disabled at the detector simulation level
by setting the jet radius $R_0$ and aggregate $\pt$ threshold to very small values.
We retain the default {\sc Delphes} efficiencies for tracks and calorimeter
deposits (including $\pt$ thresholds on low-level detector objects),
along with cell specifications and smearing (resolution) effects in the latter case.
Jet energy scale corrections are turned off (set to $1.0$) since these
are calibrated strictly for application to fully reconstructed (clustered) objects.
For purposes of comparison and validation,
we also extract information from {\sc Delphes} regarding
the leading large-radius jet ($R_0 = 1$), which is processed by 
trimming~\cite{Krohn:2009th}, pruning~\cite{Ellis:2009me}, and applying Soft-Drop.

Event analysis (including clustering) and computation of observables
are implemented with {\sc AEACuS} (cf. Appendix~\ref{sct:software}).
We begin by pre-clustering detector-level objects with
anti-$\kt$ at ($R_0 = 0.01$) to roughly mimic a characteristic
track-assisted calorimeter resolution at the LHC.
The isolation and filtering criteria described in Section~\ref{sct:filtering} are
then used in conjunction to select the subset of detector-level object
candidates retained for analysis.  Specifically, our procedure
is equivalent to keeping members gathered by the hardest isolated $N$-subjet
tree that survive filtering all the way down to the final merger.
All histograms are generated with {\sc RHADAManTHUS}~\cite{aeacus},
using {\sc MatPlotLib}~\cite{Hunter:2007} on the back end.

We begin by evaluating the fidelity of the variable large-radius SIFT
jet's directional and scale reconstruction in the context of the mono-jet sample.
Respectively, the upper-left and lower-left panels of FIG.~\ref{fig:JER_JAR} show 
distributions of the energy response $\mathcal{R}^A_B \equiv (\,\pt^{A}/\pt^{B} - 1\,)$
and angular response $\Delta R^A_B$ relative to the original 
truth-level ({\sc MadGraph}) partonic jet at various transverse boosts.
The corresponding right-hand panels feature the same two distributions
for the leading large-radius jet identified by {\sc Delphes}.
The vanishing tail of events for which the SIFT or {\sc Delphes} jet
fails ($\Delta R \leq 0.5$) relative to the partonic sum are vetoed here and throughout.

Central values and associated widths (standard deviations) are provided for the
$\mathcal{R}^A_B$ and  $\Delta R^A_B$ in TABLE~\ref{tab:JER_JAR}.
The SIFT variable-radius jet energy response is very regular,
systematically under-estimating the momentum of hard objects by about 6\%.
The {\sc Delphes} jet energy response shows more drift, transitioning
from positive values for soft objects to negative values for hard objects.
Widths of the two distributions are indistinguishable, with fluctuations amounting
to about 15\% in both cases, narrowing slightly at larger boosts.
It is anticipated from these observations that an energy calibration of
SIFT jets would be relatively straightforward, using standard techniques.
Angular performance of the two methodologies is identical,
with typical offsets and deviations both near one-tenth of a radian
(but less for hard objects and more for soft objects).

For comparison, we repeat this analysis in
FIG.~\ref{fig:JER_JAR_ZERO-ISR_GEN-JET} and TABLE~\ref{tab:JER_JAR_ZERO-ISR_GEN-JET},
using generator-level ({\sc Pythia8}) objects without detector effects
and suppressing the emission of initial-state radiation.
The distributions are substantially narrower in all cases,
with widths around a half or a third of the prior reference values.
The energy response is affected by both idealizations, but more so 
by the elimination of detector effects, whereas the angular response is
improved primarily by the elimination of initial-state radiation.
The most distinctive difference between the SIFT and large-radius
{\sc Delphes} jets at this level is that the former is bounded
from above by the partonic $\pt$, while the latter commonly
exceeds it.  The observed momentum excess is attributable to the capture of
radiation from the underlying event.  However, SIFT's filtering
stage is apparently more adept at rejecting this contaminant,
producing a reflection in the tail orientation that is
reminiscent of various approaches to grooming. 

Proceeding, we turn to attention the reconstruction of $W$-boson and top quark mass resonances,
as visualized in FIG.~\ref{fig:mass} at each of the four central simulated $\pt$ ranges. 
$M_W$ and $M_t$ are respectively recovered from di- and tri-jet samples,
by summing and squaring residual four-vector components after filtering.
A second $W$ reconstruction is obtained from decays of a $t$
by optimizing the combinatoric selection of two prongs from the ($N=3$) clustering flow.
An excess near $M_W \simeq 80$~GeV is apparent for $\pt \ge 200$~GeV,
although the top quark remains unresolved by the leading SIFT
jet at low boost, since the associated bottom is likely to be separately isolated.
The top-quark bump is clearly visible for $\pt \ge 400$~GeV,
though its centroid falls somewhat
to the left of $M_t \simeq 175$~GeV. 
The plotted distributions narrow at higher boost,
and substantially sharper peaks are observed for $\pt \ge 800$~GeV.
The systematic under-estimation of mass is consistent with effects
observed previously in the jet energy response, and it is similarly
expected to be improvable with a suitable calibration.

\section{Structure Tagging\label{sct:tagging}}

This section describes applications of the SIFT
algorithm related to structure and substructure tagging, including
discrimination of events with varying partonic multiplicities,
$N$-subjettiness axis-finding, and identification of heavy resonances.
We comparatively assess SIFT's performance on Monte Carlo
collider data against standard approaches, and quantify its
discriminating power with the aid of a Boosted Decision Tree (BDT).

Our first objective will be characterizing distinctive features in the evolution
of the SIFT measure $\delta_{AB}$ for events with different numbers of hard prongs.
We proceed by simulating pure QCD multi-jets representing LHC 
production of ($N=2\text{--}5$) gluons and/or light first-generation quarks ($u,\,d$).
Samples are generated at various partonic center-of-momentum energies, taking
$\sqrt{\hat{s}} = (\,100,\,200,\,400,\,800,\,1600,\,3200\,)$~GeV $\pm 20\%$.
In order to ensure that splittings are hard and wide
(corresponding to a number of non-overlapping
large-radius jets with more or less commensurate scales),
we require ($\,\pt \ge \sqrt{\hat{s}}\div16\,$) and ($\,\Delta R \ge 2.0\,$).
We also suppress initial-state radiation so that
consistent partonic multiplicities can be achieved,
and return to the use of generator-level ({\sc Pythia8}) objects. 
Other selections and procedures are carried forward.

In contrast to the examples in Section~\ref{sct:reconstruction},
the relevant showering products of these non-resonant systems
are not expected to be captured within a single variable large-radius jet.
Accordingly, we do not engage the isolation criterion
from Section~\ref{sct:filtering} for this application, but instead apply
exclusive clustering to the event as a whole with termination at ($N_{\rm exc}=1$).
Filtering of soft-wide radiation is retained, but values of $\delta_{AB}$
are registered only for the merger of objects surviving to the final state.
Relative to FIG.~\ref{fig:phase}, candidate object pairings
in the green and blue regions proceed to merge,
while the softer member is discarded for those in the red region.

FIG.~\ref{fig:evo} follows evolution of the SIFT measure
as it progresses from 8 down to 1 remaining objects.
Since we are considering entire events
(as opposed to a hadronic event hemisphere
recoiling off a neglected leptonic hemisphere),
attention is focused here on the upper four values of $\sqrt{\hat{s}}$
to promote closer scale alignment with prior examples.
Each of the simulated partonic multiplicities are tracked separately, represented
by the geometric mean of $\delta^N_{AB}$ over all samples at level $N$ in the clustering flow.

The relative change in the measure is larger when merging objects associated with distinct hard partons,
suggesting that the jettiness count is intrinsically imprinted on the clustering history.
Specifically, a steepening in the log-slope of the measure evolution occurs when 
transitioning past the natural object count\footnote{
If the isolated objects have dissimilar $\pt$, then
the discontinuity can be less severe, but the
increased slope may extend to ($N=1$).},
i.e., from the black markers to the white markers.
This supports the argument from Section~\ref{sct:njettree}
that the most useful halting criterion can sometimes be none at all.
In other words, it suggests that a determination of which objects should be considered resolved
might best be made after observing how those objects would otherwise recombine.

The orange bands in FIG.~\ref{fig:evo} mark the range of
$\delta^N_{AB}$ wherein all structures are fully reconstructed but not over-merged,
and the grey dashed line marks ($\delta_{AB} = 1$).
Independently of the collision energy, white points tend to land above this line and black points below it,
which helps to substantiate the isolation protocol from Section~\ref{sct:filtering}. 
Bulk features of the evolution curves are substantially similar
across the plotted examples, and practically identical above
the isolation cutoff, reflecting the scale-invariant design.
However, $\delta_{AB}^N$ ``starts'' with a smaller value from large $N$
for harder processes, and the orange band gap is expanded accordingly.
This is because the tighter collimation
(or smaller $m/\pt$, cf. Eq.~\ref{eq:drsqtilde})
associated with a large transverse boost induces
smaller values of the measure when constituents are merged.
Universality at termination is clarified by example,
considering a true dijet system with balanced $\pt$,
for which Eq.~(\ref{eq:measure_diffs}) indicates that
($\,\delta_{AB}^{N=1} \simeq \Delta R^2/2\,$).
This is consistent with illustrated values around 20 for
($\vert\Delta \eta\vert \simeq 6$) and ($\Delta \phi = \pi$). 

Our next objective will be to identify and test applications of SIFT
for resolving substructure within a narrowly collimated beam of radiation.
$N$-subjettiness represents one of the most prominent contemporary
strategies for coping with loss of structure in boosted jets.
In this prescription, one first clusters a large-radius jet, e.g.,
with ($R_0 \simeq 1.0$), which is engineered to contain all of the
products of a decaying parton such as a boosted top or $W$-boson.
For various hypotheses of the subjet count $(N = 1,\,2,\,3,\,\ldots)$,
a set of spatial axis directions are identified via a separate procedure, e.g., by reclustering
all radiation gathered by the large-radius jet with an exclusive
variant of the $\kt$ or Cambridge-Aachen
algorithms that forgoes beam isolation and forces explicit termination at $N$ jets.
One then computes a measure $\tau_N$ of compatibility with
the hypothesis, which is proportional to a sum over minimal angular separation $\Delta R$ from any of the
$N$ axes, weighted by the transverse momentum $\pt$ of each radiation component.  Maximal discrimination
of the subjet profile is achieved by taking ratios, e.g., $\tau_2/\tau_1$, or $\tau_3/\tau_2$.
This procedure will be our reference standard for benchmarking
SIFT's substructure tagging performance.

The SIFT $N$-subjet tree automatically
provides an ensemble of axis candidates at all
relevant multiplicities that are intrinsically suitable
for the computation of $N$-subjettiness.
We test this claim using the previously described
mono-, di-, and tri-jet event samples.  The axis candidates are
simply equal to the surviving objects at level $N$ in the
clustering flow.  However, this process references only  members
of the leading isolated large-radius jet,
rather than constituents of the event at large.

FIG.~\ref{fig:tau21} and FIG.~\ref{fig:tau32} respectively
exhibit distributions of $\tau_2/\tau_1$ and $\tau_3/\tau_2$
calculated in this manner at various transverse boosts. 
The intuition that $\tau_3/\tau_2$ should be effective
at separating $W$-bosons from top quarks, whereas
$\tau_2/\tau_1$ should be good for telling QCD monojets
apart from $W$'s is readily validated.
For comparison, FIG.~\ref{fig:tau2132Delphes} shows
corresponding distributions of the same two quantities
at the inner pair of $\pt$ scales,
as computed directly by {\sc Delphes} from the leading
($R_0 = 1.0$) Soft-Drop jet.
Although there are qualitative differences between the two sets
of distributions, their apparent power for substructure
discrimination is more or less similar.
This will be quantified subsequently with a BDT analysis.

Our final objective involves directly
tagging substructure with sequential
values of the SIFT measure.
Distributions of $\delta_{AB}$ at the ($N=1$) and ($N=2$) clustering stages
are plotted in FIG.~\ref{fig:siftdab1} and FIG.~\ref{fig:siftdab2} respectively,
for mono-, di-, and tri-jet samples at each of the four central simulated $\pt$ ranges. 
Clear separation between the three tested object multiplicities is observed,
with events bearing a greater count of partonic prongs tending
to aggregate at larger values of the measure, especially
after transitioning through their natural prong count.
We observe that superior substructure discrimination is achieved
by referencing the measure directly, rather than
constructing ratios in the fashion beneficial to $N$-subjettiness.
This is is connected to the fact that $\delta_{AB}$
is explicitly constructed as a ratio from the outset.

In order to concretely gauge relative performance
of the described substructure taggers,
we provide each set of simulated observables to a 
Boosted Decision Tree for training and validation.
BDTs are a kind of supervised machine learning
that is useful for discreet (usually binary) classification
in a high-dimensional space of numerical features.
In contrast to ``deep learning'' approaches based
around neural networks, where internal
operations are shrouded behind a ``black box''
and the question of ``what is learned''
may be inscrutable, the mechanics of a BDT
are entirely tractable and transparent.
While neural networks excel at extracting
hidden associations between ``low level'' features,
e.g., raw image data at the pixel level,
BDTs work best when seeded with ``high-level''
features curated for maximal information density.

At every stage of training, a BDT
identifies which feature and what transition value
optimally separates members of each class.
This creates a branch point on a decision tree, and
the procedure is iterated for samples following either fork.
Classifications are continuous, typically on the range ($0,1$),
and  are successively refined across
a deep stack of shallow trees, each ``boosted'' (reweighted)
to prioritize the correction of errors accumulated during prior stages.
Safeguards are available against over-training
on non-representative features, and scoring is always
validated on statistically independent samples.
We use 50 trees with a maximal depth of 5 levels,
a training fraction of $\sfrac{2}{3}$,
a learning rate of $\eta = 0.5$, and L2 regularization
with $\lambda = 0.1$ (but no L1 regularization). 
The BDT is implemented with {\sc MInOS}~\cite{aeacus},
using {\sc XGBoost}~\cite{Chen:2016:XST:2939672.2939785} on the backend.

\begin{figure*}[ht!]
\centering
\includegraphics[width=.54\textwidth]{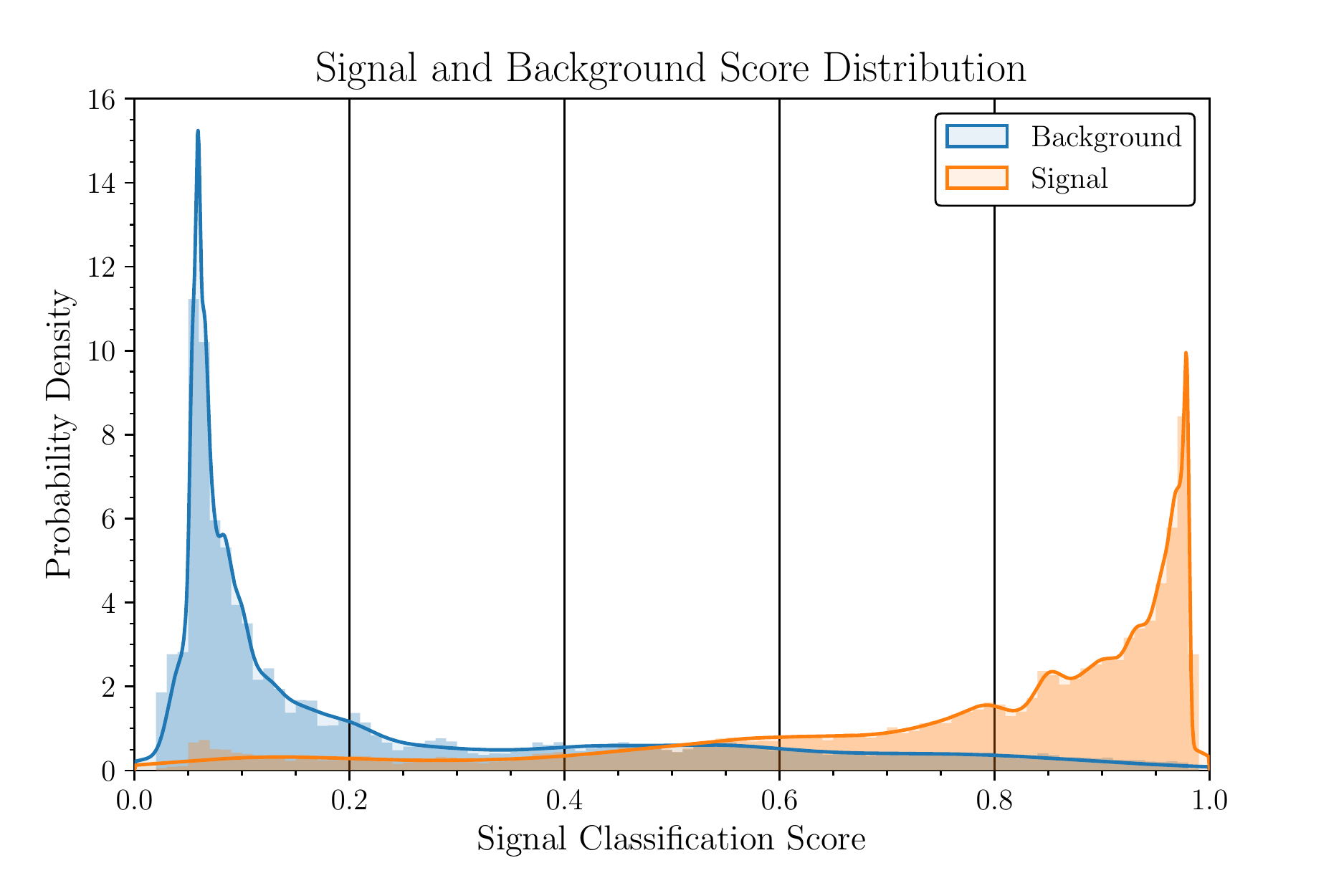} \hspace{12pt}
\includegraphics[width=.36\textwidth]{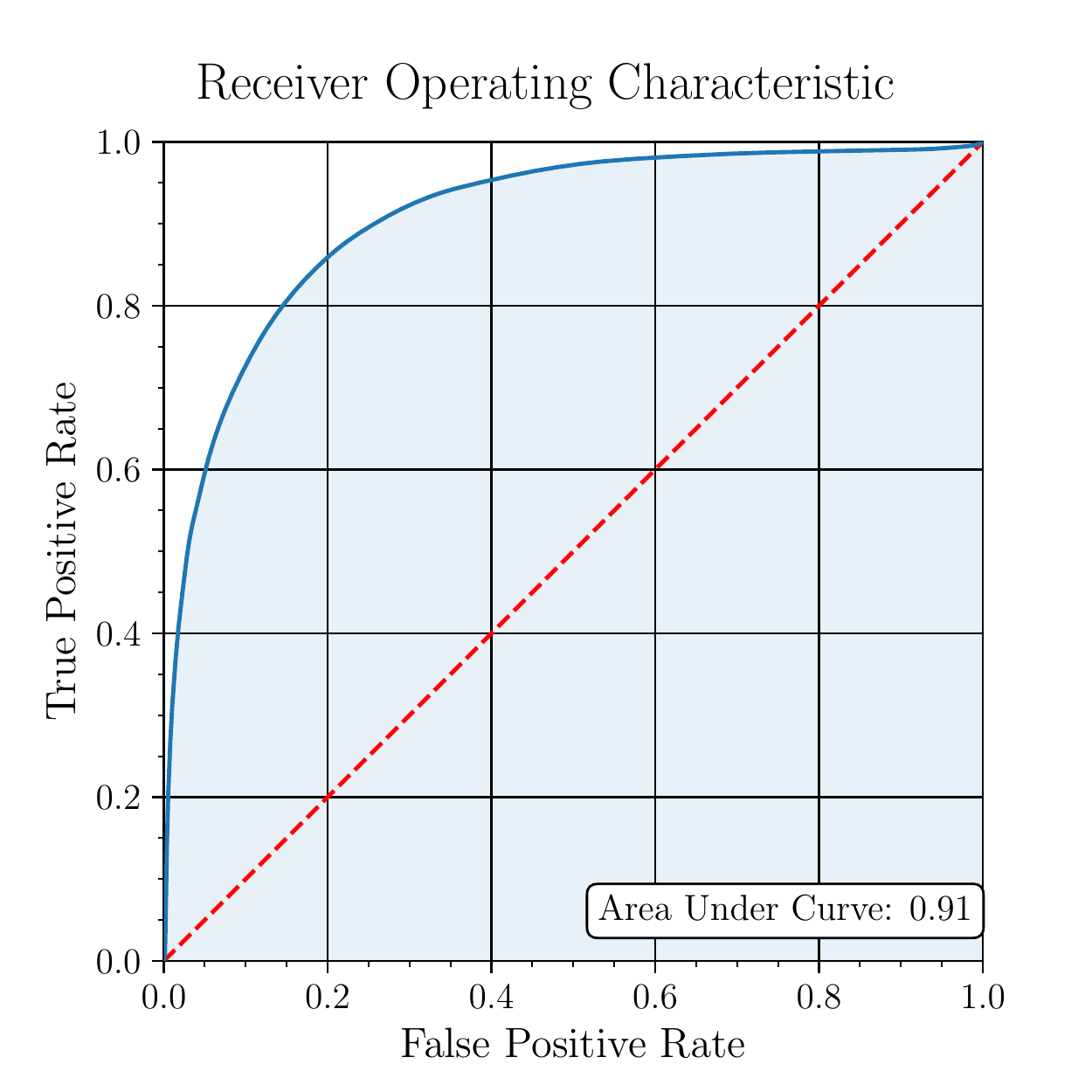}
\caption{\footnotesize
Left: Example distribution of BDT classification scores for the discrimination
of mono- and di-jet samples, respectively ``Background'' and ``Signal'', at $\pt=1600$~GeV.
Training features include the $\delta_{AB}^N$ for ($N=1\text{--}5$).
Right: Associated Receiver Operating Characteristic curve for true-positives versus false positives.}
\label{fig:distroroc}
\end{figure*}

The lefthand panel of FIG.~\ref{fig:distroroc} shows the
distribution of classification scores for mono-
and di-jet event samples at $\pt=1600$~GeV after
training on values of the SIFT measure $\delta_{AB}$
associated with the final five stages of clustering.
The two samples (plotted respectively in blue ``Background''
and orange ``Signal'') exhibit clear separation,
as would be expected from examination of the
second element of FIG.~\ref{fig:siftdab1}.
The underlying discretized sample data is represented
with translucent histograms, and the interpolation into
continuous distribution functions is shown
with solid lines.

The righthand panel of FIG.~\ref{fig:distroroc} shows the
associated Receiver Operating Characteristic (ROC)
curve, which plots the true-positive rate versus 
the false-positive rate as a function of a
sliding cutoff for the signal classification score.
The Area-Under-Curve (AUC) score, i.e., the fractional
coverage of the shaded blue region, is a good
proxy for overall discriminating power.
A score of $0.5$ indicates no separation,
whereas classifiers approaching
the score of $1.0$ are progressively ideal.

The AUC ($0.91$) from the example in FIG.~\ref{fig:distroroc}
is collected with related results in TABLE~\ref{tab:BDT_21}.
Separability of mono- and di-jet samples is quantified 
at each simulated range of $\pt$
while making various feature sets available to the BDT.
The first column uses
the four {\sc Delphes} $N$-subjettiness ratios
built from $\tau_1$ to $\tau_5$.
The next column references the same four ratios,
but as computed with objects and axes from the leading
SIFT $N$-subjet tree.  The third column provides the
BDT with the final ($N=1\text{--}5$)
values of the SIFT measure $\delta_{AB}^N$.
The last column merges information from the prior two.

The two $N$-subjettiness computations perform
similarly, but the fixed-radius {\sc Delphes} implementation
shows an advantage of a few points in the majority of trials.
The performance of $N$-subjettiness degrades at large boost,
losing more than 10 points between $\pt = 200$ and $\pt=1600$~GeV.
The SIFT $\delta_{AB}$ measure outperforms $N$-subjettiness
in five of six trials, with an average advantage (over trials)
of 7 points.  Its performance is very stable at larger boosts,
where it has an advantage of at least 10 points for $\pt \ge 800$~GeV.
Combining the SIFT measure with $N$-subjettiness generates
a marginal advantage of about 1 point relative to $\delta_{AB}$ alone.

TABLE~\ref{tab:BDT_32} represents a similar comparison of discriminating power
between resonances associated with hard 2- and 3-prong substructures.
$N$-subjettiness is less performant in this application, and the associated AUC scores
drop by around 6 points.  Performance of the SIFT measure degrades for soft events,
but it maintains efficacy for events at intermediate scales, and  
shows substantial improvement for $\pt \ge 800$~GeV, where its advantage
over $N$-subjettiness grows to around 20 points.

TABLE~\ref{tab:BDT_13} extends the comparison to resonances with hard 3- and 1-prong substructures.
The SIFT $N$-subjettiness computation is marginally
preferred here over its fixed-radius counterpart.
The $\delta_{AB}$ measure remains the best single discriminant by
a significant margin, yielding an AUC at or above $0.90$
for $\pt \ge 200$~GeV.

We conclude this section with a note on several
additional procedural variations that were tested.
Some manner of jet boundary enforcement (either via a fixed $R_0$
or the SIFT isolation criterion) is observed to be essential
to the success of all described applications.
Likewise, filtering of soft/wide radiation is vital to 
axis finding, computation of $N$-subjettiness,
and the reconstruction of mass resonances. 
Increasing the pre-clustering cone size from $0.01$ radians
to $0.1$ substantially degrades the performance of $N$-subjettiness,
whereas discrimination with $\delta_{AB}$ is more resilient to this change.
 
\section{Computability and Safety\label{sct:theory}}

This section addresses theoretical considerations associated
with computability of the SIFT observable $\delta_{AB}$.
Expressions are developed for various limits of interest.
Infrared and collinear safety is confirmed and deviations
from recursive safety are calculated and assessed.
It is suggested that SIFT's embedded filtering criterion
may help to regulate anomalous behaviors in the latter context,
improving on the Geneva algorithm.

Soft and collinear singularities drive the QCD matrix element
governing the process of hadronic showering.  In order to
compare experimental results against theoretical predictions
it is typically necessary to perform all-order resummation
over perturbative splittings.  In the context of computing
observables related to jet clustering,
the calculation must first be organized
according to an unambiguous parametric understanding of the
priority with which objects are to be merged, i.e.,~a
statement of how the applicable distance measure
ranks pairings of objects that are subject to the
relevant poles.  Specifically, cases of interest
include objects that are $i$) mutually hard but collinear,
$ii$) hierarchically dissimilar in scale,
and $iii$) mutually soft but at wide angular separation.
Pairs in the first two categories are likely to be physically
related by QCD, but those in the third are not.

In order to facilitate considerations of this type,
we outline here how the Eq.~(\ref{eq:measure_diffs}) measure
behaves in relevant limits.  The angular factor $\Delta \widetilde{R}_{AB}^2$
carries intuition for small differences by construction
(cf. Eq.~\ref{eq:drsqtilde}), and its dependence
on aggregated mass has been further clarified
in and around Eq.~(\ref{eq:rsqeffdiff}).
We turn attention then to the energy-dependent factor
$\epsilon^{AB}$, as expressed in Eq.~(\ref{eq:eps}), in
two limits of interest.  First, we take the case of
hierarchically dissimilar transverse energies, expanding
in the ratio ($\alpha \equiv \Et^A/\Et^B$) about $0$.
\begin{equation}
2\times\epsilon^{AB} \Rightarrow 2\,\alpha + \cdots
\label{eq:expandalpha}
\end{equation}
Next, we expand for small deviations
($\zeta \equiv \Et^A/\Et^B - 1$)
from matched transverse energies.
\begin{equation}
2\times\epsilon^{AB} \Rightarrow 1 - \frac{\zeta^2}{2} + \cdots
\label{eq:expandzeta}
\end{equation}

The SIFT algorithm is observed be be safe in the soft/infrared and collinear (IRC)
radiation limits, because the object separation measure explicitly
vanishes as ($\alpha \Rightarrow 0$) or ($\Delta R \Rightarrow 0$),
up to terms proportional to the daughter mass-squares
(cf. Eq.~\ref{eq:rsqeffdiff}) in the latter case.
This feature ensures that splittings at small angular
separation or with hierarchically distinct transverse energies
will be reunited during clustering at high priority.

With a clustering sequence strictly ranked by generated mass,
JADE was plagued by an ordering ambiguity
between the first and third categories described above,
which presented problems for resummation.
Geneva resolved the problem of mergers between
uncorrelated mutually soft objects at wide separation in the same
way that SIFT does, by diverging when neither
entry in the denominator carries a large energy.

\begin{figure}[ht!]
\centering
\includegraphics[width=.475\textwidth]{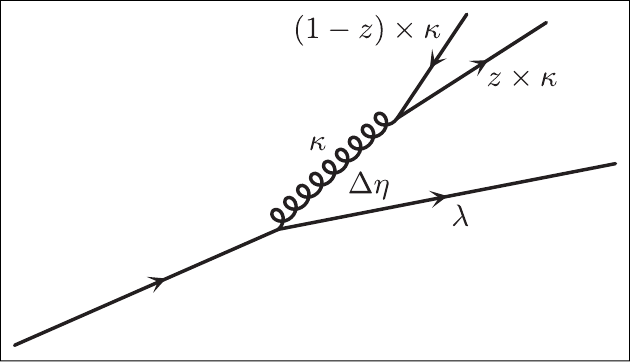}
\caption{\footnotesize
Hard object $\lambda$ emits a soft and collinear object $\kappa$ at separation $\Delta \eta$, which experiences
a secondary collinear splitting into a pair of objects with comparable hardness ($z \simeq \sfrac{1}{2}$). 
}
\label{fig:splitting}
\end{figure}

Yet, both SIFT and Geneva fall short of meeting the {\it recursive}
IRC safety conditions described in Ref.~\cite{Banfi:2004yd} at the measure level.
The challenge arises when a soft and collinear emission splits secondarily into
a very collinear pair, as first observed in Ref.~\cite{Catani:1992tm}.
This scenario is visualized in FIG.~\ref{fig:splitting}, with hard object $\lambda$
recoiling off a much softer emission $\kappa$ (having $\Et^\kappa/\Et^\lambda \ll 1$) at
a narrow pseudo-rapidity separation ($\Delta \eta \ll 1$).
Azimuthal offsets are neglected here for simplicity.  The secondary radiation
products are of comparable hardness for the situation of interest,
carrying momentum fractions ($z \simeq \sfrac{1}{2}$)
and ($1-z$) relative to their parent object $\kappa$.

It can be that the members of this secondary pair
each successively combine with the hard primary object rather than first
merging with each other. This ordering ambiguity implies that the value
of the measure $\delta_{AB}$ after the final recombination of all three
objects is likewise sensitive to the details of the secondary splitting.
However, the mismatch is guaranteed to be no more than a factor of 2.
Accordingly, this is a much milder violation than one
associated with a divergence (as for JADE).
While it does present difficulties for standard approaches to automated
computation, it does not exclude computation.

We conclude this section by sketching the relevant calculation, translating
results from Appendix~F of Ref.~\cite{Banfi:2004yd}
into the language of the current work.
The secondary splitting is characterized by a parameter
$\mu^2_\kappa \equiv \left(m_\kappa / \pt^{\hspace{0.75pt}\kappa}\right)^{2}$.
We further apply the limit ($\,\mu^2_\kappa \!\ll\! 1\,$), which implies
($\Et^{\hspace{0.75pt}\kappa} \simeq \pt^{\hspace{0.75pt}\kappa}$),
and treat the radiation products of object $\kappa$ as individually massless.
The value of the measure for merging these objects 
is readily computed with Eq.~(\ref{eq:siftmeasure}),
yielding ($\delta^z_{1-z} \simeq 2\,\mu^2_\kappa$).  Note that the
coefficient comes from the sum of squares in the measure
denominator, in the limit of a balanced splitting.
The merger of objects $\lambda$ and $\kappa$ (given prior recombination of the $\kappa$ products)
is best treated with Eq.~(\ref{eq:measure_diffs}),
defining $\mu^2_\lambda \equiv \left(m_\lambda / \pt^{\hspace{0.75pt}\lambda}\right)^{2}$,
and applying the limits in Eqs.~(\ref{eq:rsqeffdiff},~\ref{eq:expandalpha}), as follows:
\begin{equation}
\delta_\lambda^{\kappa} \simeq
\bigg( \frac{\Et^\kappa}{\Et^\lambda} \bigg)
\times
\bigg[ \, (\Delta \eta^\kappa_\lambda)^2 + \mu^2_\kappa + \mu^2_\lambda \,\bigg]
\label{eq:deltakl}
\end{equation}

However, if the remnants of object $\kappa$ instead combine in turn
with object $\lambda$, then the final value of the measure
(taking $z \ge \sfrac{1}{2}$ without loss of generality)
is instead ($\,\delta^z_\lambda \simeq z \times \delta_\lambda^{\kappa}\,$).
In addition to that overall rescaling,
the $\mu^2_\kappa$ term from Eq.~(\ref{eq:deltakl})
is absent from the analogous summation in this context.
If the $\kappa$ splitting is hierarchically imbalanced
(with $z \simeq 1$), then the secondary splittings are
less resistant to merging first and
the terminal measure value
becomes insensitive to the merging order. 

For balanced splittings, the physical showering history
will be ``correctly'' rewound if ($\,\delta^z_{1-z} < \,\delta^{1-z}_\lambda\,$).
But, there are no applicable kinematic restrictions enforcing that
condition, and SIFT's preference for associating objects
at dissimilar momentum scales actually constitutes a bias in the other direction.
On the other hand, the filtering criterion can
help curb potential ambiguities in this regime. 
Specifically, the energy scale factor ($\,2\times\epsilon^z_{1-z} \simeq 1\,$)
associated with products of object $\kappa$
will be subject here to the Eq.~(\ref{eq:expandzeta}) limit.
So, the ``wrong'' order of association is strongly correlated 
with cases where ($\,\epsilon^{1-z}_\lambda \ll 1\,$), since this
is generally required in order to overcome the tendency
for strict collinearity
($\Delta\widetilde{R}^z_{1-z} \ll \Delta\widetilde{R}^{1-z}_\lambda$)
in secondary splittings to commensurate momentum scales.
In turn, this enhances the likelihood that the Drop condition
from Eq.~(\ref{eq:iso_drop}) will veto any such merger.
A full clarification of the SIFT filtering criterion's
implications for recursive IRC safety is beyond
our current scope, but is of interest for future work.

\section{Conclusions and Summary\label{sct:conclusion}}

We have introduced a new scale-invariant jet clustering algorithm named SIFT (Scale-Invariant Filtered Tree)
that maintains the resolution of substructure for collimated decay products at large boosts.
This construction unifies the isolation of variable-large-radius jets,
recursive grooming of soft wide-angle radiation,
and finding of subjet-axis candidates into a single procedure.
The associated measure asymptotically recovers angular and kinematic behaviors of algorithms in the $\kt$-family,
by preferring early association of soft radiation with a resilient hard axis,
while avoiding the specification of a fixed cone size.
Integrated filtering and variable-radius isolation criteria resolve the halting problem
common to radius-free algorithms and block assimilation of soft wide-angle radiation.
Mutually hard structures are preserved to the end of clustering,
automatically generating a tree of subjet axis candidates
at all multiplicities $N$ for each isolated final-state object.
Excellent object identification and kinematic reconstruction are maintained without parameter tuning across
more than a magnitude order of transverse momentum scales, and superior resolution is exhibited for highly-boosted partonic systems.
The measure history captures information that is useful for tagging massive resonances,
and we have demonstrated with the aid of supervised machine learning that this observable has substantially more power
for discriminating narrow 1-, 2-, and 3-prong event shapes than the benchmark technique using $N$-subjettiness.
These properties suggest that SIFT may prove to be a useful tool for the continuing study of jet substructure.

\section*{Acknowledgements}
The authors thank Bhaskar Dutta, Teruki Kamon, William Shepherd, Andrea Banfi, Rok Medves,
Roman Kogler, Anna Albrecht, Anna Benecke, Kevin Pedro, Gregory Soyez, David Curtin, and Sander Huisman for useful discussions.
The work of AJL was supported in part by the UC Southern California Hub,
with funding from the UC National Laboratories division of the University of California Office of the President.
The work of DR was supported in part by DOE grant DE-SC0010813.
The work of JWW was supported in part by the National Science Foundation under Grant Nos. NSF PHY-2112799 and NSF PHY-1748958.
JWW thanks the Mitchell Institute of Fundamental Physics and Astronomy and the Kavli Institute for Theoretical Physics for kind hospitality.
High-performance computing resources were provided by Sam Houston State University.

\clearpage

\bibliography{bibliography}

\clearpage

\appendix

\begin{figure}[ht!]
\centering
\includegraphics[width=.45\textwidth]{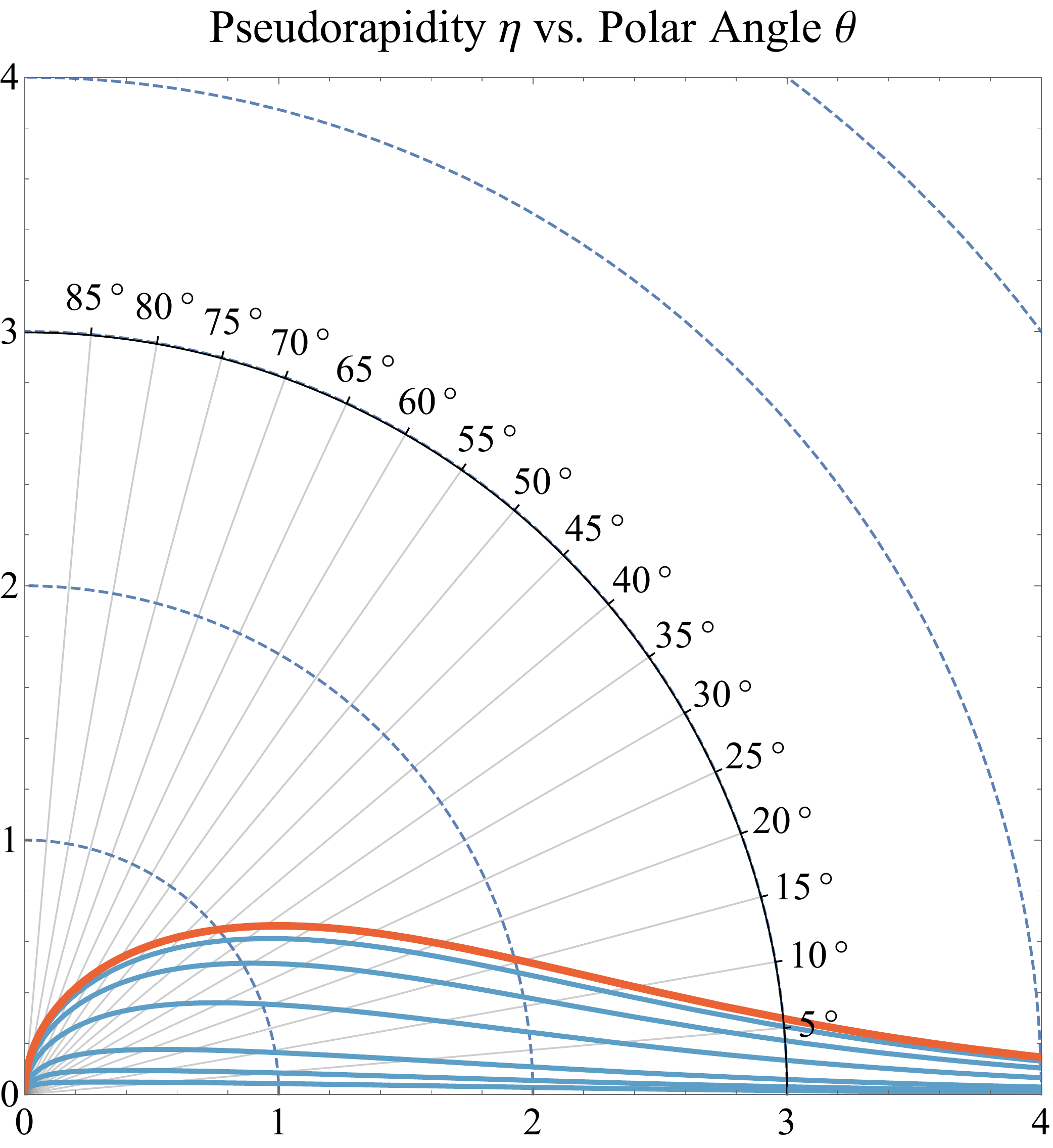}
\caption{\footnotesize
The pseudorapidity $\eta$ (bold, orange) is plotted as a function of the polar angle $\theta$.
For comparison, the longitudinal rapidity $y$ (fine, blue) is also shown for various values of ($m/\pt$),
equal to ($\sfrac{1}{2},1,2,5,10,20$) from top to bottom.}
\label{fig:eta}
\end{figure}

\section{Review of Collider Coordinates\label{sct:collidervars}}

This appendix provides a brief pedagogical review of hadron collider coordinates.
In this application, it is traditional to use a mapping from four-vector coordinates
$P_\mu \equiv \{E,\vec{p}\}$ into coordinates $\{\eta,\phi,\pt, m\}$ that are well-behaved
under Lorentz boosts along the longitudinal axis $\hat{z}$ of the beam.
The pseudo-rapidity $\eta$ (defined following) is a pure function of the zenith angle $\theta$.
\begin{equation}
\eta \equiv \frac{1}{2} \ln \left( \frac{\vert\,\vec{p}\,\vert + p_{z}}{\vert\,\vec{p}\,\vert - p_{z}} \right)
\equiv - \ln \tan \left(\frac{\theta}{2}\right)
\label{eq:eta}
\end{equation}

Forward (or backward) scattering correspond to $\eta$ equals plus (or minus) infinity,
while $\eta = 0$ represents entirely transverse scattering.
The azimuthal angle $\phi$ measures orientation about the $\hat{z}$ axis.
The transverse momentum $\pt$ is the magnitude of the 3-vector momentum
$\vec{p}$ projection perpendicular to the beam.
\begin{equation}
\pt \equiv \sqrt{p_x^2+p_y^2}
\end{equation}

The final parcel of kinematic information is the Lorentz-invariant mass $m$, which is especially important for jets representing the composition
of several lower level physical objects.  The 4-vector sum of individually massless objects may
accumulate cancellation in the three-momentum that manifests as a non-negligible mass-square in the invariant product.
\begin{equation}
p_\mu p^\mu \equiv E^2 - \vec{p}\cdot\vec{p} = m^2
\label{eq:minv}
\end{equation}

The quantity $\Delta R$ provides a radian-like measure of
the relativistic ``angular separation'' between an object pair.
\begin{equation}
\Delta R \equiv \sqrt{ {( \Delta \eta)}^2 + {( \Delta \phi )}^2}
\label{eq:deltar}
\end{equation}

The motivation for the definition in Eq.~(\ref{eq:eta}) is that differences $\Delta \eta$ in pseudorapidity
are ``nearly'' invariant under longitudinal boosts.  To be precise,
differences in the rapidity $y$ (defined following) are a strict longitudinal invariant
(as are the transverse coordinates $\pt$ and $\phi$),
and $y$ converges with $\eta$ in the relativistic $(m \ll \pt)$ limit.
\begin{align}
y &\equiv \frac{1}{2} \ln \Bigg( \frac{E + p_{z}}{E - p_{z}} \Bigg)
\label{eq:rap} \\
&= \ln \Bigg(\frac{\sqrt{\cosh^2 \eta + \frac{m^2}{\pt^2}}+\sinh \eta}{\sqrt{1+\frac{m^2}{\pt^2}}}\Bigg)
\nonumber
\end{align}

FIG.~\ref{fig:eta} provides a visualization of angular dependence of the pseudorapidity, along with
deviation from the rapidity for various amounts of transverse boost.
See Ref.~\cite{Gallicchio:2018elx} for an argument on the primacy of rapidity $y$.

\section{Software Implementations\label{sct:software}}

This appendix describes two publicly available implementations of the SIFT algorithm.
It also summarizes provided materials that facilitate reproduction of the key analyses in this manuscript.
Finally, it outlines challenges to and plans for future integration with the {\sc FastJet} contributions library.

The {\sc AEACuS}, {\sc RHADAManTHUS}, and {\sc MInOS} packages respectively
automate the processes of event analysis, visualization, and machine learning in a collider physics context.
These tools are distributed and maintained at GitHub~\cite{aeacus}
by JWW, and inquiries are welcome.
A quick-start tutorial (with a link to a video presentation) is
additionally available at Ref.~\cite{aeacustutorial}.
In brief, each of these programs is invoked from the
command line, and is interpreted with {\sc Perl 5.8+}.
Certain back-end features are implemented in {\sc Python 3}, importing
modules that include {\sc MatPlotLib} and {\sc XGBoost}, as noted previously.
All instructions regarding the computation of observables, application of event selection,
generation of plots, and application of machine learning are specified
in an associated card file, using a compact meta-language.
Control cards used in the preparation of this work are provided
with its source package on the arXiv.

These programs are designed for easy integration with the
standard {\sc MadGraph}/{\sc MadEvent}, {\sc Pythia8}, and {\sc Delphes} chain,
and cards are similarly included that document our approach
to event production, generator-level selections, and the simulation of
showering, hadronization, and detector effects.
{\sc AEACuS} auto-generates an extended-LHCO event record that
bundles parton, hadron, and detector-level information (with weights)
from the primary simulation chain for subsequent analysis.
It further facilitates a variety of jet clustering
and substructure applications, including an implementation of the SIFT
algorithm.  This usage is documented further in the example ``{\tt cut}'' cards.
The output is a space-delimited plain-text record of observables for each passing event,
which serves in turn as an input to subsequent plotting and machine-learning operations.

The second existing public implementation of the SIFT algorithm is in the
Mathematica notebook used here to produce jet clustering films and still frames.
That notebook is likewise distributed on GitHub, bundled with the tools described prior.
To run the notebook, simply place a suitable extended-LHCO event record into its working
directory and ``Evaluate Initialization Cells''.  User-adjustable parameters
are documented in the notebook, for stipulating the clustering algorithm
(members of the $\kt$ family are also supported), a cone size (as applicable),
any halting and filtering criteria, and whether ultra-soft ghost radiation should be included.
An {\tt .mp4} film is typically output within a few minutes on a laptop computer,
although running the notebook with ghost radiation enabled can be considerably more time consuming.

Additionally, a third implementation is planned that interfaces with {\sc FastJet},
facilitating broader exploration of the SIFT algorithm within existing workflows. 
The {\sc FastJet} implementation of Cacciari and Salam
is famous for reducing the na\"ive $\mathcal{O}(N^3)$ runtime required
for iterative pairwise jet clustering to $\mathcal{O}(N^2)$, or even to $\mathcal{O}(N\log N)$ in certain cases.
It assumes a pairwise clustering measure $\delta_{ij} \equiv \min [ f(\pt^i),f(\pt^j)]\times g(\Delta R_{ij})$
composed as the product of a kinematic function referencing the minimal value of $f(\pt^{i,j})$ over the pair
times a ``geometric'' measure $g(\Delta R_{ij})$ that is typically a power of the pairwise angular separation.
In contrast to the $\kt$-family of measures,
where factorization under the ``{\sc FastJet} Lemma''
reduces neighbor-finding to the $\Delta R$ plane,
the search for SIFT neighbors is necessarily active in (at least) three dimensions.
Moreover, the associated measure (cf. Eq.~\ref{eq:measure_diffs})
prioritizes \emph{furthest} neighbors along one of these axes ($u$).
Adaptations capable of confronting these unique challenges
while maintaining ``linearithmic'' $\mathcal{O}(N \log N)$ scaling
will be presented in a future work.

\end{document}